\documentclass{article}
\usepackage{amssymb}
\usepackage{amsmath}

\begin{document}

\textbf{The Lorentz transformations of the vectors }$E$, $B$, $P$, $M$
\textbf{and}

\textbf{the external electric fields from a stationary }

\textbf{superconducting wire with a steady current and from }

\textbf{a stationary permanent magnet}\bigskip

Tomislav Ivezi\'{c}

Ru\mbox
{\it{d}\hspace{-.15em}\rule[1.25ex]{.2em}{.04ex}\hspace{-.05em}}er Bo\v{s}%
kovi\'{c} Institute, P.O.B. 180, 10002 Zagreb, Croatia

E-mail: ivezic@irb.hr\textbf{\bigskip \bigskip }

In the first part of this paper we review the fundamental difference between
the usual transformations of the three-dimensional (3D) vectors of the
electric field $\mathbf{E}$, the magnetic field $\mathbf{B}$, the
polarization $\mathbf{P}$, the magnetization $\mathbf{M}$ and the Lorentz
transformations of the 4D geometric quantities, vectors\textbf{\ }$E$, $B$, $%
P$, $M$, with many additional explanations and several new results. In the
second part, we have discussed the existence of the electric field vector $E$
outside a \emph{stationary} superconducting wire with a \emph{steady current}
and also different experiments for the detection of such electric fields.
Furthermore, a fundamental prediction of the existence of the external
electric field vector $E$ from a \emph{stationary} permanent magnet is
considered. These electric fields are used for the resolution of the
\textquotedblleft charge-magnet paradox\textquotedblright\ with 4D geometric
quantities for a qualitative explanation of the Aharonov-Bohm effect in
terms of fields and not, as usual, in terms of the vector potential and for
a qualitative explanation that the particle interference is not a test of a
Lorentz-violating model of electrodynamics according to which a magnetic
solenoid generates not only a static magnetic field but also a static
electric field.\bigskip \bigskip

\noindent \textbf{1.} \textbf{Introduction\bigskip }

\noindent \textit{1.1 About this paper\bigskip }

Both, in the prerelativistic physics and in Einstein's formulation of
special relativity (SR) [1] the electric and magnetic fields are represented
by the 3-vectors $\mathbf{E(r,}t\mathbf{)}$ and $\mathbf{B(r,}t\mathbf{)}$.
The notation is as in [2], i.e. $\mathbf{E}$ and $\mathbf{B}$ are called
3-vectors and they are designated in boldface type. In the whole physical
literature after [1] the usual transformations of the 3-vectors $\mathbf{E}$
and $\mathbf{B}$,\ the last equations in \S 6., II. Electrodynamical Part,
[1] or, e.g., Eqs. (11.148) and (11.149) in [2], i.e., Eq. (\ref{JCB}) here,
are always considered to be the relativistically correct Lorentz
transformations (LT) (boosts) of $\mathbf{E}$ and $\mathbf{B}$. Here, in the
whole paper, under the name LT we shall only consider boosts. They are first
derived by Lorentz [3] and Poincar\'{e} [4] (see also two fundamental Poincar%
\'{e}'s papers with notes by Logunov [5]) and independently by Einstein [1]
and subsequently derived and quoted in almost every textbook and paper on
relativistic electrodynamics. Henceforward, these usual transformations of
the 3-vectors will be called the Lorentz-Poincar\'{e}-Einstein
transformations (LPET), according to physicists who discovered them. The
main feature of the LPET of $\mathbf{E}$ and $\mathbf{B}$ is that \emph{the
components of the transformed} $\mathbf{E}^{\prime }$ \emph{are expressed by
the mixture of components of} $\mathbf{E}$ \emph{and} $\mathbf{B}$, \emph{%
and similarly for} $\mathbf{B}^{\prime }$, Eq. (11.148) in [2]. The electric
field $\mathbf{E}$ in one frame is \textquotedblleft seen\textquotedblright\
as slightly changed electric field $\mathbf{E}^{\prime }$ and an \emph{%
induced magnetic field} $\mathbf{B}^{\prime }$ in a relatively moving
inertial frame.

However, it has recently been proved both in the tensor formalism and in the
geometric algebra formalism [6-11] that \emph{these LPET ARE NOT the LT}.
They drastically differ from the LT of the relativistically correct 4D
geometric quantities, which represent the electric and magnetic fields. In
contrast to the LPET, \emph{the correct LT always transform the 4D algebraic
object representing the electric field only to the electric field; there is
no mixing with the magnetic field. }This fundamental difference between the
LPET of the 3-vectors and the LT of the 4D geometric quantities is
considered in the first part of this paper.

It is worth mentioning that some experimentally verifiable consequences of
that fundamental difference between the LPET and the LT have been examined
in several papers. Thus, it is shown in [7] that the conventional theory
with the 3D $\mathbf{E}$ and $\mathbf{B}$ and their LPET (\ref{JCB}) yields
different values for the motional emf $\varepsilon $ for relatively moving
inertial observers, $\varepsilon =UBl$ and $\varepsilon =\gamma UBl$,
whereas the approach with 4D geometric quantities and their LT (\ref{LTE1}),
i.e., (\ref{LTE2}), always yields the same value for $\varepsilon $, which
is defined as a Lorentz scalar, $\varepsilon =\gamma UBl$. This result is
very strong evidence that the usual approach is not relativistically
correct. If the experimentalists find a way to measure the emf $\varepsilon $
for the considered problem of a conductor moving in a static magnetic field,
not only for small velocities, $U\ll c$, they should see that in the
laboratory frame $\varepsilon =\gamma UBl$ and not simply $\varepsilon =UBl$%
. That problem is of a considerable importance in practice. A similar
discussion was presented for the Faraday disk in [8]. In [12, 13] the
Trouton-Noble paradox is considered. It is shown that in the geometric
approach with 4D quantities \emph{the 4D torques} will not appear for the
moving capacitor if they do not exist for the stationary capacitor, which
means that with 4D geometric quantities the principle of relativity is
naturally satisfied and there is not the Trouton-Noble paradox. The same
conclusion holds in the low-velocity approximation $\beta \ll 1$, or $\gamma
\simeq 1$. Very similar paradox to the Trouton-Noble paradox is Jackson's
paradox. It is discussed in detail in [14]; the second paper is a simpler,
more pedagogical, version of the first one. There, in [14], it is also shown
that there is no paradox in the approach with 4D geometric quantities and
their LT.

The most important experimentally verifiable consequence of the difference
between the LPET and the LT refers to the existence of the electric field
vector outside a \emph{stationary} superconducting wire with a \emph{steady
current} and also outside a \emph{stationary} permanent magnet. The second
part of this paper, Secs. 7.1, 7.2, 8, is devoted to that problem. In Sec.
9.2, these electric fields are used for the resolution of the
\textquotedblleft charge-magnet paradox\textquotedblright\ in terms of 4D
geometric quantities without introducing some \textquotedblleft
hidden\textquotedblright\ quantities and without changing the expression for
the Lorentz force, but as a 4D geometric quantity, like the expression for
the Lorentz force density $k_{L}$ (\ref{kl}). Furthermore, in Sec. 10, these
electric fields are used for a qualitative explanation of the Aharonov-Bohm
effect in terms of fields and not, as usual, in terms of the vector
potential and in Sec. 11, for a qualitative explanation that the particle
interference is not a test of a Lorentz-violating model of electrodynamics
according to which a magnetic solenoid generates not only a static magnetic
field but also a static electric field.

The geometric approach to special relativity that is used in, e.g., [6-14],
is called the invariant special relativity (ISR). In the ISR, it is
considered that in the 4D spacetime the 4D geometric quantities are
well-defined both theoretically and \emph{experimentally;} they have an
independent physical reality. The principle of relativity is automatically
satisfied if the physical laws are expressed in terms of the 4D geometric
quantities. It is not so in the SR [1] in which it is considered that the 3D
quantities have an independent physical reality. There, the principle of
relativity is postulated and it is supposed that it holds for physical laws
expressed in terms of the 3D quantities, e.g., Maxwell's equations written
in terms of the 3-vectors $\mathbf{E}$ and $\mathbf{B}$. In the ISR,
physical quantities are represented by the abstract, coordinate-free, 4D
geometric quantities. In the papers [6-14] these quantities are treated as
either tensors as geometric objects, e.g., in [6, 8, 10], or, multivectors
in the geometric algebra formalism, e.g., in [7-9], [11-14]. If some basis
has been introduced, these coordinate-free quantities are represented as 4D
coordinate-based geometric quantities (CBGQs) comprising both components and
\emph{a basis}. Every 4D CBGQ is invariant under the passive LT; the
components transform by the LT and the basis by the inverse LT leaving the
whole CBGQ unchanged. This is the reason for the name ISR. The invariance of
a 4D CBGQ under the passive LT reflects the fact that such mathematical,
invariant, 4D geometric quantity represents \emph{the same physical quantity}
for relatively moving inertial observers, see, e.g., Eqs. (\ref{ein}), (\ref%
{enc}), (\ref{ms}) and (\ref{dv1}) here. Hence, it can be stated that in the
ISR only quantities that do not change upon the passive LT have an
independent physical reality, both theoretically and \emph{experimentally}.
In contrast to it the SR [1] deals with the Lorentz contraction, the time
dilation and the LPET of the 3D vectors $\mathbf{E}$ and $\mathbf{B}$.
However, e.g., the rest length and the Lorentz contracted length are not the
same 4D quantity for relatively moving observers, since the transformed
length $L_{0}(1-\beta ^{2})^{1/2}$ is different than the rest length $L_{0}$%
, see, e.g., Eq. (\ref{contr}) in Appendix. Rohrlich [15] named the Lorentz
contraction and other transformations which do not refer to the same 4D
quantity as the \textquotedblleft apparent\textquotedblright\
transformations (AT), whereas the transformations which refer to the same 4D
quantity as the \textquotedblleft true\textquotedblright\ transformations,
e.g., the LT. Hence, the other name for the ISR is the \textquotedblleft
True transformations relativity\textquotedblright\ (\textquotedblleft TT
relativity\textquotedblright ), which is used, e.g., in [16, 17]. As proved
in [6-11] and exposed here in Secs. 3 - 3.2 and Sec. 6, \emph{the LPET are
also the AT and not the LT.} In the 4D spacetime, as shown in detail in [16,
17], instead of the Lorentz contraction and the time dilation one has to
consider the 4D geometric quantities, the distance vector $l_{AB}$, Eq. (\ref%
{dv}) here, and the spacetime length, Eq. (\ref{sl}) here, which properly
transform under the LT, see Appendix here.\bigskip

\noindent \textit{1.2 An outline of this paper\bigskip }

An outline of the present paper is as follows. In Sec. 2, a short review of
the geometric algebra formalism is presented. For more detail see [18]. An
important result from [19] is mentioned in that section. Namely, \emph{what
is essential for the number of components of a vector field is the dimension
of its domain. Hence, the usual time-dependent} $\mathbf{E(r,}t\mathbf{)}$, $%
\mathbf{B(r,}t\mathbf{)}$ \emph{cannot be the 3-vectors, since they are
defined on the spacetime.} They are correctly defined geometric quantities,
e.g., vectors (4-vectors in the usual notation) $E(x)$, $B(x)$, where $x$ is
the position vector.

Then, in Secs. 3 - 3.2, we discuss the traditional derivation of the LPET of
the 3-vectors $\mathbf{E}$ and $\mathbf{B}$, Eq. (\ref{JCB}), and $\mathbf{P}
$ and $\mathbf{M}$, Eq. (\ref{ps}), or Eq. (\ref{psa}). As already stated,
the main feature of the LPET is that, e.g., \emph{the transformed} $\mathbf{E%
}^{\prime }$ \emph{is expressed by the mixture of the 3-vectors} $\mathbf{E}$
\emph{and} $\mathbf{B}$, \emph{and similarly for} $\mathbf{B}^{\prime }$. As
shown in Sec. 3.1, for the derivation of (\ref{JCB}) one first makes the
identification of the six independent components of $F^{\alpha \beta }$ with
six components of the 3-vectors $\mathbf{E}$ and $\mathbf{B}$, Eq. (\ref{ieb}%
). Then, it is simply argued that six independent components of $F^{\prime
a\beta }$ are the \textquotedblleft Lorentz transformed\textquotedblright\
components $E_{i}^{\prime }$ and $B_{i}^{\prime }$, Eq. (\ref{eb2}), i.e.,
\emph{the LPET of the components of} $\mathbf{E}$ \emph{and} $\mathbf{B}$
\emph{are derived assuming that they transform under the LT as the
components of} $F^{\alpha \beta }$ \emph{transform}, Eq. (11.148) in [2].
However, it is shown in that section that the identifications (\ref{ieb})
and (\ref{eb2}) depend on the chosen synchronization and that they are
meaningless for some nonstandard synchronization, e.g., the
\textquotedblleft radio\textquotedblright\ synchronization, see Eqs. (\ref%
{Fr1}) and (\ref{FEr}), which means that the LPET (\ref{JCB}) (and (\ref{ps}%
), or (\ref{psa})) are not the relativistically correct LT. In Sec. 3.2 the
same discussion is presented for the derivation of the LPET of $\mathbf{P}$
and $\mathbf{M}$, Eq. (\ref{ps}), or Eq. (\ref{psa}).

In Sec. 4, the definitions of vectors\textit{\ }$E$, $B$, Eqs. (\ref{E2})
and (\ref{E1}), and $P$, $M$, Eqs. (\ref{M1}) and (\ref{M2}), in terms of $F$%
, $v$ and $\mathcal{M}$, $u$, respectively, are examined; $v$ is the
velocity vector of the observers who measure $E$ and $B$ fields, while $u$
is the velocity vector of a moving medium. It is visible from (\ref{E1})
that \emph{in mathematically correct definitions the vectors} $E(x)$ \emph{%
and} $B(x)$ \emph{are derived from }$F$ \emph{AND} $v$, i.e.,\emph{\ they}
\emph{are defined with respect to the observer}. Similarly, it is visible
from (\ref{M2}) that $P$ and $M$ depend not only on $\mathcal{M}$ but on $u$
as well. Furthermore, the basic Lorentz invariant field equation for vacuum
with $F$, Eq.(\ref{MEF}), is written in terms of $E$ and $B$, Eq. (\ref{ebf}%
), i.e., Eqs. (\ref{A1}) and (\ref{A2}). The generalization of these field
equations to the electromagnetic field equations for moving media is
presented in [20] and also briefly considered in this section. The
generalization of (\ref{MEF}) to a moving medium is obtained simply
replacing $F$ by $F+\mathcal{M}/\varepsilon _{0}$, which yields Eqs. (\ref%
{F4}), the primary equations for the electromagnetism in moving media with
bivectors $F(x)$ and $\mathcal{M}(x)$. Then, these equations are written
with vectors $E(x)$, $B(x)$, $P(x)$ and $M(x)$, Eqs. (\ref{Ej})\ and (\ref%
{MP2}).\ As stated in [20], Eq. (\ref{F4}), i.e.,\ Eqs. (\ref{Ej})\ and (\ref%
{MP2}),\ comprise and generalize all usual Maxwell's equations with
3-vectors for moving media. The equations (\ref{Ej})\ and (\ref{MP2})
contain both the velocity vector $u$ of a moving medium and the velocity
vector $v$ of the observers who measure $E$ and $B$ fields. They are first
reported in [20] and do not appear in the previous literature.

In Sec. 5, the LT of vectors $E$ and $B$, as 4D geometric quantities, are
examined and compared with Minkowski's results. Note that Minkowski,
Sec.11.6 in his famous paper [21], was the first who introduced vectors (in
the usual notation 4-vectors) of the electric and magnetic fields and
correctly defined their LT. It is shown that the LT of vectors $E$ and $B$
are obtained by a mathematically correct procedure in the 4D spacetime. As
explained in [11], \emph{Minkowski, in Sec. }11.6\emph{\ in }[21]\emph{,
showed that both factors of the vector} $E$, \emph{as the product of one
bivector and one vector, has to be transformed by the LT. }That fundamental
Minkowski's result is reinvented and generalized in [6-11]. Thus, $E$ from (%
\ref{E1}), $E=F\cdot v/c$, transforms under the active LT, e.g., Eqs. (\ref%
{LTR}) and (\ref{RM}), in such a manner that both $F$ and the velocity of
the observer $v$ are transformed by the LT, Eq. (\ref{LM3}). These
coordinate-free LT yield how vector $E$ transforms under the active LT, Eq. (%
\ref{LTE1}). If these transformations are written in the standard basis then
the transformations of the components are obtained, Eq. (\ref{LTE2}). The
most important result is that \emph{under the relativistically correct LT
the electric field vector }$E$ \emph{transforms again to the electric field
vector }$E^{\prime }$\emph{; there is no mixing with the magnetic field} $B$.

In Sec. 6, the LPET of the components of the 3-vectors $\mathbf{E}$ and $%
\mathbf{B}$\ are retrieved using the geometric algebra formalism, i.e., the
4D geometric quantities. If in the transformation of $E=F\cdot v/c$ \emph{%
only} $F$ \emph{is transformed by the LT}, \emph{but not the velocity of the
observer }$v$, then the LPET of the electric field vector $E$ are obtained,
Eqs. (\ref{EP1}) and (\ref{J1}). These coordinate-free LPET are also written
in the standard basis, Eq. (\ref{J2}), and it is visible that \emph{the
components of the transformed} $E_{F}^{\prime }$ \emph{are expressed by the
mixture of components of} $E$ \emph{and} $B$. As seen from Eq. (\ref{B}),
the same result is obtained for the magnetic field vector $B$. The
comparison of the relation for the LPET of the components of $E$ (\ref{J2})
with the LPET for the components $E_{x,y,z}$ of the 3-vector $\mathbf{E}$,
which are given, e.g. by Eq. (11.148) in [2], explicitly shows that \emph{%
they are exactly the same transformations. }But, the LPET of the vector $E$,
(\ref{EP1}) and (\ref{J1}), are obtained by a mathematically incorrect
procedure (\emph{only} $F$ \emph{is transformed}), which means that \emph{%
they are not the relativistically correct LT} and consequently, contrary to
the general opinion, \emph{the LPET of the 3-vector} $\mathbf{E}$ (and $%
\mathbf{B}$, $\mathbf{P}$, $\mathbf{M}$) \emph{ARE\ NOT THE\ LT but the AT}.

In Sec. 7.1, the second-order electric fields outside a stationary conductor
with steady current are considered. In the usual approaches, e.g., [22-26],
there is a magnetic field 3-vector outside a stationary (superconducting)
wire with steady current, but, according to the LPET (\ref{JCB}), there are
both, the slightly changed magnetic field and an induced \emph{second-order
external electric field 3-vector for the same but moving wire with steady
current. }Similarly, e.g. [22, 23], it is argued that a neutral stationary
current loop has only a magnetic moment 3-vector. According to the LPET for
the 3-vectors $\mathbf{p}$ and $\mathbf{m}$, which are the same as (\ref{ps}%
), that current loop acquires an electric dipole moment (\ref{plp}) as well,
if it is moving with uniform 3-velocity $\mathbf{U}$ ($\mathbf{\beta =U}/c$%
). However, \emph{in the 4D spacetime, the electric and magnetic fields and
the dipole moments are not the 3-vectors but the 4D vectors} $E$, $B$, $p$, $%
m$, \emph{which transform under the LT} (\ref{LTE1}), \emph{i.e.,} (\ref%
{LTE2}) \emph{and not under the LPET of the 3-vectors} (\ref{JCB}) \emph{and}
(\ref{ps}), \emph{or} (\ref{psa}). The electric field vector $E$ (the same
for $B$, $p$, $m$) transforms by the LT again to the electric field vector
without mixing with $B$ and therefore if $E$ exists for a moving wire with a
steady current, or a moving current loop, it must exist for the same but
\emph{stationary} wire or current loop. The determination of the electric
field vector for the stationary current-carrying conductor is investigated
in detail in [27] and here it is briefly reviewed with some additional
explanations. The expression for the current density vector in the rest
frame of the wire is given by Eq. (\ref{jotmi}), whereas the expression for
the external second-order electric field vector of the stationary wire with
steady current is given by Eq. (\ref{eovi}). Observe that in the second
paper in [27], the incorrect quadrupole field of the stationary current loop
from the published version is replaced by the dipole field. Therefore,
henceforward, if referred to [27] I mean that the corrected version has to
be taken into account.

In Sec. 7.2, the experiments for the detection of the second-order electric
fields outside a stationary conductor with steady current are discussed.
This is a new consideration that is not reported in my previous papers. In
the measurements [28, 29] a direct contact with the superconducting coil is
used and because of that they cannot either support or disprove the theory
presented in [27]. In [30], a non-contact method of measuring is used, but
in order to \textquotedblleft see\textquotedblright\ the external
second-order electric fields the coil used in their experimental setup would
need to be a superconducting coil. Recently, [31], the most promising method
is proposed and it deals with cold ions. The theory presented in [31] is
essentially the same as in my paper [32]. However, both theories, [32] and
[31], explicitly use the Lorentz contraction in the derivation of the
expression (\ref{eovi}) for the external second-order electric field and as
such they are not the relativistically correct theories. In [27] the
relativistically correct theory is presented, but it seems that Folman,
[31], either was not aware of [27] or more believed in the usual approach
with the Lorentz contraction and the LPET (\ref{JCB}) than to the
mathematically correct 4D geometric approach.

In Sec. 8, an essentially new prediction is presented that \emph{a
stationary permanent magnet possesses an intrinsic polarization, which
induces the external electric field.} This prediction is supported both in
the usual Amp\`{e}rian approach in which a permanent magnet is an assembly
of small current loops and also using the recent fundamentally new results
from [33], i.e., the relations (\ref{gu}) and (\ref{dm}), which show that
any fundamental particle has not only the intrinsic magnetic dipole moment
(MDM) $m$ but also the intrinsic electric dipole moment (EDM) $p$. Then, in
the same way as the MDMs determine the magnetization $M$ of a stationary
permanent magnet the EDMs determine its polarization $P$, which induces an
electric field outside a permanent magnet (moving or \emph{stationary}). We
suggest that the experimental setup from [31] could be also used for the
measurement of that electric field outside a stationary permanent magnet.

In Secs. 9 - 9.2 the \textquotedblleft charge-magnet
paradox\textquotedblright\ from [34] is discussed in detail together with
the highlight of it from [35] and different resolutions\ from [36]. The
paradox, that in a static electric field a MDM $\mathbf{m}$ is subject to a
torque $\mathbf{N}$ in some frames and not in others is stated to be
resolved\ in [34] replacing the conventional Lorentz force (density), Eq.
(5) in [34], by Einstein-Laub law, Eq. (6) in [34], which predicts no torque
in all frames. In [34], all quantities $\mathbf{E}$, $\mathbf{B}$, $\mathbf{P%
}$, $\mathbf{M}$, $\mathbf{F}$, $\mathbf{N}$, etc. are the 3-vectors and
their transformations are given by the LPET (\ref{JCB}) for $\mathbf{E}$ and
$\mathbf{B}$ and (\ref{ps}), or (\ref{psa}) for $\mathbf{P}$ and $\mathbf{M}$%
, and the same for EDM $\mathbf{p}$ and MDM $\mathbf{m}$, including (\ref%
{plp}). All other approaches from [36] also deal with the 3D quantities and
their AT and often introduce some \textquotedblleft
hidden\textquotedblright\ quantities. Moreover, the resolutions from [34-36]
\emph{depend on the chosen synchronization} and they are meaningless \emph{%
if only the Einstein synchronization is replaced by the} \textquotedblleft
\emph{radio}\textquotedblright\ \emph{synchronization}, as can be concluded
from Eqs. (\ref{Fr1}) and (\ref{FEr}) in Sec. 3.1. This means that from the
viewpoint of the ISR the resolutions from [34-36] are not relativistically
correct resolutions. All these treatments from [34-36] are objected in Sec.
9.1.

The treatment of the interaction between a static electric field and a
permanent magnet that is presented in Sec. 9.2 differs in two important
respects relative to [34-36]. As explicitly shown in the very similar
treatments of the Trouton-Noble paradox [12, 13] and Jackson's paradox [14]
in the approach with 4D geometric quantities there is no paradox and the
same is shown for the \textquotedblleft charge-magnet
paradox.\textquotedblright\ Every 4D geometric quantity, e.g., the Lorentz
force vector, the torque bivector is the same quantity for all relatively
moving inertial observers and for all bases chosen in them. However, in
contrast to the Trouton-Noble paradox [12, 13] and Jackson's paradox [14] in
which the torque bivector is zero in all frames, in this case the torque
bivector $N$ is different from zero even in the common rest frame of the
considered charge and magnet from [34]. The reason for that important
difference is that according to the 4D geometric approach, i.e., the ISR, a
stationary permanent magnet possesses not only the magnetization vector $M$
but the polarization vector $P$ as well. In the formulation with the torque
bivector $N$ the relations (\ref{tn}), (\ref{m}) and (\ref{enc}) show that
the principle of relativity is naturally satisfied and that there is no
paradox. As can be seen from (\ref{tst}), (\ref{st}) and (\ref{RMP}), (\ref%
{ms}), the same holds for the formulation with the \textquotedblleft
space-space\textquotedblright\ torque $N_{s}$ and the \textquotedblleft
time-space\textquotedblright\ torque $N_{t}$. It is visible from (\ref{tst})
and (\ref{st}) that the torque $N_{s}$ is always determined by the
interaction of the EDM $p$ of the permanent magnet and the electric field $E$%
, whereas $N_{t}$ is determined by the interaction of the MDM $m$ and $E$.

In Sec. 10, it is suggested that the existence of the electric fields from a
stationary permanent magnet yields the possibility for the explanation of
the Aharonov-Bohm effect in terms of fields and not, as usual, in terms of
the vector potential.

In Sec. 11, the results from [37] are discussed. In [37] it is argued that a
static electric field generated by a magnetic solenoid is due to
Lorentz-violating electrodynamics and that the particle interference is a
test of that Lorentz-violating model of electrodynamics. However, according
to Secs. 7.1 and 8 from this paper there is a static electric field
generated by a magnetic solenoid, see Eq. (\ref{eovi}), that is obtained in
a completely Lorentz-covariant approach with the 4D geometric quantities,
which means that the particle interference is not a test of a
Lorentz-violating model of electrodynamics.

In Sec. 12, a discussion and conclusions are presented.

In Appendix, it is briefly described the essential difference between the 4D
geometric approach, i.e., the ISR, with the spacetime length and Einstein's
definition of the Lorentz contraction for a moving rod. \textit{\bigskip }

\noindent \textbf{2. A brief summary of geometric algebra}\bigskip

The calculations in this paper are performed in the geometric algebra
formalism. Physical quantities are represented by the abstract multivectors,
or, if some basis has been introduced, these abstract quantities are
represented as 4D CBGQs comprising both components and \emph{a basis}. For
simplicity and for easier understanding mainly the standard basis, described
below, will be used.

Here, for the reader's convenience, we provide a brief summary of geometric
algebra. Usually Clifford vectors are written in lower case ($a$) and
general multivectors (Clifford aggregate) in upper case ($A$). The space of
multivectors is graded and multivectors containing elements of a single
grade, $r$, are termed homogeneous and usually written $A_{r}.$ The
geometric (Clifford) product is written by simply juxtaposing multivectors $%
AB$. A basic operation on multivectors is the degree projection $%
\left\langle A\right\rangle _{r}$ which selects from the multivector $A$ its
$r-$ vector part ($0=$ scalar, $1=$ vector, $2=$ bivector, ....). The
geometric product of a grade-$r$ multivector $A_{r}$ with a grade-$s$
multivector $B_{s}$ decomposes into $A_{r}B_{s}=\left\langle AB\right\rangle
_{\ r+s}+\left\langle AB\right\rangle _{\ r+s-2}...+\left\langle
AB\right\rangle _{\ \left\vert r-s\right\vert }$. The inner and outer (or
exterior) products are the lowest-grade and the highest-grade terms
respectively of the above series; $A_{r}\cdot B_{s}\equiv \left\langle
AB\right\rangle _{\ \left\vert r-s\right\vert }$ and $A_{r}\wedge
B_{s}\equiv \left\langle AB\right\rangle _{\ r+s}$. For vectors $a$ and $b$
we have: $ab=a\cdot b+a\wedge b$, where $a\cdot b\equiv (1/2)(ab+ba)$, $%
a\wedge b\equiv (1/2)(ab-ba)$.

In this paper the notation will not be the same as in the above mathematical
presentation. Some vectors will be denoted in lower case, like $u$, $v$ (the
velocities), $x$ (the position vector), whereas some others in upper case,
like vectors of the electric and magnetic fields $E$ and $B$ respectively,
the Lorentz force $K_{L}$. Bivectors will be denoted in upper case but
without subscript that denotes the grade. Thus, for example, the
electromagnetic field $F$ is a bivector.

In, e.g., [18], one usually introduces the standard basis. The generators of
the spacetime algebra (the Clifford algebra generated by Minkowski
spacetime) are taken to be four basis vectors $\left\{ \gamma _{\mu
}\right\} ,\mu =0...3,$ satisfying $\gamma _{\mu }\cdot \gamma _{\nu }=\eta
_{\mu \nu }=diag(+---).$ This basis, the standard basis, is a right-handed
orthonormal frame of vectors in the Minkowski spacetime $M^{4}$ with $\gamma
_{0}$ in the forward light cone. The $\gamma _{k}$ ($k=1,2,3$) are spacelike
vectors. The $\gamma _{\mu }$ generate by multiplication a complete basis
for the spacetime algebra: $1$, $\gamma _{\mu }$, $\gamma _{\mu }\wedge
\gamma _{\nu }$, $\gamma _{\mu }\gamma _{5}$, $\gamma _{5}$ ($2^{4}=16$
independent elements). $\gamma _{5}$ is the right-handed unit pseudoscalar, $%
\gamma _{5}=\gamma _{0}\wedge \gamma _{1}\wedge \gamma _{2}\wedge \gamma
_{3} $. Any multivector can be expressed as a linear combination of these 16
basis elements of the spacetime algebra. For all mathematical details
regarding the spacetime algebra reader can consult [18]. It is worth noting
that the standard basis $\left\{ \gamma _{\mu }\right\} $ corresponds, in
fact, to the specific system of coordinates, i.e., to Einstein's system of
coordinates. In Einstein's system of coordinates the standard, i.e.,
Einstein's synchronization [1] of distant clocks and Cartesian space
coordinates $x^{i}$ are used in the chosen inertial frame. However different
systems of coordinates are allowed in an inertial frame and they are all
equivalent in the description of physical phenomena. For example, in [16]
two very different, but physically completely equivalent, systems of
coordinates, Einstein's system of coordinates and the system of coordinates
with a nonstandard synchronization, the \textquotedblleft
everyday,\textquotedblright\ i.e., the \textquotedblleft
radio\textquotedblright\ (\textquotedblleft r\textquotedblright ),
synchronization, are exposed and exploited throughout the paper. The
\textquotedblleft r\textquotedblright , synchronization is also used and
explained in [38], [17], [27], [33], see also Sec. 3.1 here. For simplicity
and for easier understanding we shall mainly deal with the standard basis,
but remembering that the approach with 4D geometric quantities holds for any
choice of basis in $M^{4}$. Observe that the usual covariant approach, e.g.,
from the well-known textbooks [2], [26] exclusively deals with \emph{%
components} implicitly taken in a specific basis, the standard basis.

Here, it is worth mentioning an important result regarding the usual
formulation of electromagnetism (as in [2], [26]), which is presented in
[19]. This is also mentioned in [11]. It is explained in [19] that an
individual vector has no dimension; the dimension is associated with the
vector space and with the manifold where this vector is tangent. Hence, what
is essential for the number of components of a vector field is the number of
variables on which that vector field depends, i.e., the dimension of its
domain. This means that the usual time-dependent $\mathbf{E(r,}t\mathbf{)}$,
$\mathbf{B(r,}t\mathbf{)}$ cannot be the 3-vectors, since they are defined
on the spacetime. That fact determines that such vector fields, when
represented in some basis, have to have four components (some of them can be
zero). Therefore, we use the term \textquotedblleft
vector\textquotedblright\ for the correctly defined geometric quantity,
which is defined on the spacetime. However, an incorrect expression, the
3-vector or the 3D vector, will still remain for the usual $\mathbf{E(r,}t%
\mathbf{)}$, $\mathbf{B(r,}t\mathbf{)}$ from [2], see Eq. (\ref{JCB}%
).\bigskip

\noindent \textbf{3. The LPET of the 3-vectors} $\mathbf{E}$ \textbf{and} $%
\mathbf{B}$, $\mathbf{P}$ \textbf{and} $\mathbf{M}$\bigskip

\noindent \textit{3.1 The 3-vectors }$\mathbf{E}$ \textit{and} $\mathbf{B}$\
\textit{and their LPET\bigskip }

Firstly, we discuss and object the derivation of the LPET of the 3-vectors $%
\mathbf{E}$ and $\mathbf{B}$ by the identification of the components of $%
\mathbf{E}$ and $\mathbf{B}$ with the components (implicitly taken in the
standard basis) of the electromagnetic field tensor $F^{\alpha \beta }$, as,
e.g., in the usual covariant approach [2]. Einstein's derivation [1] of the
LPET of the components of $\mathbf{E}$ and $\mathbf{B}$ is discussed and
objected in Sec. 5.3 in [16]. (In [16] the LPET are called the
\textquotedblleft apparent\textquotedblright\ transformations.) In the usual
covariant approach, e.g., [2], the field-strength tensor $F^{\alpha \beta }$
(only components in the standard basis and not the whole tensor as a 4D
geometric quantity) is introduced and defined in terms of the vector
potential $A^{\mu }$, Eq. (11.136) in [2].

However, as already stated, the components are coordinate quantities and
they do not contain the whole information about the physical quantity. They
do not completely represent a physical quantity that is defined on the 4D
spacetime, since a basis of the spacetime is not included. Furthermore, in
such formulation the gauge dependent vector potential $A^{\mu }$ (not
measurable quantity) is considered to be the primary quantity from which the
field-strength tensor $F^{\alpha \beta }$ (a measurable quantity) is
derived! In contrast to such usual approach, e.g., [2], it is shown in [12]
that in the 4D spacetime the electromagnetic field, the bivector $F=F(x)$,
can be taken as the primary quantity for the whole electromagnetism and
there is no need for the electromagnetic potentials.

Then, in the usual approaches, the covariant form of the Maxwell equations
is written with $F^{\alpha \beta }$ and its dual $^{\ast }F^{\alpha \beta }$
\begin{equation}
\partial _{\alpha }F^{a\beta }=j^{\beta }/\varepsilon _{0}c,\quad \partial
_{\alpha }\ ^{\ast }F^{\alpha \beta }=0,  \label{maxco}
\end{equation}%
where $^{\ast }F^{\alpha \beta }=(1/2)\varepsilon ^{\alpha \beta \gamma
\delta }F_{\gamma \delta }$. In order to get the component form of the
Maxwell equations with the 3D\textbf{\ }$\mathbf{E}$\textbf{\ }and $\mathbf{B%
}$
\begin{eqnarray}
\partial _{k}E_{k}-j^{0}/c\varepsilon _{0} &=&0,\quad -\partial
_{0}E_{i}+c\varepsilon _{ikj}\partial _{j}B_{k}-j^{i}/c\varepsilon _{0}=0,
\notag \\
\partial _{k}B_{k} &=&0,\quad c\partial _{0}B_{i}+\varepsilon _{ikj}\partial
_{j}E_{k}=0  \label{j3}
\end{eqnarray}%
from Eq. (\ref{maxco}) one simply makes \emph{the identification of }the six
independent components of $F^{\alpha \beta }$ with six components of the
3-vectors $\mathbf{E}$ and $\mathbf{B}$. These identifications are
\begin{equation}
E_{i}=F^{i0},\qquad B_{i}=(1/2)\varepsilon _{ijk}F_{kj}  \label{ieb}
\end{equation}%
(the indices $i$, $j$, $k$, $...=1,2,3$), Eq. (11.137) in [2]. The
components of the 3D fields $\mathbf{E}$ and $\mathbf{B}$ are written with
lowered (generic) subscripts, since they are not the spatial components of
the 4D quantities. This refers to the third-rank antisymmetric $\varepsilon $
tensor too. The super- and subscripts are used only on the components of the
4D quantities. Then the 3D $\mathbf{E}$ and $\mathbf{B}$, as \emph{geometric
quantities in the 3D space}, are constructed from these six independent
components of $F^{\mu \nu }$ and \emph{the unit 3D vectors }$\mathbf{i},$ $%
\mathbf{j},$ $\mathbf{k,}$ e.g., $\mathbf{E=}F^{10}\mathbf{i}+F^{20}\mathbf{j%
}+F^{30}\mathbf{k}$.

It is worth noting that such an identification of the components of $\mathbf{%
E}$ and $\mathbf{B}$ with the components of $F^{\alpha \beta }$ is
synchronization dependent as explicitly shown in [16]. This is also
discussed in [33]. There, it is shown that the mentioned identifications are
meaningless in the \textquotedblleft r\textquotedblright\ synchronization,
i.e., in the $\left\{ r_{\mu }\right\} $ basis, in which only the Einstein
synchronization is replaced by an asymmetric synchronization, the
\textquotedblleft radio\textquotedblright\ synchronization. At the end of
this section, the \textquotedblleft r\textquotedblright\ synchronization is
explained in more detail. As explained in [16] (and [33]), in the
\textquotedblleft r\textquotedblright\ synchronization
\begin{equation}
F_{r}^{10}=E_{1}+cB_{3}-cB_{2}.  \label{Fr1}
\end{equation}%
Hence, the identification $E_{1r}=F_{r}^{10}$, as in (\ref{ieb}), shows that
the component $E_{1r}$ in the $\left\{ r_{\mu }\right\} $ basis is expressed
as the combination of $E_{i}$ and $B_{i}$ components from the $\left\{
\gamma _{\mu }\right\} $ basis%
\begin{equation}
E_{1r}=F_{r}^{10},\quad E_{1r}=E_{1}+cB_{3}-cB_{2}.  \label{FEr}
\end{equation}%
This means that if the \textquotedblleft r\textquotedblright\
synchronization is used, i.e., if the appropriate metric is used, then it is
not possible to make the usual identifications (\ref{ieb}). It follows that
the usual identifications, Eq. (11.137) in [2], are meaningful \emph{only}
when the Minkowski metric, e.g., $diag(1,-1,-1,-1)$, is used. Thus, these
identifications depend on the chosen synchronization, i.e., the metric.
\emph{But,} \emph{different synchronizations are nothing else than different
conventions and physics must not depend on conventions.}

In the usual covariant approach, e.g., [2], one transforms by the passive LT
the covariant Maxwell equations (\ref{maxco}) and finds
\begin{equation}
\partial _{\alpha }^{\prime }F^{\prime a\beta }=j^{\prime \beta
}/\varepsilon _{0}c,\quad \partial _{\alpha }^{\prime }\ ^{\ast }F^{\prime
\alpha \beta }=0.  \label{cme}
\end{equation}%
Under the passive LT the set of components, e.g., $j^{\mu }$ from the $S$
frame transforms to $j^{\prime \mu }$ in the relatively moving inertial
frame of reference $S^{\prime }$, $j^{\prime \mu }=L_{\ \nu }^{\mu }j^{\nu }$%
, where, for the boost in the $\gamma _{1}$ direction, $L_{\ 0}^{0}=L_{\
1}^{1}=\gamma $, $L_{\ 1}^{0}=L_{\ 0}^{1}=-\beta \gamma $, $L_{\ 2}^{2}=L_{\
3}^{3}=1$ and all other components are zero. Then, the same identification
as in Eq. (\ref{ieb}) is assumed to hold for the transformed components $%
E_{i}^{\prime }$ and $B_{i}^{\prime }$
\begin{equation}
E_{i}^{\prime }=F^{\prime i0},\quad B_{i}^{\prime }=(1/2c)\varepsilon
_{ijk}F_{kj}^{\prime }.  \label{eb2}
\end{equation}%
For example,
\begin{equation}
F^{\prime 10}=F^{10},\ F^{\prime 20}=\gamma (F^{20}-\beta F^{21}),\
F^{\prime 30}=\gamma (F^{30}-\beta F^{31}),  \label{fe}
\end{equation}%
which yields (by Eqs. (\ref{ieb}) and (\ref{eb2})) that
\begin{equation}
E_{1}^{\prime }=E_{1},\ E_{2}^{\prime }=\gamma (E_{2}-\beta cB_{3}),\
E_{3}^{\prime }=\gamma (E_{3}+\beta cB_{2}),  \label{ee}
\end{equation}%
see Sec. 11.10 and Eq. (11.148) in [2]. The same remark about the (generic)
subscripts holds also here. Thus in the usual covariant approach the
components $F^{a\beta }$ are transformed by the passive LT into $F^{\prime
a\beta }$ and then it is simply argued that six independent components of $%
F^{\prime a\beta }$ are the \textquotedblleft Lorentz
transformed\textquotedblright\ components $E_{i}^{\prime }$ and $%
B_{i}^{\prime }$, i.e., \emph{the LPET of the components of} $\mathbf{E}$
\emph{and} $\mathbf{B}$ \emph{are derived assuming that they transform under
the LT as the components of} $F^{\alpha \beta }$ \emph{transform}, Eq.
(11.148) in [2]. Then, $\mathbf{E}^{\prime }$ and $\mathbf{B}^{\prime }$ are
constructed in $S^{\prime }$ in the same way as in $S$, i.e. multiplying the
components $E_{x,y,z}^{\prime }$ and $B_{x,y,z}^{\prime }$ by the unit
3-vectors $\mathbf{i}^{\prime }$, $\mathbf{j}^{\prime }$, $\mathbf{k}%
^{\prime }$. This yields the LPET of the 3-vectors $\mathbf{E}$ and $\mathbf{%
B}$\textbf{, }Eq. (11.149) in [2], i.e. Eq. (\ref{JCB}) here

\begin{eqnarray}
\mathbf{E}^{\prime } &=&\gamma (\mathbf{E}+\mathbf{\beta \times }c\mathbf{B)-%
}(\gamma ^{2}/(1+\gamma ))\mathbf{\beta (\beta \cdot E),}  \notag \\
\mathbf{B}^{\prime } &=&\gamma (\mathbf{B}-(1/c)\mathbf{\beta \times E)-}%
(\gamma ^{2}/(1+\gamma ))\mathbf{\beta (\beta \cdot B),}  \label{JCB}
\end{eqnarray}%
where $\mathbf{E}^{\prime }$, $\mathbf{E}$, $\mathbf{\beta }$ and $\mathbf{B}%
^{\prime }$, $\mathbf{B}$ are all 3-vectors.

Observe that there are no LT, or any other transformations, that transform
the unit 3-vectors $\mathbf{i}$, $\mathbf{j}$, $\mathbf{k}$ into the unit
3-vectors $\mathbf{i}^{\prime }$, $\mathbf{j}^{\prime }$, $\mathbf{k}%
^{\prime }$. It is seen from Eqs. (11.148) and (11.149) in [2], i.e. from
Eq. (\ref{JCB}) here, that \emph{the transformed} $\mathbf{E}^{\prime }$
\emph{is expressed by the mixture of the 3-vectors} $\mathbf{E}$ \emph{and} $%
\mathbf{B}$, \emph{and similarly for} $\mathbf{B}^{\prime }$. The electric
field $\mathbf{E}$ in one frame is \textquotedblleft seen\textquotedblright\
as slightly changed electric field $\mathbf{E}^{\prime }$ and an \emph{%
induced magnetic field} $\mathbf{B}^{\prime }$ in a relatively moving
inertial frame.

This type of the derivation of (\ref{JCB}) was first presented in Sec. 3 in
[21]. There, and in section 7.2 as well, Minkowski made the same
identification of the components of $F^{\alpha \beta }$ with components of
the 3-vectors $\mathbf{E}$ and $\mathbf{B}$ (his $\mathbf{M}$), as in
equation (11.137) in [2]. The equations (11.148) in [2] are nothing else but
the equations (6) and (7) from Sec. 3 in Minkowski's paper [21]. Later, the
same derivation is used in numerous textbooks and papers treating
relativistic electrodynamics.

Minkowski's identifications, i.e., Eq. (11.137) in [2], refer, as already
stated, only to the components implicitly taken in the standard basis, which
means that they are not generally valid. Namely, in the 4D spacetime,
physical quantities are represented by the abstract 4D geometric quantities
that are basis independent. In some basis, as already mentioned, they are
represented as CBGQs which contain \emph{both} components \emph{and the}
\emph{basis vectors} (4-vectors in the usual notation). The principle of
relativity is naturally satisfied for physical laws written with such 4D
geometric quantities, whereas in Einstein's formulation with the 3-vectors
or with their components it has to be postulated outside the mathematical
formulation of the theory. Components taken alone are only a part of a
physical quantity; another, equally important, part are the basis vectors.
The LPET (\ref{JCB}) are derived using synchronization dependent
identifications of \emph{components} of $F^{\alpha \beta }$ with \emph{%
components} of the 3-vectors $\mathbf{E}$ and $\mathbf{B}$ in both
relatively moving inertial frames of reference. \emph{This shows that both
the 3-vectors }$\mathbf{E}$ \emph{and} $\mathbf{B}$ \emph{and their LPET} (%
\ref{JCB}) \emph{are determined in a relativistically incorrect way; the
quantities entering into that derivation are not properly defined in the 4D
spacetime.}

In almost all textbooks and papers on relativistic electrodynamics the LPET (%
\ref{JCB}) (or (\ref{ps}), which are given below) are simply employed
without any physical interpretation. It seems that the majority of
physicists believe that it is physically justified to have, e.g., only
magnetic field $\mathbf{B}$ in one frame, in $S$, which transforms into a
slightly changed magnetic field $\mathbf{B}^{\prime }$ and a \emph{new
electric field} $\mathbf{E}^{\prime }$ in a relatively moving $S^{\prime }$
frame. In some textbooks the authors tried to give a physical
\textquotedblleft explanation\textquotedblright\ for the appearance of that
\emph{electric field} $\mathbf{E}^{\prime }$ for the observers in $S^{\prime
}$. Thus, for example, Rosser [22] asked the reader (Problem 6.13) to
interpret the origin of the electric field present in $S^{\prime }$. Let us
assume that the external magnetic field in $S$ is due to a permanent magnet
at rest in $S$. Then, as in Sec. 3.2 here, according to the LPET (\ref{ps})
((\ref{psa})) a moving magnet has an electric polarization $\mathbf{P}$ (\ref%
{Pl1}), which gives an electric field outside the moving magnet. However, as
discussed below, in the relativistically correct 4D geometric approach the
relation (\ref{Pl1}) for the polarization $\mathbf{P}$, which is induced by
the movement of a permanent magnetization $\mathbf{M}^{\prime }$, does not
hold. It is not derived in a relativistically correct manner; the quantities
entering into that derivation are not properly defined in the 4D spacetime.

Here, for the reader's convenience, we explain here the \textquotedblleft
r\textquotedblright\ synchronization. As explained, e.g., in [16], different
systems of coordinates, including different synchronizations, are allowed in
an inertial frame and they are all equivalent in the description of physical
phenomena. Thus in [16], both, Einstein's synchronization [1] and the
\textquotedblleft r\textquotedblright\ synchronization are exposed in
detail. The \textquotedblleft r\textquotedblright\ synchronization is
commonly used in everyday life and not Einstein's synchronization. \emph{In
the \textquotedblleft r\textquotedblright\ synchronization there is an
absolute simultaneity.} Hence, contrary to the common opinion, \emph{the
relativity of simultaneity is not an intrinsic relativistic effect.} As
stated in [38]: \textquotedblleft For if we turn on the radio and set our
clock by the standard announcement \textquotedblright ...\textquotedblleft
at the sound of the last tone, it will be 12 o'clock,\textquotedblright\
then we have synchronized our clock with the studio clock according to the
\textquotedblleft r\textquotedblright\ synchronization. In order to treat
different systems of coordinates on an equal footing it is presented, Eq.
(4) in [16], the transformation matrix that connects Einstein's system of
coordinates with another system of coordinates in the same reference frame.
Furthermore, Eq. (2) in [27], Eq. (1) in [16], it is derived such form of
the LT, which is independent of the chosen system of coordinates, including
different synchronizations. The unit vectors in the standard basis $\left\{
\gamma _{\mu }\right\} $ and the $\{r_{\mu }\}$ basis, i.e., with the
\textquotedblleft r\textquotedblright\ synchronization, [16], are connected
as
\begin{equation}
r_{0}=\gamma _{0},\quad r_{i}=\gamma _{0}+\gamma _{i}.  \label{re}
\end{equation}%
Hence, the components $g_{\mu \nu ,r}$ of the metric tensor are $g_{ii,r}=0$%
, and all other components are $=1$. Remember that in the $\left\{ \gamma
_{\mu }\right\} $ basis $g_{\mu \nu }=diag(1,-1,-1,-1)$. (Note that in [16]
and [17] the Minkowski metric is $g_{\mu \nu }=diag(-1,1,1,1)$.) Then,
according to (4) from [16], one can use $g_{\mu \nu ,r}$ to find the
transformation matrix $R_{\;\nu }^{\mu }$ that connects the components from
the $\left\{ \gamma _{\mu }\right\} $ basis with the components from the $%
\{r_{\mu }\}$ basis. The only components that are different from zero are%
\begin{equation}
R_{\;\mu }^{\mu }=-R_{\;i}^{0}=1.  \label{er}
\end{equation}%
The inverse matrix $(R_{\;\nu }^{\mu })^{-1}$ connects the \textquotedblleft
old\textquotedblright\ basis, $\left\{ \gamma _{\mu }\right\} $, with the
\textquotedblleft new\textquotedblright\ one, $\{r_{\mu }\}$. Hence, the
components of the position vector $x$ are connected as
\begin{equation}
x_{r}^{0}=x^{0}-x^{1}-x^{2}-x^{3},\quad x_{r}^{i}=x^{i}.  \label{ptr}
\end{equation}%
Observe that vector $x$ can be decomposed in both bases and it holds that%
\begin{equation}
x=x^{\mu }\gamma _{\mu }=x_{r}^{\mu }r_{\mu }.  \label{rx}
\end{equation}%
Obviously, the components of any vector transform in the same way as in (\ref%
{ptr}), e.g., for the components of the electric field vector $E$ it holds
that
\begin{equation}
E_{r}^{0}=E^{0\text{ }}-E^{1}-E^{2}-E^{3},\quad E_{r}^{i}=E^{i}.  \label{er1}
\end{equation}%
In the same way as in (\ref{rx}), it can be written for the vector $E$ that $%
E=E^{\mu }\gamma _{\mu }=E_{r}^{\mu }r_{\mu }$.

It is visible from Eqs. (\ref{re}) and (\ref{ptr}), i.e., from the fact that
the metric tensor $g_{\mu \nu ,r}$ is not diagonal, in the $\{r_{\mu }\}$
basis it is not possible to make the separation of the 4D spacetime into the
time and the 3D space as it is possible in the standard basis $\left\{
\gamma _{\mu }\right\} $. Thus \emph{the space-time split is not possible in
the} $\{r_{\mu }\}$ \emph{basis.} In the first and the second paper (in that
second paper the \textquotedblleft r\textquotedblright\ synchronization is
used as well) in [17] some of the well-known experiments: the
\textquotedblleft muon\textquotedblright\ experiment, the Michelson-Morley
type experiments, the Kennedy-Thorndike type experiments and the
Ives-Stilwell type experiments are analyzed using Einstein's formulations of
SR, which deals with the Lorentz contraction and the time dilation, and the
ISR, i.e., the approach with 4D geometric quantities, the position vector,
the distance vector between two events and the spacetime length. It is shown
that all experiments are in a complete agreement, independently of the
chosen synchronization, with the 4D geometric approach,\ i.e., with the ISR,
whereas it is not the case with the Einstein's approach with the Lorentz
contraction and the time dilation if the \textquotedblleft
r\textquotedblright\ synchronization is used. In the third paper in [17] the
same is shown considering in detail the Michelson-Morley experiment.\bigskip

\noindent \textit{3.2 The 3-vectors }$\mathbf{P}$ \textit{and} $\mathbf{M}$\
\textit{and their LPET\bigskip }

The LPET of the polarization and the magnetization 3-vectors $\mathbf{P}$
and $\mathbf{M}$ are also often derived from the covariant formulation using
the mentioned, synchronization dependent, identifications of components
(implicitly taken in the standard basis) of the magnetization-polarization
tensor $\mathcal{M}^{\alpha \beta }$ with components of the 3-vectors $%
\mathbf{P}$ and $\mathbf{M}$ in both relatively moving inertial frames of
reference. Thus, in $S$, these identifications are $P_{i}=\mathcal{M}^{i0}$,
$M_{i}=(c/2)\varepsilon _{ijk}\mathcal{M}_{jk}$ and the same identifications
hold in the relatively moving inertial frame of reference $S^{\prime }$, $%
P_{i}^{\prime }=\mathcal{M}^{\prime i0}$, $M_{i}^{\prime }=(c/2)\varepsilon
_{ijk}\mathcal{M}_{jk}^{\prime }$, see, e.g., Secs. 18-5 and 18-6 in [23].
The same remark about the (generic) subscripts holds also here. This
procedure yields

\begin{eqnarray}
\mathbf{P} &=&\gamma (\mathbf{P}^{\prime }+\mathbf{\beta \times M}^{\prime
}/c\mathbf{)-}(\gamma ^{2}/(1+\gamma ))\mathbf{\beta (\beta \cdot P}^{\prime
}\mathbf{),}  \notag \\
\mathbf{M} &=&\gamma (\mathbf{M}^{\prime }-\mathbf{\beta \times }c\mathbf{P}%
^{\prime }\mathbf{)-}(\gamma ^{2}/(1+\gamma ))\mathbf{\beta (\beta \cdot M}%
^{\prime }\mathbf{),}  \label{ps}
\end{eqnarray}%
see e.g., Eqs. (18-68) - (18-71) in [23], or Eqs. (4.2) in [39], or Eqs.
(6.78a) and (6.81a) in [22], etc. In the mentioned equations the
transformations (\ref{ps}) are written in an equivalent form as
\begin{eqnarray}
\mathbf{P}_{\parallel } &=&\mathbf{P}_{\parallel }^{\prime }\mathbf{,\quad P}%
_{\perp }=\gamma (\mathbf{P}^{\prime }+\mathbf{\beta \times M}^{\prime }/c%
\mathbf{)}_{\perp }  \notag \\
\mathbf{M}_{\parallel } &=&\mathbf{M}_{\parallel }^{\prime }\mathbf{,\quad M}%
_{\perp }=\gamma (\mathbf{M}^{\prime }\mathbf{-}c\mathbf{\beta \times P}%
^{\prime }\mathbf{)}_{\perp }.  \label{psa}
\end{eqnarray}%
The inverse relations are obtained in the usual way by the exchange of the
primed and unprimed quantities and by the replacement $\mathbf{\beta
\rightarrow }-\mathbf{\beta }$. The main feature of the LPET of $\mathbf{P}$
and $\mathbf{M}$ (\ref{ps}), or (\ref{psa}), is the same as for the LPET of $%
\mathbf{E}$ and $\mathbf{B}$, i.e., \emph{the components of the transformed}
$\mathbf{P}^{\prime }$ \emph{are expressed by the mixture of components of} $%
\mathbf{P}$ \emph{and} $\mathbf{M}$,\emph{\ and similarly for} $\mathbf{M}%
^{\prime }$.

Using completely the same procedure with the identifications of components
one can derive the LPET of the 3-vectors of the electric $\mathbf{p}$ and
magnetic $\mathbf{m}$ dipole moments from the tensor of the dipole moments $%
D^{a\beta }$. Hence, these LPET for the 3-vectors $\mathbf{p}$ and $\mathbf{m%
}$ are the same as (\ref{ps}), or (\ref{psa}), but with $\mathbf{p}$ and $%
\mathbf{m}$ replacing $\mathbf{P}$ and $\mathbf{M}$, respectively.

The interpretation and the derivation of the transformations (\ref{ps}) (or (%
\ref{psa})) in terms of simplified classical models is presented in, e.g.
[23] and [22]. It is stated in Sec. 18-6 in [23] that the relation $\mathbf{P%
}_{\parallel }^{\prime }=\mathbf{P}_{\parallel }$ is expected,
\textquotedblleft since $\mathbf{P}_{\parallel }$ is the product of an
(invariant) charge and a distance divided by a volume, \emph{both contracted
in the same ratio}.\textquotedblright\ (my emphasis) The calculation is
given in section 6.7.1. in [22]. The $S^{\prime }$ frame is taken to be the
rest frame of the material. The classical model assumes that in $S^{\prime }$
the dielectric consists of $n^{\prime }$ stationary dipoles/m$^{3}$. If $%
\mathbf{P}^{\prime }$ is parallel to $\mathbf{\beta }$, then $P^{\prime
}=n^{\prime }p^{\prime }=n^{\prime }(ql_{0})$, $P=np=(\gamma n^{\prime
})(\gamma ^{-1}p^{\prime })=P^{\prime }$; $\gamma n^{\prime }$ is \emph{due
to the contraction of the volume}, whereas $\gamma ^{-1}p^{\prime }$ is
\emph{due to the contraction of a distance}, $l=\gamma ^{-1}l_{0}$. If the
atomic electric dipoles are perpendicular to $\mathbf{\beta }$, then the
first term $\gamma \mathbf{P}_{\perp }^{\prime }$ simply follows from $%
p=ql=ql_{0}=p^{\prime }$ (\emph{there is no Lorentz contraction if} $\mathbf{%
p}^{\prime }\perp \mathbf{\beta }$), but $n=\gamma n^{\prime }$, and thus
the atomic electric dipoles give rise to a polarization $\gamma \mathbf{P}%
_{\perp }^{\prime }$ in $S$.

It is argued, both in [23] and [22], that the extra term $(\mathbf{\beta
\times M}^{\prime }/c\mathbf{)}_{\perp }$ in (\ref{psa}) has no
non-relativistic counterpart. In the classical model, for purposes of
calculating $\mathbf{M}^{\prime }$, the magnetic dipoles are considered as
little current loops. In section 18-4 in [23], see Fig. 18-4, and in Sec.
6.5 in [22], see Fig. 6.4 a,b, it is argued that a neutral stationary
current loop, which has a magnetic moment $\mathbf{m}^{\prime }$ in its rest
frame $S^{\prime }$, acquires an electric dipole moment
\begin{equation}
\mathbf{p}=\mathbf{\beta \times m}^{\prime }/c  \label{plp}
\end{equation}%
if it is moving with uniform 3-velocity $\mathbf{U}$ ($\mathbf{\beta =U}/c$)
relative to the laboratory frame $S$. The result (\ref{plp}) also follows
from the LPET for the 3-vectors $\mathbf{p}$ and $\mathbf{m}$, which are, as
already stated, the same as (\ref{ps}) with $\mathbf{p}$ and $\mathbf{m}$
replacing $\mathbf{P}$ and $\mathbf{M}$, respectively. It is taken that
\emph{in the rest frame of the neutral current loop the electric moment} $%
\mathbf{p}^{\prime }$ \emph{is zero.}

In the Amp\`{e}rian approach a permanent magnet is essentially an assembly
of current loops. Hence, if (\ref{plp}) holds for each atomic magnetic
dipole in the moving magnet, then one has for the electric dipole moment per
unit volume in $S$, $n\mathbf{p}=\mathbf{\gamma }n^{\prime }\mathbf{U\times m%
}^{\prime }/c^{2}=\gamma (\mathbf{U\times M}^{\prime }/c^{2}\mathbf{)}%
_{\perp }$. Thus, if a permanent magnetization $\mathbf{M}^{\prime }$ is
viewed from a moving frame it produces an electric polarization
\begin{equation}
\mathbf{P=\gamma U\times M}^{\prime }/c^{2}.  \label{Pl1}
\end{equation}%
In other words, \emph{according to all usual approaches, if an observer
moves with a 3-velocity} $\mathbf{U}$ \emph{relative to a medium of
magnetization} $\mathbf{M}^{\prime }$ \emph{that observer will observe an
equivalent electric polarization given by }$\mathbf{P}$, (\ref{Pl1}). Adding
this term to the term $\gamma \mathbf{P}_{\perp }^{\prime }$ yields $\mathbf{%
P}_{\perp }$ from (\ref{psa}).

It can be seen from the mentioned textbooks, [23], [22], that the relations (%
\ref{plp}) and (\ref{Pl1}) are obtained using \emph{the Lorentz contraction}
and \emph{the relativity of simultaneity.}

However, as shown in [16] and in the comparison with well-known experiments
that test special relativity [17], \emph{the relativity of simultaneity, the
Lorentz contraction and the time dilation are not well-defined in the 4D
spacetime. They are not intrinsic relativistic effects, because they depend
on the chosen synchronization. }A clear presentation of the relativistic
incorrectness of the Lorentz contraction is already given in Sec. 2.2 in
[27]. Already in the year 1966, Rohrlich [15] clearly explained that \emph{%
the Lorentz contraction }is not a true relativistic transformation, i.e., it
\emph{has nothing to do with the Lorentz transformation. }Similarly, in the
next year, Gamba [40] stated for the Lorentz contraction: \textquotedblleft
Although it is a completely useless concept in physics, it will probably
continue to remain in the books as an historical relic for the fascination
of the layman.\textquotedblright\ In the geometric approach, the ISR, in
[16], [17], [27] it is \emph{proved} that in the 4D spacetime two relatively
moving observers cannot compare spatial lengths taken alone, which are
synchronously determined for the observer. For the reader's convenience the
relativistic incorrectness of the Lorentz contraction is explicitly shown in
Appendix here. Also, it is proved in [16] and [17] that it is not correct to
compare the temporal distances taken alone, since they are not well-defined
quantities in the 4D spacetime. The properly defined quantities are the
distance vector between two events $A$ and $B$ with the position vectors $%
x_{A}$ and $x_{B}$ and the spacetime length, which is a Lorentz scalar.
Obviously, Gamba [40] was wrong with his statement about the Lorentz
contraction. The papers [15] and [40] as well as my papers [16], [17], [27]
remained almost completely overlooked and still the Lorentz contraction and
the time dilation are treated in the leading physical journals and in the
well-known textbooks, including [2] and [26], as intrinsic relativistic
effects.

This consideration reveals in another way that the LPET of $\mathbf{P}$ and $%
\mathbf{M}$ (\ref{ps}), or (\ref{psa}), are not relativistically correct
transformations, i.e., (\ref{ps}), \emph{or} (\ref{psa}), \emph{are not the
LT but the AT.} \bigskip

\noindent \textbf{4. The definitions of vectors}\textit{\ }$E$, $B$ \textbf{%
and} $P$, $M$ \textbf{in terms of} $F$, $v$ \textbf{and}

$\mathcal{M}$, $u$, \textbf{respectively. The Lorentz invariant field
equations }

\textbf{for vacuum and for a magnetized and polarized moving medium}\bigskip

Instead of dealing with quantities that are not well-defined in the 4D
spacetime, like the 3-vectors $\mathbf{E}$ and $\mathbf{B}$ and their LPET (%
\ref{JCB}), or with components implicitly taken in the standard basis as in
the usual covariant approaches, [2], [26], we deal with 4D geometric
quantities, which are properly defined in the 4D spacetime. Moreover, it is
shown, particularly in [12], that the bivector $F=F(x)$, which represent the
electromagnetic field, can be taken as the primary quantity for the whole
electromagnetism and the field equation for $F$

\begin{equation}
\partial F=j/\varepsilon _{0}c,\quad \partial \cdot F+\partial \wedge
F=j/\varepsilon _{0}c  \label{MEF}
\end{equation}%
is the basic equation. As shown in [12], the bivector field $F$ yields the
complete description of the electromagnetic field and, in fact, there is no
need to introduce either the field vectors or the potentials. For the given
sources the Clifford algebra formalism enables one to find in a simple way
the electromagnetic field $F$, see Eqs. (7) and (8) in [12]. However, if one
introduces the electric and magnetic fields, then they can be represented by
different algebraic objects. \emph{These fields are not determined by the
usual identifications of the components, Eqs.} (\ref{ieb}) \emph{and} (\ref%
{eb2}), \emph{but they are derived in a mathematically correct way from} $F$%
, \emph{as in Eqs.} (\ref{E2}) \emph{and} (\ref{E1}) \emph{here.}

In this geometric approach the electric and magnetic fields are represented
by vectors $E(x)$ and $B(x)$.\ We deal with such representations of the
electric and magnetic fields because they are simple and much closer to the
classical representation of the electric and magnetic fields by the 3D
vectors $\mathbf{E}$ and $\mathbf{B}$ than, e.g. the representations by
bivectors, which are used in [18]. The decomposition of $F$ in terms of
vectors $E$, $B$ and $v$ is given as
\begin{equation}
F=(1/c)E\wedge v+(IB)\cdot v,  \label{E2}
\end{equation}%
and $E$ and $B$ are determined as
\begin{equation}
E=(1/c)F\cdot v,\quad B=-(1/c^{2})I(F\wedge v).  \label{E1}
\end{equation}%
There is no rest frame for the field $F$, that is, for $E$ and $B$, and
therefore the vector $v$ in the decomposition (\ref{E2}) is interpreted as
the velocity vector of the observers who measure $E$ and $B$ fields. Then $%
E(x)$ and $B(x)$ are defined with respect to $v$, i.e., with respect to the
observer. From (\ref{E2}) and (\ref{E1}) it also holds that $E\cdot v=B\cdot
v=0$; only three components of $E$ and three components of $B$ are
independent since $F$ is antisymmetric. The unit pseudoscalar $I$ from (\ref%
{E2}) and (\ref{E1}) is defined algebraically without introducing any
reference frame, as in Sec. 1.2. in the second reference in [18]. We choose $%
I$ in such a way that when $I$ is represented in the $\left\{ \gamma _{\mu
}\right\} $ basis it becomes $I=\gamma _{0}\wedge \gamma _{1}\wedge \gamma
_{2}\wedge \gamma _{3}=\gamma _{5}$. With such choice for $I$, $\left\{
\gamma _{1},\gamma _{2},\gamma _{3}\right\} $ form a right-handed
orthonormal set, as usual for a 3D Cartesian frame. The LT do not change the
orientation of the spacetime.

The equations that correspond to equations (\ref{E1}) (and (\ref{E2})), but
in the tensor formalism, with abstract indices $a$, $b$, $c$, .. , are $%
E^{a}=(1/c)F^{ab}v_{b},\ B^{a}=(1/2c^{2})\varepsilon ^{abcd}F_{bc}v_{d}$ and
$F^{ab}=(1/c)(E^{a}v^{b}-E^{b}v^{a})+\varepsilon ^{abcd}v_{c}B_{d}$, e.g.,
Eqs. (39) and (40) in [16]. They are based on the theorem that any second
rank antisymmetric tensor can be decomposed into two vectors and a unit
time-like vector (the velocity vector/c). These equations show that in the
tensor formalism too both the electric and magnetic fields can be
represented by vectors.

Let us introduce the frame of \textquotedblleft fiducial\textquotedblright\
observers as the frame in which the observers who measure fields $E$ and $B$
are at rest. That frame with the standard basis $\{\gamma _{\mu }\}$ in it
is called the $\gamma _{0}$-frame. In the $\gamma _{0}$-frame $v=c\gamma
_{0} $ and therefore $E$ from (\ref{E1}) becomes $E=F\cdot \gamma _{0}$ and $%
B=-(1/c)\gamma _{5}(F\wedge \gamma _{0})$. Similarly, the decomposition (\ref%
{E2}) becomes $F=E\wedge \gamma _{0}+c(\gamma _{5}B)\cdot \gamma _{0}$. All
these quantities can be written as CBGQs in the standard basis $\left\{
\gamma _{\mu }\right\} $. This yields for $E$ and $B$

\begin{eqnarray}
E &=&E^{\mu }\gamma _{\mu }=0\gamma _{0}+F^{i0}\gamma _{i},  \notag \\
B &=&B^{\mu }\gamma _{\mu }=0\gamma _{0}+(1/2c)\varepsilon
^{0ijk}F_{kj}\gamma _{i}.  \label{gnl}
\end{eqnarray}%
Note that $\gamma _{0}=(\gamma _{0})^{\mu }\gamma _{\mu }$ with $(\gamma
_{0})^{\mu }=(1,0,0,0)$. The components of $F$ in the $\left\{ \gamma _{\mu
}\right\} $ basis give rise to the tensor (components) $F^{\mu \nu }=\gamma
^{\nu }\cdot (\gamma ^{\mu }\cdot F)=(\gamma ^{\nu }\wedge \gamma ^{\mu
})\cdot F$. It can be easily checked that in the $\gamma _{0}$-frame $E\cdot
\gamma _{0}=B\cdot \gamma _{0}=0$, which means that $E$ and $B$ are
orthogonal to $\gamma _{0}$. Hence, the temporal components of $E$ and $B$
are zero $E^{0}=B^{0}=0$ and only the spatial components remain
\begin{equation}
E^{i}=F^{i0},\quad B^{i}=(1/2c)\varepsilon ^{0ijk}F_{kj}.  \label{sko}
\end{equation}%
Thus, $E$ and $B$ actually refer to the 3D subspace orthogonal to the
specific timelike direction $\gamma _{0}$. In the $\gamma _{0}$-frame the
remaining spatial components of $E$ and $B$ from (\ref{sko}) are the same as
the components of the usual 3D $\mathbf{E}$ and $\mathbf{B}$, Eq. (\ref{ieb}%
), which are obtained by the usual identification of the components $F^{\mu
\nu }$ (implicitly taken in the standard basis) with the components of the
3D vectors $\mathbf{E}$ and $\mathbf{B}$. However, there is a very important
difference between the identifications (\ref{ieb}) and the above spatial
components of $E$ and $B$ in the $\gamma _{0}$-frame, Eq. (\ref{sko}), which
explicitly reveals that the usual procedure with the identifications (\ref%
{ieb}) is not correct in the 4D spacetime. As explained above, the
components of the 3D fields $\mathbf{E}$ and $\mathbf{B}$ in (\ref{ieb}) are
not the spatial components of the 4D quantities. They transform according to
the LPET, Eq. (11.148) in [2]. Also, the antisymmetric $\varepsilon $ tensor
is a third-rank antisymmetric tensor. On the other hand, the components of $%
E $ and $B$ in (\ref{gnl}), i.e., in (\ref{sko}), are the spatial components
of the 4D geometric quantities that are taken in the standard basis. They
transform according to the LT, which are given below, Eq. (\ref{LTE2}).
Also, the antisymmetric $\varepsilon $ tensor in (\ref{gnl}) is a
fourth-rank antisymmetric tensor. In the usual covariant approaches one
forgets about the temporal components $E^{0}$ and $B^{0}$ and simply makes
the identification of six independent components of $F^{\mu \nu }$ with
three components $E_{i}$ and three components $B_{i}$ according to the
relations (\ref{ieb}).

In Eq. (12) in [20], in the same way as in (\ref{E2}), the generalized
magnetization-polarization bivector $\mathcal{M(}x\mathcal{)}$ is decomposed
into two vectors, the polarization vector $P(x)$ and the magnetization
vector $M(x)$ and the unit time-like vector $u/c$
\begin{equation}
\mathcal{M}=P\wedge u/c+(MI)\cdot u/c^{2}.  \label{M1}
\end{equation}%
There is the rest frame for a medium, i.e., for $\mathcal{M}$, or $P$ and $M$%
, and therefore the vector $u$ in the decomposition (\ref{M1}) is identified
with bulk velocity vector of the medium in spacetime. Then, $P(x)$ and $M(x)$
are defined with respect to $u$ as
\begin{equation}
P=\mathcal{M}\cdot u/c,\quad M=cI(\mathcal{M}\wedge u/c)  \label{M2}
\end{equation}%
and it holds that $P\cdot u=M\cdot u=0$; only three components of $P$ and
three components of $M$ are independent since $\mathcal{M}$ is
antisymmetric. As in the case with $F$ in (\ref{E2}) and (\ref{E1}), it is
visible from (\ref{M2}) that $P$ and $M$ depend not only on $\mathcal{M}$
but on $u$ as well.

Here, we briefly examine the Lorentz invariant field equations for vacuum
and for a magnetized and polarized moving medium. Inserting (\ref{E2}) into
the field equation for $F$ (\ref{MEF}) one finds the field equation in terms
of $E$ and $B$;
\begin{equation}
\partial (E\wedge (v/c)+(IB)\cdot v)=j/\varepsilon _{0}c.  \label{ebf}
\end{equation}%
As explained in [8], Eq. (\ref{ebf}) represents the Lorentz invariant
generalization of the usual Maxwell equations. That form (\ref{ebf}) is the
most general form of the field equations with electric and magnetic fields
as properly defined quantities in the 4D spacetime.

The equation (\ref{ebf}) with the geometric product can be divided into the
vector part (with sources)
\begin{equation}
\partial \cdot (E\wedge v/c+(IB)\cdot v)=j/\varepsilon _{0}c  \label{A1}
\end{equation}%
and the trivector part (without sources)%
\begin{equation}
\partial \wedge (E\wedge v/c+(IB)\cdot v)=0.  \label{A2}
\end{equation}

In that form it is clear that \emph{it is not possible to separate the field
equation with sources for the} $E$ \emph{field from that one for the }$B$
\emph{field.} Thus, \emph{in the 4D spacetime}, \emph{the generalizations
with 4D geometric quantities of the usual Amp\`{e}re-Maxwell law and Gauss's
law are inseparably connected in one law} - \emph{Eq. }(\ref{A1}).
Similarly, \emph{in Eq. }(\ref{A2}), \emph{Faraday's law and the law that
expresses the absence of magnetic charge are also inseparably connected in
one law, which is expressed in terms of the 4D geometric quantities.} This
is an essential difference relative to Maxwell's equations with the
3-vectors $\mathbf{E}$ and $\mathbf{B}$.

The mathematical reason for such an inseparability is that, e.g., the
gradient operator $\partial $ is a vector field defined on the 4D spacetime.
If represented in some basis then its vector character remains unchanged
only when \emph{all its components together with associated basis vectors}
are taken into account in the considered equation. The same holds for other
vectors $E$, $B$, $j$, etc. and multivectors like $F$, $\mathcal{M}$, ... .
For example, in general, in the 4D spacetime, the current density vector $j$
is a well-defined physical quantity, but it is not the case with the usual
charge density $\rho $ and the usual current density $\mathbf{j}$ as a
3-vector. Similarly, in general, the gradient operator $\partial $ cannot be
divided into the usual time derivative and the spatial derivatives, e.g., in
the $\left\{ r_{\mu }\right\} $ basis with the \textquotedblleft
r\textquotedblright\ synchronization. In the 4D spacetime, an independent
physical reality is attributed to the position vector $x$, the gradient
operator $\partial $, the current density vector $j$, the vectors of the
electric and magnetic fields $E$ and $B$, respectively, etc., but not to the
3-vector $\mathbf{r}$ and the time $t$, to the 3-vectors $\mathbf{j}$, $%
\mathbf{E}$, $\mathbf{B}$, etc.

The generalization of (\ref{MEF}) to a moving medium is presented in [20]
and it is obtained simply replacing $F$ by $F+\mathcal{M}/\varepsilon _{0}$,
which yields the primary equations for the electromagnetism in moving media%
\begin{equation}
\partial (\varepsilon _{0}F+\mathcal{M})=j^{(C)}/c;\quad \partial \cdot
(\varepsilon _{0}F+\mathcal{M})=j^{(C)}/c,\ \partial \wedge F=0,  \label{F4}
\end{equation}%
where $j^{(C)}$ is the conduction current density of the \emph{free} charges
and $j^{(\mathcal{M})}=-c\partial \cdot \mathcal{M}$ is the
magnetization-polarization current density of the \emph{bound} charges. The
total current density vector $j$ is $j=j^{(C)}+j^{(\mathcal{M})}$. Inserting
the decomposition (\ref{M1}) into $j^{(\mathcal{M})}$ it follows that
\begin{equation}
j^{(\mathcal{M})}=-(\partial \cdot P)u+(u\cdot \partial )P+(1/c)[u\wedge
(\partial \wedge M)]I.  \label{jm}
\end{equation}%
In the standard basis $\{\gamma _{\mu }\}$ and in the rest frame of the
medium, $u=c\gamma _{0}$, $P^{0}=M^{0}=0$,
\begin{equation}
j^{(\mathcal{M})\mu }=(-c\partial _{k}P^{k},c\partial _{0}P^{i}-\varepsilon
^{0ijk}\partial _{j}M_{k}).  \label{jms}
\end{equation}%
In the usual formulation with the 3-vectors these components correspond to,
e.g., Eq. (18-61) in [23], i.e., $\rho =-\nabla \mathbf{P}$, $\mathbf{%
j=\partial P/}\partial t+\nabla \mathbf{\times M}$. In (\ref{jms}), the
components $P^{i}$, $M^{i}$ with the upper indices correspond to the
components of the 3-vectors. However, \emph{in the 4D spacetime it is not
correct to write the components of the properly defined vectors in terms of
the 3-vectors and operations with them. There are no 3-vectors in the 4D
spacetime.}

In most materials $\mathcal{M}$ is a function of the field $F$ and this
dependence is determined by the constitutive relations. In that case (\ref%
{F4}) are well-defined equations for $F$. Recently, the constitutive
relations and the magnetoelectric effect for moving media are investigated
in detail in [41].

Then, in [20], the general form of the field equation for a magnetized and
polarized moving medium expressed in terms of $E(x)$, $B(x)$, $P(x)$ and $%
M(x)$ is obtained by the insertion of Eqs. (\ref{E2}) and (\ref{M1}) into
the field equation (\ref{F4}). It is Eq. (15) in [20], which is the
generalization of Eq. (\ref{ebf}) (with the geometric product) to the moving
media. The generalizations of Eq. (\ref{A1}) to the moving media is given by
Eq. (16) in [20], the vector part (with sources), i.e., in the
\textquotedblleft source representation\textquotedblright\ by the equation

\begin{equation}
\partial \cdot \{\varepsilon _{0}[E\wedge v/c+(IB)\cdot
v]\}=j^{(C)}/c-\partial \cdot \lbrack P\wedge u/c+(1/c^{2})(MI)\cdot u],
\label{Ej}
\end{equation}%
according to which the sources of $E$ and $B$ fields are $j^{(C)}$ and $P$
and $M$. Obviously, from (\ref{Ej}), \emph{it is not possible to separate
the field equation with sources for the} $E$ \emph{field from that one for
the }$B$ \emph{field.} This is an essential difference relative to Maxwell's
equations with the 3-vectors $\mathbf{E}$, $\mathbf{B}$, $\mathbf{P}$ and $%
\mathbf{M}$. The field equation without sources, the trivector part, remains
unchanged relative to the corresponding equation for vacuum (\ref{A2})%
\begin{equation}
\partial \wedge \lbrack E\wedge v/c+(IB)\cdot v]=0.  \label{MP2}
\end{equation}%
In Eq. (\ref{Ej}), i.e., in Eq. (18) in [20], there are two different
velocities $u$ and $v$ and such an equation is not previously reported in
the physics literature.

As stated in [20], \emph{Eq.} (\ref{F4}), i.e.,\emph{\ Eqs. }(\ref{Ej})\emph{%
\ and }(\ref{MP2})\emph{\ comprise and generalize all usual Maxwell's
equations (with 3-vectors) for moving media.}

Again, as in the above discussion for the vacuum, \emph{in the 4D spacetime}%
, in contrast to the usual formulation of electromagnetism with the
3-vectors $\mathbf{E}$, $\mathbf{B}$, $\mathbf{P}$, $\mathbf{M}$,\emph{\ }$%
\mathbf{j}$, ... , there are no two laws, the Amp\`{e}r-Maxwell law and
Gauss's law, but \emph{only one law}, that is expressed by Eq. (\ref{Ej})
and the same for other two laws and Eq. (\ref{MP2}).

In the same way as in [12] one can derive the expression for the Lorentz
force density $k_{L}$,
\begin{equation}
k_{L}=F\cdot j/c=(1/c)F\cdot (j^{(C)}-c\partial \cdot \mathcal{M}).
\label{kj}
\end{equation}%
Inserting the decompositions (\ref{E2}) and (\ref{M1}), i.e., (\ref{E2}) and
(\ref{jm}), into $k_{L}$ (\ref{kj}) one can find $k_{L}$ expressed in terms
of $E(x)$, $B(x)$, $P(x)$ and $M(x)$. This is discussed in Sec. 9 in
connection with the \textquotedblleft charge-magnet
paradox.\textquotedblright \bigskip

\noindent \textbf{5. The LT of vectors }$E$ \textbf{and} $B$\textbf{; both }$%
F$ \textbf{and}

\textbf{the observer are transformed}\textit{\bigskip }

As seen from (\ref{E2}) and (\ref{E1}) all quantities $F$, $E$, $B$ and $v$
are abstract 4D geometric quantities. If these geometric quantities from (%
\ref{E2}) and (\ref{E1}) are represented in some basis then they contain
both components and basis vectors. In his fundamental work, Minkowski, in
Sec. 11.6\ in\emph{\ }[21], wrote the relation (55) that corresponds to (\ref%
{E2}), but he considered that the quantities $w$, $\Phi $ and $\Psi $, which
correspond to our $v$, $E$ and $B$, are $1\times 4$ matrices and that $F$ is
a $4\times 4$ matrix. Their components are implicitly determined in the
standard basis. In Sec. 11.6 in [21], the next paragraph below Eq. (44),
Minkowski described how $w$ and $F$ separately transform under the LT $A$
(the matrix of the LT is denoted as $A$ in [21]) and then how the product $%
wF $ transforms. Thus, he wrote
\begin{equation}
w^{\prime }=wA  \label{w}
\end{equation}%
for the LT of the velocity vector $w$ and
\begin{equation}
F^{\prime }=A^{-1}FA  \label{fcr}
\end{equation}%
for the LT of the field-strength tensor. Then, the mathematically correct LT
of $\Phi =wF$ are
\begin{equation}
\Phi =wF\longrightarrow \Phi ^{\prime }=(wA)(A^{-1}FA)=(wF)A=\Phi A,
\label{M3}
\end{equation}%
which means that under the LT both quantities, the velocity $w$ and $F$ are
transformed and their product transforms as any other vector (i.e., in [21],
as an $1\times 4$ matrix) transforms. The most important thing is that \emph{%
the electric field vector }$\Phi $ \emph{transforms by the LT again to the
electric field vector }$\Phi ^{\prime }$\emph{; there is no mixing with the
magnetic field} $\Psi $.

These correct LT of the electric and magnetic fields are reinvented and
generalized in terms of 4D geometric quantities in [6-11]. In the geometric
algebra, the LT (the active ones)\ are described by rotors $R$, $R\widetilde{%
R}=1$, where the reverse $\widetilde{R}$ is defined by the operation of
reversion according to which $\widetilde{AB}=\widetilde{B}\widetilde{A}$,
for any multivectors $A$ and $B$, $\widetilde{a}=a$, for any vector $a$, and
it reverses the order of vectors in any given expression. For boosts in an
arbitrary direction the rotor $R$ is given by Eq. (8) in [7, 9], or Eq. (10)
in [11], as
\begin{equation}
R=(1+\gamma +\gamma \gamma _{0}\beta )/(2(1+\gamma ))^{1/2},  \label{LTR}
\end{equation}%
where $\gamma =(1-\beta ^{2})^{-1/2}$, the vector $\beta $ is $\beta =\beta
n $, $\beta $ on the r.h.s. of that equation is the scalar velocity in units
of $c$ and $n$ is not the basis vector but any unit space-like vector
orthogonal to $\gamma _{0}$. Then, \emph{any multivector} $N$ \emph{%
transforms by active LT in the same way,} i.e., as
\begin{equation}
N\rightarrow N^{\prime }=RN\widetilde{R}.  \label{RM}
\end{equation}%
Hence, vector $E$ transforms by the LT $R$ as $E\longrightarrow E^{\prime
}=RE\widetilde{R}$. In the $\gamma _{0}$-frame, $v=c\gamma _{0}$ is taken in
(\ref{E1}). Then $E$ becomes $E=F\cdot \gamma _{0}$ and it transforms under
the LT in the same manner as in Minkowski's relation (\ref{M3}), i.e., that
both $F$ and $v$ are transformed by the LT $R$ as
\begin{equation}
E=F\cdot \gamma _{0}\longrightarrow E^{\prime }=R(F\cdot \gamma _{0})%
\widetilde{R}=(RF\widetilde{R})\cdot (R\gamma _{0}\widetilde{R}).
\label{LM3}
\end{equation}%
These correct LT give that

\begin{equation}
E^{\prime }=E+\gamma (E\cdot \beta )\{\gamma _{0}-(\gamma /(1+\gamma ))\beta
\}.  \label{LTE1}
\end{equation}%
\emph{In the same way vector }$B$ \emph{transforms and vectors} $P$, $M$
\emph{as well}, \emph{but for} $P$ \emph{and} $M$ \emph{the LT,} \emph{like}
(\ref{LTE1}), \emph{are the transformations from the rest frame of the
medium }($u=c\gamma _{0}$). For boosts in the direction $\gamma _{1}$ one
has to take that $\beta =\beta \gamma _{1}$ (on the l.h.s. is vector $\beta $
and on the r.h.s. $\beta $ is a scalar) in the above expression for the
rotor $R$ (all in the standard basis). Hence, in the $\left\{ \gamma _{\mu
}\right\} $ basis and when $\beta =\beta \gamma _{1}$ Eq. (\ref{LTE1})
becomes
\begin{equation}
E^{\prime \nu }\gamma _{\nu }=-\beta \gamma E^{1}\gamma _{0}+\gamma
E^{1}\gamma _{1}+E^{2}\gamma _{2}+E^{3}\gamma _{3}.  \label{LTE2}
\end{equation}%
As already mentioned in Sec. 1, the relations (\ref{LTE1}) and (\ref{LTE2})
are the fundamental results, which show that \emph{under the
relativistically correct LT the electric field vector }$E$ \emph{transforms
again to the electric field vector }$E^{\prime }$\emph{; there is no mixing
with the magnetic field} $B$. The same happens with vectors $P$ and $M$. The
same fundamental result can be obtained if electric and magnetic fields are
represented, e.g., by bivectors as in [9]. In general, it can be stated that
\emph{the LT always transform the 4D algebraic object (vector, bivector)
representing the electric field only to the electric field, and similarly
for the magnetic field.}

It is important to note that $E^{\prime }$ (\emph{and} $B^{\prime }$) \emph{%
from} (\ref{LTE1}) \emph{and} (\ref{LTE2}) \emph{are not orthogonal to} $%
\gamma _{0},$ i.e., \emph{they} \emph{have} \emph{temporal components} $\neq
0.$ They do not belong to the same 3D subspace as $E$ and $B$, but they are
in the 4D spacetime spanned by the whole standard basis $\left\{ \gamma
_{\mu }\right\} $.

The same components as in (\ref{LTE2}) would be obtained for $\Phi ^{\prime
}=\Phi A$ in Minkowski's relation (\ref{M3}) if the components of $w$ are $%
(0,0,0,ic)$ in his notation, which corresponds to $v=c\gamma _{0}$ in our
formulation. It is worth noting that only in Sec. 11.6 in [21] Minkowski
dealt with vectors (only components) $w$, $\Phi $ and $\Psi $, but in the
rest of [21] he exclusively dealt with the usual 3-vectors $\mathbf{v}$, $%
\mathbf{E}$ and $\mathbf{B}$ (our notation) and not with correctly defined
vectors $w$, $\Phi $ and $\Psi $.

In Sec. 11 under the title \textquotedblleft Minkowski in 1908, and Ivezi%
\'{c} Since 2003: Lorentz Covariance\textquotedblright\ in the third paper
in [19] Oziewicz, from the mathematical point of view, nicely explains the
results obtained in my papers [6-11]. (The references in the quoted part
refer to the mentioned Oziewicz's paper.) He states:

\textquotedblleft Ivez\'{\i}\'{c} observed the logical and mathematical
inconsistency of textbook treatments of the Lorentz-covariance since 2003.
He noted that it is illogical to consider a closed differential biform F to
be Lorentz-covariant, and at the same time, keep observer's time-like vector
field, a `4-velocity', P$\simeq $(1,0,0,0), to be
Lorentz-invariant-absolute. For example, compare how an absolute observer is
hidden in calculations presented in (Misner, Thorne \&Wheeler [31], Chapt.
3).

Minkowski [1], and then Ivezi\'{c} [7-10], observed correctly that if a
Lorentz transformation is an isomorphism of a vector space, then the entire
algebra of tensor fields must be Lorentz-covariant. Every vector is
Lorentz-covariant, and an observer-monad timelike vector field, also must be
Lorentz-covariant. All tensor fields, F and P, must be Lorentz-covariant. An
active Lorentz transformation must act on all tensor fields, including an
observer's time-like vector field. Hence electromagnetic field F, potential
A , and Paul P, must be Lorentz-covariant (Ivezi\'{c} [7-10]).

Instead of Fock's and Jackson's transformations (10.2) - (10.3), (our Eqs. (%
\ref{EP1}) and (\ref{J1}), my remark) Ivezi\'{c} defined the
Lorentz-covariance for the compound electric and magnetic fields, (7.2), (it
corresponds to our Eq. (\ref{E1}), my remark) exactly as defined by
Minkowski in [1], \S 11.6, just before formula (46). We stress that
Minkowski in [1] does not in practice use his definition of
Lorentz-covariance. Instead of (10.2) - (10.3), the Lorentz transformation
of electric and magnetic concomitant vector fields according to the
Minkowski and Ivezi\'{c} definition of Lorentz covariance is:
...\textquotedblright\ given by Oziewicz's equations (11.4) - (11.9). His
relation (11.5) is our Eq. (\ref{LTE1}).

Here, it is at place to give an interesting remark regarding Oziewicz's
papers [19]. It has often been argued that it cannot be that the 3-vectors $%
\mathbf{E}$, $\mathbf{B}$, $\mathbf{P}$ and $\mathbf{M}$ and their
transformations (\ref{JCB}) and (\ref{ps}), or (\ref{psa}), have to be
replaced by the 4D geometric quantities, e.g., vectors $E$, $B$, $P$ and $M$
and by their mathematically and relativistically correct LT (\ref{LTE1}) and
(\ref{LTE2}). On the other hand Oziewicz, in difference to all others,
correctly considers from the outset that \emph{there are no 3-vectors in the
4D spacetime.} But, he incorrectly considers that, e.g., the transformations
of the 3-vectors $\mathbf{E}$ and $\mathbf{B}$, Eq. (\ref{JCB}), i.e.,
Jackson's Eq. (11.149) in [2], are his equations (10.2) - (10.3), i.e., our
Eqs. (\ref{EP1}) and (\ref{J1}). His equations (10.2) - (10.3) are the
equations with 4D geometric quantities and they correspond to our Eqs. (\ref%
{EP1}) and (\ref{J1}) given below, whereas Eq. (\ref{JCB}) contains \emph{%
only the 3-vectors} $\mathbf{E}$, $\mathbf{B}$, $\mathbf{E}^{\prime }$, $%
\mathbf{B}^{\prime }$ \emph{and the velocity 3-vector}. Thus, his equations
(10.2) - (10.3) are not \textquotedblleft Fock's and Jackson's
transformations.\textquotedblright

There is a very important consequence of the LT (\ref{LTE1}) and (\ref{LTE2}%
), or the same for $P$ and $M$. As mentioned above, \emph{under the
relativistically correct LT the polarization vector }$P$ \emph{transforms
again to the polarization vector }$P^{\prime }$\emph{; there is no mixing
with the magnetization vector} $M$. On the other hand, according to (\ref%
{JCB}) if there is an external, static, magnetic field (3-vector) outside,
e.g., a stationary current loop then there is the magnetic field \emph{and a
static electric field (3-vector) }outside the same current loop which moves
with uniform 3-velocity $\mathbf{U}$.

According to the LT (\ref{LTE1}) and (\ref{LTE2}), if there is an electric
field outside \emph{moving} magnet \emph{it would necessary need to exist
outside the same but stationary magnet}. Thus, in the Amp\`{e}rian approach,
there is a polarization vector $P$ (remember that $P$ is a 4D geometric
quantity and not a 3-vector) for a stationary permanent magnet as an
assembly of small current loops. As explained in Sec. 7.1 below, this
happens because every current loop behaves like an electric dipole at points
far from that loop. Then, that $P$ induces an external electric field. This
will be discussed in much more detail in Sec. 8 below.\bigskip

\noindent \textbf{6. The usual transformations of vectors }$E$ and $B$;

\textbf{only }$F$ \textbf{is transformed but not the observer\bigskip }

Let us examine what will be obtained if in the transformation of $E=F\cdot
\gamma _{0}$ only $F$ is transformed by the LT $R$, Eq. (\ref{LTR}), but not
the velocity of the observer $v=c\gamma _{0}$. Of course, it will not be the
LT of $E=F\cdot \gamma _{0}$, because they are given by Eq. (\ref{LM3}).
Thus
\begin{equation}
E=F\cdot \gamma _{0}\longrightarrow E_{F}^{\prime }=(RF\widetilde{R})\cdot
\gamma _{0}=F^{\prime }\cdot \gamma _{0}.  \label{EP1}
\end{equation}%
This procedure yields that

\begin{equation}
E_{F}^{\prime }=\gamma \{E+(\beta \wedge \gamma _{0}\wedge cB)I\}+(\gamma
^{2}/(1+\gamma ))\beta (\beta \cdot E).  \label{J1}
\end{equation}%
If Eqs. (\ref{EP1}) and (\ref{J1}) are written in the standard basis and if
it is taken that $\beta =\beta \gamma _{1}$, then they become
\begin{eqnarray}
E_{F}^{\prime } &=&F^{\prime }\cdot \gamma _{0}=0\gamma _{0}+F^{\prime
i0}\gamma _{i}=E_{F}^{\prime \nu }\gamma _{\nu }  \notag \\
&=&E^{1}\gamma _{1}+\gamma (E^{2}-c\beta B^{3})\gamma _{2}+\gamma
(E^{3}+c\beta B^{2})\gamma _{3}.  \label{J2}
\end{eqnarray}%
Similarly, we find for $B_{F}^{\prime }$%
\begin{eqnarray}
B_{F}^{\prime } &=&-(1/c)\gamma _{5}(F^{\prime }\wedge \gamma _{0})=0\gamma
_{0}+(1/2c)\varepsilon ^{0ijk}F_{kj}^{\prime }\gamma _{i}=B_{F}^{\prime \nu
}\gamma _{\nu }  \notag \\
&=&B^{1}\gamma _{1}+(\gamma B^{2}+\beta \gamma E^{3}/c)\gamma _{2}+(\gamma
B^{3}-\beta \gamma E^{2}/c)\gamma _{3}.  \label{B}
\end{eqnarray}%
From the transformations (\ref{J2}) and (\ref{B}) one simply finds the
transformations of the spatial components $E_{F}^{\prime i}$ and $%
B_{F}^{\prime i}$%
\begin{equation}
E_{F}^{\prime i}=F^{\prime i0},\quad B_{F}^{\prime i}=(1/2c)\varepsilon
^{0ijk}F_{kj}^{\prime },  \label{sk1}
\end{equation}%
which is the relation (\ref{sko}) but with the primed quantities.

It is seen from (\ref{J2}) that \emph{the components of the transformed} $%
E_{F}^{\prime }$ \emph{are expressed by the mixture of components of} $E$
\emph{and} $B$. \emph{The same conclusion follows for} $B_{F}^{\prime }$
\emph{from} (\ref{B}).

The transformation (\ref{J1}) can be compared with the LPET for the 3-vector
$\mathbf{E}$ that are given by the first equation in (\ref{JCB}), and Eq. (%
\ref{J2}) can be compared with Eq. (11.148) in [2], i.e., with Eq. (\ref{ee}%
) here. Remember that in Eq. (\ref{JCB}) $\mathbf{E}^{\prime }$, $\mathbf{E}$%
, $\mathbf{B}^{\prime }$, $\mathbf{B}$ and $\mathbf{\beta }$ are all the
usual 3-vectors. The comparison of Eq. (\ref{J2}) with Eq. (11.148) in [2]
shows that \emph{the transformations of components (taken in the standard
basis) of} $E_{F}^{\prime }$ \emph{are exactly the same as the
transformations of }$E_{x,y,z}$ \emph{from Eq. }(11.148)\emph{\ in }[2]\emph{%
.} The same conclusion holds for the comparison of Eq. (\ref{B}) and $%
B_{x,y,z}$ from Eq. (11.148) in [2]. The result that the components in (\ref%
{J2}) are the same as the components of $\mathbf{E}^{\prime }$ from (\ref%
{JCB}) is completely understandable. Namely, (\ref{J1}) and (\ref{J2}) are
obtained by the application of the LT \emph{only} to $F$. On the other hand,
it is already stated in Sec. 3.1 that the LPET of the components of $\mathbf{%
E}$ and $\mathbf{B}$ are derived assuming that they transform under the LT
as the components of $F^{\alpha \beta }$ transform, Eqs. (\ref{ieb}) and (%
\ref{eb2}).

In contrast to the LT of $E$ (\ref{LTE2}) (and the same for $B$), it is
visible from (\ref{J2}), (\ref{B}) and (\ref{sk1}) that $E_{F}^{\prime }$
\emph{and} $B_{F}^{\prime }$ \emph{are again in the 3D subspace of the} $%
\gamma _{0}$ \emph{-} \emph{observer,} as it holds for $E$ and $B$ in the $%
\gamma _{0}$-frame, Eqs. (\ref{gnl}) and (\ref{sko}). Thus for the
transformed $E_{F}^{\prime }$ \emph{and} $B_{F}^{\prime }$ again hold that $%
E_{F}^{\prime 0}=B_{F}^{\prime 0}=0$, i.e., that $E_{F}^{\prime }\cdot
\gamma _{0}=B_{F}^{\prime }\cdot \gamma _{0}=0$ as for $E$ and $B$ in the $%
\gamma _{0}$-frame. This shows in another way that \emph{the LPET} (\ref{J2}%
), (\ref{B}) \emph{and} (\ref{sk1})\emph{\ are not the LT, since the LT
cannot transform some quantity from the 3D subspace again only to the 3D
subspace.}

The transformations (\ref{EP1}) and (\ref{J2})-(\ref{sk1}) are first
discussed in detail in [6-11] and compared with the LPET (11.148) and
(11.149) from [2], whereas the general form of $E_{F}^{\prime }$, Eq. (\ref%
{J1}), is first given in [19].

We now point out another difference between the LT and the LPET. If instead
of the active LT we consider the passive LT then, e.g. the vector $E=E^{\nu
}\gamma _{\nu }=E^{\prime \nu }\gamma _{\nu }^{\prime }$ will remain
unchanged, because the components $E^{\nu }$ transform by the LT and the
basis vectors $\gamma _{\nu }$ by the inverse LT leaving the whole $E$
invariant under the passive LT. Of course, the same holds for all bases
including those with nonstandard synchronizations, as discussed, e.g., in
[16] and [33]. For the $\left\{ r_{\mu }\right\} $ basis, this can be easily
proved using $R_{\;\nu }^{\mu }$, i.e., Eqs. (\ref{re}) and (\ref{er1});
\begin{equation}
E=E^{\nu }\gamma _{\nu }=E^{\prime \nu }\gamma _{\nu }^{\prime }=E_{r}^{\nu
}r_{\nu }=E_{r}^{\prime \nu }r_{\nu }^{\prime }.  \label{ein}
\end{equation}%
The primed quantities in both bases are the Lorentz transforms of the
unprimed ones. For the general form of the LT, that is independent of the
chosen system of coordinates, including different synchronizations, see,
e.g., Eq. (1) in [16]. The LT in the $\{r_{\mu }\}$ basis are given in the
same paper by Eq. (2) or Eq. (21) in [33].

\emph{This invariance of} $E$ \emph{means that the electric field} $E$ \emph{%
is the same physical quantity for all relatively moving observers and for
all bases used by them. }In the same way this requirement has to be
fulfilled for any well-defined 4D quantity.

It is not so with the 3-vector $\mathbf{E}$ and its LPET, or, equivalently,
with $E_{F}^{\prime }$. Namely, $\mathbf{E=}E_{x}\mathbf{i}+E_{y}\mathbf{j}%
+E_{z}\mathbf{k}$ is completely different than $\mathbf{E}^{\prime }$, $%
\mathbf{E}^{\prime }\mathbf{=}E_{x}^{\prime }\mathbf{i}^{\prime
}+E_{y}^{\prime }\mathbf{j}^{\prime }+E_{z}^{\prime }\mathbf{k}^{\prime }$
from (\ref{JCB}), $\mathbf{E}\neq \mathbf{E}^{\prime }$, and the same holds
for $E_{F}^{\prime }$, i.e., $E^{\nu }\gamma _{\nu }\neq E_{F}^{\prime \nu
}\gamma _{\nu }^{\prime }$. This means that although $\mathbf{E}$ and $%
\mathbf{E}^{\prime }$ are measured by different observers \emph{they are not
the same quantity for such relatively moving observers. }The observers are
not looking at the same physical object, here the electric field vector, but
at two different objects. \emph{Every observer makes measurement of its own
3-vector field,} $\mathbf{E}$ \emph{and} $\mathbf{E}^{\prime }$, \emph{and
such measurements are not related by the LT.} As far as relativity is
concerned the quantities, e.g., $\mathbf{E}$ and $\mathbf{E}^{\prime }$,
i.e., $E^{\nu }\gamma _{\nu }$ and $E_{F}^{\prime \nu }\gamma _{\nu
}^{\prime }$, are not related to one another. Their identification is a
typical case of \emph{mistaken identity}.

In the 4D geometric approach, i.e., in the ISR, different relatively moving
inertial 4D observers can compare only 4D quantities, here $E^{\nu }\gamma
_{\nu }$ and $E^{\prime \nu }\gamma _{\nu }^{\prime }$, because they are
connected by the LT. The experimentalists have to measure \emph{all
components} of 4D quantities, here of $E$, in both frames $S^{\prime }$ and $%
S$. The observers in $S^{\prime }$ and $S$ are able to compare only such
complete set of data which corresponds to the \emph{same} 4D geometric
quantity. Hence, from the point of view of the ISR the transformations for $%
E_{F.}^{\prime i}$ and $B_{F.}^{\prime i}$, Eq. (\ref{sk1}), i.e., Eq.
(11.148) in [2], are not the LT. Therefore, contrary to the general belief,
\emph{it is not true from the geometric approach viewpoint that, }e.g., Sec.
11.10 in [2]: \textquotedblleft A purely electric or magnetic field in one
coordinate system will appear as a mixture of electric and magnetic fields
in another coordinate frame.\textquotedblright \bigskip

\noindent \textbf{7. Clausius' hypothesis and the second-order electric
field outside}

\textbf{a stationary superconductor with steady current}\textit{\bigskip }

\noindent \textit{7.1 The second-order electric field outside a stationary
superconductor }

\textit{with steady current\bigskip }

As stated in Sec. 3.2, in the classical model, the magnetic dipoles are
considered as little current loops. In Sec. 18-4 in [23] and Sec. 6.5. in
[22], and in all other usual approaches, it is assumed that a stationary
current loop \emph{with steady current} is globally and \emph{locally}
charge neutral (Clausius' hypothesis, (1877)), i.e., \emph{it is simply
supposed }that in the ions' rest frame $S$ the charge density of the moving
electrons is $\rho _{-}=-\rho _{0}$, where $\rho _{0}$ is the positive
charge density for the wire at rest but without a current. Clausius (see
Ref. [4] in the first paper in [28]) stated that hypothesis in another but
equivalent way, i.e., he stated, as an \textquotedblleft experimental
assumption\textquotedblright\ that a \textquotedblleft closed current in a
stationary conductor exerts no force on stationary
electricity.\textquotedblright\ But, it is also assumed that the same
current loop does not remain locally neutral when observed from another
inertial frame. Namely, in [23], [22], ..., it is argued that the legs
parallel to the 3-velocity $\mathbf{U}$ will carry the charges equal in
magnitude but opposite in sign, because they are Lorentz contracted, whereas
the legs perpendicular to $\mathbf{U}$ remain uncharged, because there is no
Lorentz contraction for them. Thus, an electric dipole moment given by the
expression (\ref{plp}) is obtained. Hence, there is an external, static,
magnetic field (3-vector) outside the stationary current loop, but there is
the magnetic field \emph{and a static electric field (3-vector) }outside the
same current loop which moves with uniform 3-velocity $\mathbf{U}$. Note
that the legs of the current loop are treated as that they are infinite
wires with steady currents. It can be seen from the mentioned textbooks,
[23], [22], that such result is obtained using \emph{the Lorentz contraction}
and \emph{the relativity of simultaneity. }In all these derivations it is
also used the conventional definition of charge in terms of 3D quantities,
\begin{equation}
Q=(1/c)\int_{V(t)}j^{0}(\mathbf{r},t)dV.  \label{q1}
\end{equation}%
In that definition the volume $V(t)$ is taken at a particular coordinate
time $t$ and it is stationary in some inertial frame of reference $S.$ The
values of the charge density $\rho (\mathbf{r},t)=j^{0}(\mathbf{r},t)/c$ are
taken simultaneously for all $\mathbf{r}$ in $V(t)$. It is supposed in all
usual treatments (see, particularly, the well-known Purcell's textbook [24])
that the volume elements $dV^{\prime }$ are Lorentz \textquotedblleft
contracted\textquotedblright\ in a relatively moving inertial frame of
reference $S^{\prime }$ and all of them, i.e., the whole volume $V^{\prime
}(t^{\prime })$, are taken simultaneously at some arbitrary $t^{\prime }$ in
$S^{\prime }$. \emph{The coordinate time} $t^{\prime }$\emph{\ in }$%
S^{\prime }$\emph{\ is not connected in any way with }$t$\emph{\ in }$S$.
Also, it is assumed that $j^{0}$ from $S$ is transformed (using the Lorentz
\textquotedblleft contraction\textquotedblright ) only to $j^{\prime 0}$ in $%
S^{\prime }$ and all $j^{\prime 0}$ are taken simultaneously at the same $%
t^{\prime }$ in $S^{\prime }$. The new%
\begin{equation}
Q^{\prime }=(1/c)\int_{V^{\prime }(t^{\prime })}j^{\prime 0}(\mathbf{r}%
^{\prime },t^{\prime })dV^{\prime }  \label{qc}
\end{equation}%
in $S^{\prime }$ is considered to be equal to the charge $Q$ in $S$, $%
Q^{\prime }=Q$, (the total charge is invariant). But we remark that the
charge $Q$ defined in such a manner cannot be invariant under the LT. The LT
cannot transform one component $j^{0}$ from an inertial frame of reference $%
S $ to the same component $j^{\prime 0}$ in $S^{\prime }.$ Also, if all $%
j^{0}$ values are taken simultaneously at some $t$ in $S$ then the LT cannot
transform them to the values $j^{\prime 0}$ which are again all
simultaneous, but now \emph{at some arbitrary} $t^{\prime }$ in $S^{\prime }$%
. This consideration shows that such usual definition of charge cannot be
the relativistically correct definition.

In the 4D spacetime, two relatively moving observers cannot compare spatial
lengths taken alone, which are synchronously determined for the observer,
see Appendix here. Similarly, in the 4D spacetime, the temporal distances
taken alone are not well-defined and two relatively moving observers cannot
compare them. For the thorough discussion of these usual definitions of the
Lorentz contraction and the time dilation see, e.g., Secs. 4. - 4.2 and
Figs. 3 and 4 in [16]. Also, a clear presentation of the relativistic
incorrectness of the Lorentz contraction is already given in Sec. 2.2 in
[27]. In the ISR, the properly defined quantity is the distance vector and
the spacetime length, which is a Lorentz scalar, see, e.g., Secs. 3 - 3.2
and Figs. 1 and 2 in [16] and also Sec. 2.1 in [27].

Furthermore,\emph{\ }the above mentioned conventional definition of charge
in terms of 3D quantities is objected in Sec. 3 in [27] and Sec. 5.3 in
[16]. There, it is shown that such definition with 3D quantities has to be
replaced by the definition in terms of 4D geometric quantities, i.e., the
charge is defined as a Lorentz scalar. The total electric charge $Q$ in a
three-dimensional hypersurface $H$ (with two-dimensional boundary $\delta H$%
), as a Lorentz scalar, is defined by the equation%
\begin{equation}
Q_{\delta H}=(1/c)\int_{H}j\cdot ndH,  \label{qh}
\end{equation}%
where $j$ is the current density vector and the vector $n$ is the unit
normal to $H$.

In Secs. 3 and 3.1 in [27], the external electric field for an infinite wire
with a steady current and the Clausius hypothesis are examined in detail.

In the prerelativistic physics and in Einstein's formulation of the
relativistic physics the charge density is well-defined quantity both for
charges at rest and for the moving charges. As discussed in Sec. 3 in [27],
it is considered in the usual approaches, e.g., [24-26], that the charge
density of the moving charges is properly defined; it is enhanced by $\gamma
=(1-\beta ^{2})^{-1/2}$ relative to the proper charge density due to \emph{%
the Lorentz contraction of the moving volume.}\ Therefore, both in the
prerelativistic physics and in Einstein's formulation of the relativistic
physics the Clausius hypothesis is meaningful, i.e., it can be properly
formulated. Hence, in the usual approaches, \emph{there is no external
electric field for a stationary current-carrying conductor.} Of course, the
whole consideration refers to an ideal conductor or to a superconductor,
because for a stationary \emph{resistive} conductor carrying constant
current there is always an external static magnetic field \emph{and a time
independent external electric field that is proportional to the current and
which is caused by the distribution of surface charges on the conductor, }%
see, e.g., [42] and references therein.

On the other hand, as already mentioned several times, see, e.g., Sec. 3. in
[27] and Sec. 4 here, in the 4D spacetime, only the current density vector $%
j $ is a well-defined physical quantity, but not the usual charge density $%
\rho $ and the usual current density $\mathbf{j}$ as a 3-vector. \emph{In
the 4D spacetime it is not possible to give a definite physical meaning to
the charge density of moving charges.} As discussed in Appendix, the Lorentz
contracted length is meaningless in the 4D spacetime. As shown in Secs. 3 -
3.3 in [27], for an infinite wire with a steady current (the wire is
situated along the $x^{1}$ axis) and if the standard basis is introduced,
one can take that in $S^{\prime }$, in which the drift velocity 3-vector of
the electrons is zero, the current density vector of the electrons (in one
spatial dimension) is
\begin{equation}
j_{-}^{\prime }=(-c\rho _{0})\gamma _{0}+0\gamma _{1},  \label{ronula}
\end{equation}%
i.e., as that the proper charge density $\rho _{-}^{\prime }$ of the
electrons ($j^{\prime i}{}_{-}=0$) is equal to $-\rho _{0}$. This is
completely different than the Clausius hypothesis. Then, by means of (\ref%
{ronula}) and the LT one finds the current density vectors in $S$, the rest
frame of the wire, i.e., the lab frame, as (only components)%
\begin{equation}
j_{-}^{\mu }=(-c\gamma \rho _{0},-c\gamma \beta \rho _{0}),\mathrm{\quad }%
j^{\mu }=(c(1-\gamma )\rho _{0},-c\gamma \beta \rho _{0}),  \label{jotmi}
\end{equation}%
where $j=j^{\mu }\gamma _{\mu }$ is the total current density vector in $S$,
i.e., (components in the standard basis) $j^{\mu }=j_{-}^{\mu }+j_{+}^{\mu }$
and $j_{+}^{\mu }=(c\rho _{0},0)$. Observe that it holds that $j=j^{\mu
}\gamma _{\mu }=j^{\prime \mu }\gamma _{\mu }^{\prime }$, where the primed
quantities are the Lorentz transforms of the unprimed ones. The equations (%
\ref{ronula}) and (\ref{jotmi}) are Eqs. (11) and (12) in [27], respectively.

The same equations, (\ref{ronula}) (i.e., only components) and (\ref{jotmi})
were already obtained in [32]. There, in contrast to the usual approach, the
Lorentz contraction is introduced not only for the mean spacing between
moving ions in the $S^{\prime }$ frame, but also it is assumed that there is
a Lorentz contraction of the mean spacing between moving electrons in the
lab frame, i.e., in the stationary wire with steady current. This may seem
surprising that the same equations exist in [32] in which the Lorentz
contraction is used and in [27] and here, where the 4D geometric quantities
are used. But, the results obtained in [32] are not actually based on the
use of the Lorentz contraction, than on the \emph{assumption} that in the
electrons' rest frame $S^{\prime }$ the electrons' charge density $\rho
_{-}^{\prime }$ is $=-\rho _{0}.$ The Lorentz contraction of the mean
distance between moving electrons was only taken as the interpretation for
the assumption that $\rho _{-}^{\prime }=-\rho _{0}$. In the 4D geometric
approach from [27] and here Eq. (\ref{ronula}) is neither \emph{hypothesis}
as in the traditional approach with the 3D quantities and the Lorentz
contraction, nor the \emph{assumption} as in [32], but it is a consequence
of the covariant definition of an invariant charge (\ref{qh}) and of the
invariance of the rest length (see Appendix), i.e., it is the result of the
use of the correctly defined 4D geometric quantities.

The components in the standard basis of the electric and magnetic fields are
determined from the known current density vector $j^{\mu }$ (\ref{jotmi}) in
Secs. 3.2 - 3.3 in [27]. Taking that the rest frame of the wire is the $%
\gamma _{0}$-frame, i.e., that $E^{0}=0,\ E^{i}=F^{i0}$, then the external
electric field is $E=E^{\mu }\gamma _{\mu }=0\gamma _{0}+E^{i}\gamma _{i}$,
where the components are given by Eq. (22) in [27]; that field is in the
plane orthogonal to the wire and in the radial direction in that plane (the
wire is situated along the $x^{1}$ axis)
\begin{equation}
E^{0}=E^{1}=0,\ E^{2}=2k(1-\gamma )\rho _{0}ya^{-2},\ E^{3}=2k(1-\gamma
)\rho _{0}za^{-2},  \label{eovi}
\end{equation}%
where $k=1/4\pi \varepsilon _{0}$, $a^{2}=y^{2}+z^{2}$, $%
(x^{1},x^{2},x^{3})=(x,y,z)$. Then, it is concluded in Sec. 3.3. in [27]:
\textquotedblleft \emph{The equation }(\ref{eovi}) \emph{shows that the
observer who is at rest relative to a wire with steady current will see,
i.e., measure, the second-order electric field outside such a
current-carrying conductor.}\textquotedblright\ (\textquotedblleft the
second-order electric field\textquotedblright\ means that $E^{i}\propto
U^{2}/c^{2}$, where $U$ is the magnitude of the drift velocity 3-vector of
the electrons.) Note that such fields, but as the 3-vectors, are first
predicted on different grounds in [32], see the above discussion about the
assumption from [32] that $\rho _{-}^{\prime }=-\rho _{0}$. As already
mentioned, the second-order electric field (\ref{eovi}) exists in a
resistive wire with a constant current as well, but there it is much smaller
but the contribution to the external electric field that is caused by the
quasistatic surface charges.

Recently, the same treatment with the Lorentz contraction and the same
results as in [32] are presented in [31]. That work, [31], from the
theoretical point of view is almost the same as the treatment in [32], i.e.,
it is not with 4D geometric quantities and thus it is not a mathematically
and relativistically correct treatment. Several results from [32] are
incorrectly understood and interpreted in [31]. This will not be discussed
here since both papers deal with the Lorentz contraction. But [31] is an
important progress in the investigation of the existence of the second-order
electric field, because it presents, as asserted in [31]: \textquotedblleft
a new analysis of the experimental sensitivity required to observe the
hypothesized effect and analyzes the feasibility of several novel
experimental methods to make such an observation.\textquotedblright\ This
will be discussed below in Sec. 7.2 together with the discussion of the
already performed experiments and some suggested experiments.

Under the passive LT it holds that $E=E^{\mu }\gamma _{\mu }=E^{\prime \mu
}\gamma _{\mu }^{\prime }$. This essential feature of the approach with 4D
geometric quantities, i.e., of the ISR, shows that \emph{if the electric
field vector exists in one inertial frame of reference, say in the rest
frame of the electrons, the} $S^{\prime }$ \emph{frame, as in all usual
approaches, then it must necessary exist in the rest frame of the ions,
i.e., of the wire, the} $S$ \emph{frame. }

Similarly, it is obtained in [27] that the components in the standard basis
of the magnetic field are $B^{0}=0$ and $B^{i}$, which are the same as for
the usual expression for the magnetic field of an infinite straight wire
with current (only the components $j^{i}$ are $\gamma $ times bigger).
\begin{equation}
B^{0}=B^{1}=0,\ B^{2}=\gamma \mu _{0}Iya^{-2},\ B^{3}=\gamma \mu
_{0}Iza^{-2},  \label{be}
\end{equation}%
where $I=\rho _{0}AU$, $A$ is the cross-sectional area. The vectors of the
electric and magnetic fields in some relatively moving frame, e.g., the rest
frame of the electrons, can be obtained using (\ref{eovi}), (\ref{be}) and
the LT (\ref{LTE1}) and (\ref{LTE2}). Again, as for $E$, it holds that under
the passive LT $B$ is unchanged, $B=B^{\mu }\gamma _{\mu }=B^{\prime \mu
}\gamma _{\mu }^{\prime }$, as can be easily checked.

In Sec. 4 in [27], the same consideration is presented for a current loop
and it is shown that \emph{the second-order external electric field exists}
\emph{not only for a moving current loop}, as in the usual approaches, but
\emph{for the stationary current loop as well}. There are opposite charges
on opposite sides of a square loop with current, but the total charge of
that loop is zero. All these charges are invariant charges, which means that
they are the same for both, moving and stationary current loop. They are
defined as the Lorentz scalars, i.e., as in Eq. (\ref{qh}) for $Q_{\delta H}$%
. \emph{At points far from that current loop it behaves like an electric
dipole, but as a 4D geometric quantity.} It is incorrectly asserted in the
published paper in [27] that such a distribution of charges behaves like an
electric quadrupole. This is corrected in the v2 of the second paper in
[27]. However, it is worth noting that, in the same way as in the usual
approaches, the legs of the current loop are treated as that they are
infinite wires with steady currents.

Thus, in the 4D geometric approach, i.e., in the ISR, the relation (\ref{plp}%
), which is derived by the use of the Lorentz contraction and the time
dilation,\emph{\ }does not hold for the current loop and consequently the
relation (\ref{Pl1}) for the polarization $\mathbf{P}$, which is induced by
the movement of a permanent magnetization $\mathbf{M}^{\prime }$, does not
hold as well. The vector of the electric field outside a moving current loop
is not caused by \emph{the 3D polarization} $\mathbf{P}$, (\ref{Pl1}), but
it could be determined by the LT, the same as (\ref{LTE1}) and (\ref{LTE2}),
of \emph{the vector of the electric field of the same but stationary current
loop.\bigskip }

\noindent \textit{7.2 The experiments for the detection of the second-order
electric fields\bigskip }

The external second-order electric fields from a stationary,
superconducting, current loop, i.e., coil, have not yet detected. However,
in [28], an $I^{2}$ - dependent potential resulting from constant current in
closed superconducting coils has been reported and the same happened in the
first variation of experiments in [29], but not in the second one. It is
worth mentioning that all these experiments are sensitive only to a monopole
field and thus they cannot either support or disprove the theory presented
in [27] and here, which predicts a dipole field; very small (second-order)
external electric field. This happens because a direct contact with the
superconducting coil is used in the measurements [28], [29] of the $I^{2}$ -
dependent potential. In such measurements it can be only \textquotedblleft
seen\textquotedblright\ if some charge is created or destroyed. In order to
directly measure the external electric fields it is necessary to use a
non-contact method of measuring. Such a method is recently presented in
[30]. The experiments [30] are \textquotedblleft based essentially on the
detection of a non-zero force between a circular steady current and a
charge, both at rest in the Earth frame.\textquotedblright\ For the
experimental setup see Fig. 1. in both papers in [30]. Observe that a
Helmholtz coil $\gamma $ that is used in the experiments is a \emph{normal
metal with finite resistivity}. The authors of [30] considered that they
obtained a positive evidence for a non-zero force and that such results
\textquotedblleft show that \emph{local Lorentz invariance could in fact be
broken even in electromagnetic experiments} .. .\textquotedblright\ However,
their results have nothing to do with the breakdown of the local Lorentz
invariance because, as mentioned above, the standard Maxwell theory predicts
that there are always static magnetic fields and \emph{a time independent
electric field outside a normal conductor with a steady current. }Such an%
\emph{\ external electric field }causes\emph{\ }a non-zero force on a
stationary charge $q$ in the experiments in [30].

In order to \textquotedblleft see\textquotedblright\ the existence of the
external second-order electric fields the coil used in the experimental
setup in [30] would need to be a superconducting coil. Hence, \emph{we
propose to experimentalists to make the similar measurements as in }[30]%
\emph{, but using a superconducting coil.} It would be an important test of
the validity of the relations (\ref{ronula}) and (\ref{jotmi}), i.e., of the
relation (\ref{eovi}), or the validity of the usual approaches which assume
the Clausius hypothesis. Namely, it is often declared that the classical
electromagnetic theory predicts a zero external electric field for a
stationary superconducting coil. But, it is not true. Maxwell equations
enable one to find fields in the case that the sources are known. Hence,
Maxwell equations will give a zero external electric field only if one
supposes that in a stationary superconductor with a steady current the local
charge density (in the ions' rest frame and in the standard basis) is
everywhere zero. Inside the classical electromagnetic theory this statement
that the local charge density is everywhere zero is merely \emph{a hypothesis%
}, the Clausius hypothesis. This means that the possible existence of the
second-order external electric fields from steady currents in a stationary
superconductor is not at all in contradiction with classical
electromagnetism. However, as stated in [27] and mentioned above, in
contrast to the usual approaches in which there is no either physical or
mathematical justification for the Clausius hypothesis or for some other
choice, e.g., $\rho _{-}^{\prime }=-\rho _{0}$, in the 4D geometric approach
Eqs. (\ref{ronula}), (\ref{jotmi}) and (\ref{eovi}) are the consequences of
the use of correctly defined quantities in the 4D spacetime, i.e., the 4D
geometric quantities. The fields $E$ and $B$ with components in the standard
basis given by Eqs. (\ref{eovi}) and (\ref{be}), respectively, correctly
transform under the LT (\ref{LTE1}) and (\ref{LTE2}), which is not the case
with the 3-vectors $\mathbf{E}$ and $\mathbf{B}$ obtained in all previous
approaches.

Recently, an interesting possibility to experimentally investigate these
second-order external electric fields is proposed in [31]. As already
mentioned above, the theory from [31] is not relativistically correct, but
the proposed experiments with the cold ions could be the right way to detect
such small electric fields. Moreover, the estimated size of that electric
field could be even bigger than that one found in [31] if the experiments
would be made using a superconducting coil that is wound using a large
number of bifilar pairs.

Another possibility to study these second-order external electric fields by
the use of a non-contact method of measuring could be similar to that one
used to measure the Casimir force in [43], i.e., by the use of a torsion
balance.

As stated in Sec. 3.3 in [27], the second-order electric fields could play
an important role in many physical phenomena with steady currents,
particularly in tokamaks and astrophysics, where high currents exist, and in
superconductors, where the electric fields that are caused by surface
charges are absent.

Regarding the role of the second-order electric fields in tokamaks, we have
to mention that recently, [44] and references [1], [5] and [7] therein, the
existence of the radial electric field $E_{r}(r)$ in quasi-neutral tokamak
plasma is examined taking into account the Lorentz contraction of
\textquotedblleft an \textquotedblleft electron ring\textquotedblright\
circumference in steady state tokamak plasma rotating in toroidal direction
with current velocity $V_{e}(r)$ ... .\textquotedblright\ The similar
consideration with the Lorentz contraction of an \textquotedblleft electron
ring\textquotedblright\ was already reported in [45]. But, as we mentioned
several times, such theories with the Lorentz contraction are not the
relativistically correct theories. It would be very important for physics
that the experimentalists find the best way for the direct and precise
measurements of the second-order electric fields that are predicted in [27]
and here by the relativistically correct approach with 4D geometric
quantities.\bigskip

\noindent \textbf{8. The electric field outside a stationary permanent magnet%
}\textit{\bigskip }

An interesting consequence of the above consideration refers to the
existence of \emph{the electric field outside a stationary permanent magnet.}
It was mentioned at the end of [20]. Namely, as already stated, in the Amp%
\`{e}rian approach a permanent magnet is essentially an assembly of current
loops. However, as discussed above, \emph{the second-order external electric
field exists} \emph{not only for every moving current loop, but also for the
same stationary current loop}, which yields that \emph{the electric field
would need to exist not only outside a moving permanent magnet but outside a
stationary permanent magnet as well.} This conclusion is also supported by
the following argument. According to the LT (\ref{LTE1}) and (\ref{LTE2}),
if there is an electric field outside the \emph{moving} magnet \emph{it
would necessary need to exist outside the same but stationary magnet}. Thus,
in the Amp\`{e}rian approach, there is a polarization vector $P$ (remember
that $P$ is a 4D geometric quantity and not a 3-vector) for a stationary
permanent magnet as an assembly of small current loops. As explained above,
this happens because \emph{every current loop behaves like an electric
dipole at points far from that loop.} Then, that $P$ induces an external
electric field.

If we abandon the Amp\`{e}rian approach then, nevertheless, there is another
explanation for the possibility that a stationary permanent magnet possesses
an intrinsic polarization.

According to the well-known Uhlenbeck-Goudsmit hypothesis there is a
connection between the 3-vectors of the magnetic moment $\mathbf{m}$ of an
electron and its spin $\mathbf{S}$, $\mathbf{m}=\gamma _{S}\mathbf{S}$.
However, in the 4D spacetime, i.e., in the approach with 4D geometric
quantities, the 3-vectors $\mathbf{m}$ and $\mathbf{S}$ are not properly
defined quantities. In the 4D spacetime, as explained in [33], and in
[46-48], the primary quantity (with independent physical reality) for the
dipole moments is the dipole moment bivector $D$ (four-tensor $D^{ab}$ in
[33], [46-49]) of a fundamental particle. It is decomposed into the EDM
vector $p$, the MDM vector $m$ and the unit time-like vector $u/c$, where $u$
is the velocity vector of the particle. Then, $p$ and $m$ are derived from $%
D $ and the velocity vector of the particle $u$ according to the equations
\begin{eqnarray}
D &=&(1/c)[p\wedge u+(mI)\cdot u/c],  \notag \\
p &=&D\cdot u/c,\quad m=I(D\wedge u),  \label{d}
\end{eqnarray}%
Eq. (2) in [33] (but in the tensor notation). It holds that $p\cdot u=m\cdot
u=0$; only three components of $p$ and three components of $m$ are
independent since $\mathcal{M}$ is antisymmetric. In the particle's rest
frame (the $K^{\prime }$ frame) and the standard basis $\{\gamma _{\mu
}^{\prime }\}$, $u=c\gamma _{0}^{\prime }$, and using (\ref{d}), it follows
that $p^{\prime 0}=m^{\prime 0}=0$, $p^{\prime i}=D^{\prime i0}$, $m^{\prime
i}=(c/2)\varepsilon ^{0ijk}D_{jk}^{\prime }$. Therefore $p$ and $m$ can be
called the \textquotedblleft time-space\textquotedblright\ part and the
\textquotedblleft space-space\textquotedblright\ part, respectively, of the
dipole moment bivector $D$. But, these parts are written with quotation
marks because in all other relatively moving frames and in all other bases
the above identifications of the components of $p$ and $m$ with the
components of $D$ do not hold.

Similarly, it is shown in [33] (earlier, in [48] and Ref. [3] in [48]) that
the primary quantity \emph{with} \emph{definite physical reality} for the
\emph{intrinsic} angular momenta is the spin bivector $\mathcal{S}$
(four-tensor $S^{ab}$ in [33], [48]), which is decomposed into the usual
\textquotedblleft space-space\textquotedblright\ intrinsic angular momentum $%
S$, the \textquotedblleft time-space\textquotedblright\ intrinsic angular
momentum $Z$ and the unit time-like vector $u/c$, where $u$ is the velocity
vector of the particle%
\begin{eqnarray}
\mathcal{S} &=&(1/c)[Z\wedge u+(SI)\cdot u],  \notag \\
Z &=&\mathcal{S}\cdot u/c,\quad S=I(\mathcal{S}\wedge u),  \label{e}
\end{eqnarray}%
Eq. (8) in [33]. It holds that $Z\cdot u=S\cdot u=0$; only three components
of $Z$ and three components of $S$ are independent since $\mathcal{S}$ is
antisymmetric. $S$ and $Z$ depend not only on $\mathcal{S}$ but on $u$ as
well. Only in the particle's rest frame, the $K^{\prime }$ frame, and the $%
\{\gamma _{\mu }^{\prime }\}$ basis, $u=c\gamma _{0}^{\prime }$ and $%
S^{\prime 0}=Z^{\prime 0}=0$, $S^{\prime i}=(1/2c)\varepsilon ^{0ijk}%
\mathcal{S}_{jk}^{\prime }$, $Z^{\prime i}=\mathcal{S}^{\prime i0}$. \emph{%
According to Eq.} (\ref{e}), \emph{a new \textquotedblleft
time-space\textquotedblright\ spin} $Z$ \emph{is introduced and it is a
physical quantity in the same measure as it is the usual \textquotedblleft
space-space\textquotedblright\ spin} $S$.

Then, in [33], it is suggested that instead of the connection between the
3-vectors $\mathbf{m}$ and $\mathbf{S}$ we need to have the connection
between the dipole moment bivector $D$ and the spin bivector $\mathcal{S}$,
which is formulated in the form of the generalized Uhlenbeck-Goudsmit
hypothesis
\begin{equation}
D=g_{S}\mathcal{S},  \label{gu}
\end{equation}%
Eq. (9) in [33]. Hence, inserting the decompositions of $D$ (\ref{d}) and $%
\mathcal{S}$ (\ref{e}) into Eq. (\ref{gu}) we find the connections between
the dipole moments $m$ and $p$ and the corresponding intrinsic angular
momenta $S$ and $Z$, respectively,
\begin{equation}
m=cg_{S}S,\ p=g_{S}Z,  \label{dm}
\end{equation}%
Eq. (10) in [33]. In the particle's rest frame, the $K^{\prime }$ frame, and
the $\{\gamma _{\mu }^{\prime }\}$ basis, $u=c\gamma _{0}^{\prime }$ and $%
p^{\prime 0}=m^{\prime 0}=0$, $p^{\prime i}=g_{S}Z^{\prime i}$, $m^{\prime
i}=cg_{S}S^{\prime i}$. Comparing this last relation with $\mathbf{m}=\gamma
_{S}\mathbf{S}$, we see that $g_{S}=\gamma _{S}/c$. In contrast to all
previous approaches with the 3-vectors in which both the MDM $\mathbf{m}%
^{\prime }$ and the EDM $\mathbf{p}^{\prime }$ of an elementary particle are
determined by the usual spin $\mathbf{S}^{\prime }$, we find that the
intrinsic MDM vector $m$ of an elementary particle is determined by the
\textquotedblleft space-space\textquotedblright\ spin vector $S$,\ whereas
\emph{the intrinsic EDM vector} $p$ \emph{is determined by the
\textquotedblleft time-space\textquotedblright\ spin vector} $Z$. \emph{The
relations} (\ref{gu}) \emph{and} (\ref{dm}) \emph{show that any fundamental
particle has not only the intrinsic MDM} $m$, \emph{but also the intrinsic
EDM} $p$ \emph{whose magnitude is} $(1/c)$ \emph{of that for} $m$. The EDM $%
p $ (see (\ref{dm})) emerges from the connection with the intrinsic angular
momentum $Z$, i.e. from (\ref{gu}) and (\ref{d}), (\ref{e}) in the same way
as the MDM $m$ emerges from the connection with the intrinsic angular
momentum $S$. The EDM $p$ is an intrinsic property of a fundamental particle
in completely the same way as it is the intrinsic MDM $m$. As stated in
[33]: \textquotedblleft The EDM obtained in this way is of quite different
physical nature than in the elementary particle theories, e.g., in the
standard model and in SUSY. There, an EDM is obtained by a dynamic
calculation and it stems from an asymmetry in the charge distribution inside
a fundamental particle, which is thought of as a charged
cloud.\textquotedblright

These fundamental results for the generalized Uhlenbeck-Goudsmit hypothesis (%
\ref{gu}), i.e., for the new spin $Z$ and the associated EDM $p$ of a
fundamental particle (\ref{dm}) are used in [33] and [49] for the discussion
of the shortcomings in experimental searches for a permanent EDM of
particles. In all these searches the AT of $\mathbf{E}$ and $\mathbf{B}$ (%
\ref{JCB}) and the AT for the 3D EDM $\mathbf{p}$ and MDM $\mathbf{m}$ (\ref%
{plp}) are always considered to be relativistically correct, i.e., that they
are the LT. Furthermore, it is supposed in the elementary particle theories,
that not only the 3D MDM $\mathbf{m}$ is proportional to the usual 3D spin $%
\mathbf{S}$ ($\mathbf{m}=m(\mathbf{S}/S)$, the Uhlenbeck-Goudsmit
hypothesis), but the 3D EDM $\mathbf{d}$ as well. The results from [33] and
[49] explicitly show that from the ISR viewpoint this basic assumption is
not true and that it has to be replaced by the relation (\ref{dm}) which
deal with 4D geometric quantities that are correctly defined in the 4D
spacetime.

If the spins and the dipole moments are quantized, i.e., if they become
operators, then, in [48], the commutation relations for the components in
the standard basis of the intrinsic angular momentums $S$ and $Z$ are given
by Eq. (4) in [48]

\begin{eqnarray}
\lbrack S^{\mu },S^{\nu }] &=&(i\hslash /c)\varepsilon ^{\mu \nu \alpha
\beta }S_{a}u_{\beta },\ [Z^{\mu },Z^{\nu }]=(-i\hslash /c)\varepsilon ^{\mu
\nu \alpha \beta }S_{a}u_{\beta },  \notag \\
\lbrack S^{\mu },Z^{\nu }] &=&(i\hslash /c)\varepsilon ^{\mu \nu \alpha
\beta }Z_{a}u_{\beta }.  \label{sz}
\end{eqnarray}%
Taking into account the relation (\ref{dm}) the commutation relations for
the components of $m$ and $p$, $m^{\mu }$ and $p^{\mu }$ respectively, are
expressed in terms of those for $S^{\mu }$ and $Z^{\mu }$, Eq. (\ref{sz}),
and they are given by Eq. (5) in [48]

\begin{equation}
\lbrack m^{\mu },m^{\nu }]=c^{2}g_{S}^{2}[S^{\mu },S^{\nu }],\ [p^{\mu
},p^{\nu }]=g_{S}^{2}[Z^{\mu },Z^{\nu }],\ [m^{\mu },p^{\nu
}]=cg_{S}^{2}[S^{\mu },Z^{\nu }].  \label{1}
\end{equation}

As explained above, an electron possesses both intrinsic angular momentums,
spins, $S$ and $Z$ and, according to (\ref{dm}), the associated dipole
moments $m$ and $p$, respectively. All these quantities, $S$ and $Z$, $m$
and $p$ are vectors, properly defined geometric quantities in the 4D
spacetime. They transform according to the LT, the same as (\ref{LTE1}) and (%
\ref{LTE2}). This means that \emph{under the LT vector} $m$ \emph{transforms
again to the magnetic dipole moment} and, contrary to the LPET of the
3-vectors $\mathbf{m}$ and $\mathbf{p}$ (the same as (\ref{ps})) \emph{there
is no mixing with} $p$. The same holds for vectors $S$ and $Z$.

At this point we present a simple discussion, only qualitative arguments,
about the existence of an electric field outside a stationary permanent
magnet and about the experimental detection of that field. It can be
concluded that in the same way as the MDMs determine the magnetization $M$
of a stationary permanent magnet the EDMs determine its polarization $P$,
which induces an electric field outside a permanent magnet (moving or \emph{%
stationary}). Note that according to the LT, the same as (\ref{LTE1}) and (%
\ref{LTE2}), if there is a polarization vector $P$ in one inertial frame of
reference it will exist in all other relatively moving inertial frames of
reference. As $g_{S}=\gamma _{S}/c$, the EDM $p$ is in magnitude $1/c$ of
the MDM $m$. Consequently, the external electric field is much smaller than
the magnetic field outside a stationary permanent magnet. It can be one of
the reasons why such electric fields are not yet experimentally detected.
Another, equally important reason, is that the experimentalists never looked
for such fields because it was no theory that predicts them.

In order to directly measure the external electric fields from a stationary
permanent magnet it is necessary to use a non-contact method of measuring. A
possible experimental setup for the detection of that field could be again
with the cold ions as in [31]. Another way of measuring such electric fields
from a stationary permanent magnet would be some modification of torsion
balance setup that is used in [43] for the detection of the Casimir force.%
\emph{\bigskip }

\noindent \textbf{9. \textquotedblleft Charge-magnet
paradox\textquotedblright\ and its resolution with the 4D torques}\textit{%
\bigskip }

\noindent \textit{9.1 Criticism of the \textquotedblleft
resolutions\textquotedblright\ [34-36]} \textit{of Mansuripur's paradox
\bigskip }

A nice example which clearly illustrates the essential difference between
the 4D geometric approach, i.e., the ISR, with relativistically correct LT
and Einstein's formulation of SR with the AT of the 3-vectors refers to the
recently reported [34] \textquotedblleft charge-magnet
paradox,\textquotedblright\ to the highlight of it [35] and different
\textquotedblleft resolutions\textquotedblright\ [36]. In [34], it is argued
that in the presence of the magnetization $\mathbf{M}$ and the electric
polarization $\mathbf{P}$ the usual expression for the Lorentz force with
the 3-vectors fails to accord with the principle of relativity, because it
leads to an apparent paradox involving a MDM $\mathbf{m}$ in the presence of
an electric field $\mathbf{E}$; in a static electric field a MDM $\mathbf{m}$
is subject to a torque $\mathbf{T}$ in some frames and not in others.
(Henceforward we shall denote the usual 3D torque by $\mathbf{T}$.)
Mansuripur [34] argues that the conventional Lorentz force (density), Eq.
(5) in [34], should be replaced by the Einstein-Laub law, Eq. (6) in [34],
which predicts no torque $\mathbf{T}$ in all frames. The paradox is
reminiscent of the Trouton-Noble paradox [50] and of the Jackson paradox
[51]. For the resolution of these paradoxes with the 4D geometric
quantities, i.e., in the ISR, see [12], [13] and [14], respectively. The
paper [34] and the paper [35] that highlighted it and almost all papers in
[36] that objected it deal with the 3-vectors and their LPET (\ref{JCB}) for
$\mathbf{E}$ and $\mathbf{B}$, (\ref{ps}), or (\ref{psa}) for $\mathbf{P}$
and $\mathbf{M}$, and the same for EDM $\mathbf{p}$ and MDM $\mathbf{m}$,
i.e., (\ref{plp}) or (\ref{Pl1}), as that these transformations are the
relativistically correct LT. For example, in [34], it is stated:
\textquotedblleft Lorentz transformation to the $xyz$ frame then yields a
pair of point dipoles, ..,\textquotedblright\ see also Eqs. (9 - 10b) and
Eqs. (11 - 12b). Furthermore, Eq. (\ref{plp}), or (\ref{Pl1}), is taken as
the main point in the majority of the expositions in [34-36]. But, these
transformations are the AT and not the LT. This means that the requirement
like (\ref{ein}) does not hold either for the conventional Lorentz force or
the Einstein-Laub expression and also for the corresponding 3D torques. All
these quantities are not well-defined quantities in the 4D spacetime.
Similarly happens for all papers from [34-36]. Hence, as will be shown in
this section, from the ISR viewpoint, the resolutions of the
\textquotedblleft charge-magnet paradox\textquotedblright\ that are proposed
in [34-36] are not the relativistically correct resolutions.

In [34], in the discussion of another example, not only that LPET (\ref{JCB}%
) and (\ref{ps}) are considered to be the LT, but the equation $\mathbf{F=}d%
\mathbf{p}/dt$ is considered to be \textquotedblleft the relativistic
version of Newton's law.\textquotedblright\ There, it is stated
\textquotedblleft .. its relativistic momentum $\mathbf{p}$ increases with
time, not because of a change of velocity but because of a change of
mass.\textquotedblright\ It is shown in the preceding sections, i.e., in
[6-11], that the LPET ARE NOT the LT, and as shown in detail in, e.g., [14],
[10], any 3D quantity cannot correctly transform under the LT; it is not the
same quantity for relatively moving observers in the 4D spacetime. \emph{The
equation} $\mathbf{F=}d\mathbf{p}/dt$ \emph{with} $\mathbf{p=}m\gamma _{u}%
\mathbf{u}$ \emph{is not the relativistic equation of motion} since,
contrary to the common assertions, e.g., [2], [22], [26], it does not retain
the same form in two relatively moving inertial frames $S$ and $S^{\prime }$%
, i.e., it is not covariant under the LT, see, e.g., the discussions in Sec.
3 in the first paper in [14] and particularly in [10]. In that equation the
primed 3D quantities are not obtained by the LT from the unprimed ones, but
they are obtained in terms of the AT for the 3D force $\mathbf{F}$, e.g.,
Eqs. (12.66) and (12.67) in [26], and the 3D momentum $\mathbf{p}$, i.e.,
the 3D velocity $\mathbf{u}$, Eq. (11.31) in [2]. The requirement like (\ref%
{ein}) does not hold for any of the mentioned 3D quantities. Instead of the
equation with the 3D quantities one has to use the equation of motion with
4D geometric quantities, Eq. (10) in [14], $K=dp/d\tau $, $p=mu$, where $p$
is the proper momentum vector and $\tau $ is the proper time, a Lorentz
scalar. For the definition of the Lorentz force $K_{L}$%
\begin{equation}
K_{L}=(q/c)F\cdot u=(q/c)\left[ (1/c)E\wedge v+(IB)\cdot v\right] \cdot u,
\label{KEB}
\end{equation}%
see, e.g., the first paper in [14]. The quantities $K$ ($K_{L}$), $p$, $u$
\emph{transform in the same way, like any other vector, i.e., according to
the LT, the same as} (\ref{LTE1}), (\ref{LTE2}) and not according to the
awkward, AT of the 3-force $\mathbf{F}$ and the 3-momentum $\mathbf{p}$,
i.e., the 3-velocity $\mathbf{u}$. Moreover, only the rest mass is well
defined quantity in the 4D spacetime and thus there is not \textquotedblleft
a change of mass.\textquotedblright\

In [35], it is tried to explain why the moving magnet appears to be
electrically polarized. The explanation is completely the same as it is the
derivation of the relations (\ref{plp}) and (\ref{Pl1}) in [23] and [22] and
in Sec. 3.2 here, i.e., it explicitly uses \emph{the Lorentz contraction}.
But, as can be seen, e.g., from Appendix here, the Lorentz contraction is an
AT, i.e., it is ill-defined in the 4D spacetime. Similar consideration with
the use of \emph{the Lorentz contraction} is presented in Unnikrishnan's
paper in [36].

In some papers from [36], e.g., Griffiths and Hnizdo (GH), McDonald,
Saldanha, the model of the \textquotedblleft hidden\textquotedblright\
momentum and the \textquotedblleft hidden\textquotedblright\ angular
momentum is invoked in order to show that, contrary to the claims in [34],
the usual Lorentz force law with the 3-vectors is consistent with SR. Their
approach is also with the 3-vectors and the LPET, i.e., the AT, and,
particularly, in all three papers it is taken that the moving magnetic
dipole, i.e., an Amp\`{e}rian loop, acquires an electric dipole moment given
by Eq. (\ref{plp}). Let us explain in some detail the approach from GH [36].
They find that if the magnetic dipole is considered to be an electric
current loop then the net torque $\mathbf{T}$ (their $\mathbf{N}$) on the
dipole, Eq. (4) in GH [36], is non-zero in the lab frame, i.e., the
principle of relativity is violated. Observe that $\mathbf{T}$ from their
Eq. (4) is obtained from the term $\mathbf{p\times E}$, where $\mathbf{p}$
is the induced electric dipole moment, their Eq. (3), i.e., our Eq. (\ref%
{plp}), which means that \emph{the violation of the principle of relativity
is a direct consequence of the AT for the 3-vectors} $\mathbf{p}$ \emph{and}
$\mathbf{m}$. To resolve the problem they introduce two sorts of the angular
momentums \textquotedblleft .. $\mathbf{L}_{o}$ is the \textquotedblleft
overt\textquotedblright\ angular momentum (associated with actual rotation
of the object), and $\mathbf{L}_{h}$ is the \textquotedblleft
hidden\textquotedblright\ angular momentum (so called because it is \emph{not%
} associated with any overt rotation of the object),\textquotedblright\ $%
\mathbf{L}=\mathbf{L}_{o}+\mathbf{L}_{h}$, where $\mathbf{L}_{h}$ is given
by their Eq. (10). Furthermore, they define the torque $\mathbf{T}$ by their
Eq. (12), $\mathbf{T}=d\mathbf{L}/dt=d\mathbf{L}_{o}/dt+d\mathbf{L}_{h}/dt$
and state \textquotedblleft \emph{Physically} $d\mathbf{L}_{h}/dt$ \emph{is
not a torque}, (my emphasis) but (the rate of change of) a piece of the
angular momentum .. .\textquotedblright\ In order to get the agreement with
the principle of relativity they argue that it is not $\mathbf{T}$, which
has to be zero in both frames, but it is the \textquotedblleft
effective\textquotedblright\ torque $\mathbf{T}_{\text{eff}}$ given by their
Eq. (13). Thus, they assert: \textquotedblleft .. it is the
\textquotedblleft overt\textquotedblright\ torque (given by Eq. (13), my
remark) that must vanish to resolve the paradox, since the dipole is not
rotating .. .\textquotedblright\

Several objections to their treatment are at place already here. (i) First
of all, as already mentioned, their approach deals with the 3-vectors and
their AT and not with the 4D quantities and their LT. The requirement like (%
\ref{ein}) is not fulfilled for any 3D quantity. (ii) In such an approach
with 3-vectors the introduction of the \textquotedblleft
hidden\textquotedblright\ quantities is very artificial and it is not
justified either mathematically or physically. (iii) The derivatives in Eqs.
(11-13) are over the \emph{coordinate time} and not over the proper time,
which means that neither of them can properly transform under the LT. (iv)
Their assertion about $d\mathbf{L}_{h}/dt$ is meaningless. It cannot be that
in an equation (their Eq. (13)) with physical quantities $\mathbf{T}$ is a
torque and $d\mathbf{L}_{o}/dt$ is also a torque, but $d\mathbf{L}_{h}/dt$
\emph{is not a torque}. Obviously, in the usual formulation with the
3-vectors there is no proper physical interpretation for $d\mathbf{L}_{h}/dt$%
, i.e., for different \textquotedblleft hidden\textquotedblright\ quantities.

\emph{It can be seen from the resolution of Jackson's paradox }[14],\emph{\
the Trouton-Noble paradox }[12, 13]\emph{\ and from Sec. 9.2 here that if
the physical reality is attributed to the 4D geometric quantities which
properly transform under the LT and not, as usual, to the 3D quantities
which transform according to the AT, then all quantities are mathematically
and physically correctly defined and there is no \textquotedblleft
hidden\textquotedblright\ quantity, i.e., the quantity for which there is
not some proper physical interpretation in the 4D spacetime.} For example,
the angular momentum is not the 3-vector $\mathbf{L}=\mathbf{r}\times
\mathbf{p}$ and the torque is not $\mathbf{T}=\mathbf{r}\times \mathbf{F}$
with $\mathbf{T}=d\mathbf{L}/dt$, but they are the abstract 4D quantities,
here the bivectors,
\begin{equation}
J=x\wedge p,\ N=x\wedge K;\quad N=dJ/d\tau ,  \label{MKN}
\end{equation}%
where $x$ is the position vector, $p$ is the proper momentum vector, $p=mu$,
$u$ is the proper velocity vector $u=dx/d\tau $ of a charge $q$ (it is
defined to be the tangent to its world line), $\tau $ is the proper time and
$K$ is the force vector. If $J$ and $N$ are written as CBGQs in the $%
\{\gamma _{\mu }\}$ basis they are given by Eq. (12) in the first paper in
[14],%
\begin{eqnarray}
J &=&(1/2)J^{\mu \nu }\gamma _{\mu }\wedge \gamma _{\nu },\ J^{\mu \nu
}=m(x^{\mu }u^{\nu }-x^{\nu }u^{\mu }),  \notag \\
N &=&(1/2)N^{\mu \nu }\gamma _{\mu }\wedge \gamma _{\nu },\ N^{\mu \nu
}=x^{\mu }K^{\nu }-x^{\nu }K^{\mu }.  \label{mn}
\end{eqnarray}%
The components $J^{\mu \nu }$ from (\ref{mn}) are identical to the usual
covariant angular momentum four-tensor and similarly for $N^{\mu \nu }$.

Cross [36], in the Comment on Mansuripur's paper, states: \textquotedblleft
The torque density is not a vector, but the antisymmetric tensor ..
\textquotedblright\ and also \textquotedblleft .. the angular momentum,
which is the second rank tensor .. .\textquotedblright\ So, he correctly
notices that the angular momentum and the torque density are not the 3D
vectors. However, his quantities $J^{\alpha \beta }$ and $T^{\alpha \beta }$
also are not tensors, but components implicitly taken in the standard basis
as in the usual covariant approaches, e.g., from [2], [22], [26]. They
correspond to $J^{\mu \nu }$ and $N^{\mu \nu }$ from (\ref{mn}). Tensors are
abstract, geometric quantities, e.g., in the abstact index notation $M^{ab}$%
, Eq. (4) in [33], which corresponds to the bivector $J$ in (\ref{MKN}) in
the geometric algebra formalism. If tensors are represented in some basis
then they become the CBGQs that contain components \emph{and the basis
vectors} as in Eq. (\ref{mn}). As already mentioned several times, if one
does not use Einstein's synchronization but, e.g., the \textquotedblleft
r\textquotedblright\ synchronization then it is not possible to make the
identification of the components of the 3-vectors $\mathbf{E}$ and $\mathbf{B%
}$ with the components $F^{\alpha \beta }$ of the electromagnetic field $F$,
or the components of the 3-vectors $\mathbf{R}$ and $\mathbf{T}$ (in Cross's
notation) with the components $T^{\alpha \beta }$ of the torque tensor, etc.
This means that the physical laws have to be written with the abstract
quantities and then to represent these quantities in some basis as the
CBGQs. Furthermore, he states, \textquotedblleft For short we write $M=(%
\mathbf{P},-\mathbf{M})$ in terms of the defining
3-vectors.\textquotedblright\ Similarly, he writes \textquotedblleft $F=(-%
\mathbf{E},-\mathbf{B})$ is the Faraday tensor ..\textquotedblright\ and
also \textquotedblleft $m=(\mathbf{p},-\mathbf{m})$ is the tensor of
moments.\textquotedblright\ Hence, in his covariant approach with
\textquotedblleft tensors,\textquotedblright\ in the same way as in all
traditional approaches, it is considered that the 3-vectors are the primary
quantities with a definite physical reality, whereas $M^{\alpha \beta }$, $%
F^{\alpha \beta }$, $m^{\alpha \beta }$, etc. are only auxiliary
mathematical quantities that are defined by the components of the 3-vectors.

The situation is completely different in the ISR in which the primary
quantities are the abstract quantities, here the bivectors, thus not only
components, $\mathcal{M}$, $F$, $D$ (our notation), etc. As seen from Eqs. (%
\ref{M1}) and (\ref{M2}), Eqs. (\ref{E2}) and (\ref{E1}), Eq. (\ref{d}),
etc. $P$ and $M$ are derived from $\mathcal{M}$ and they depend not only on $%
\mathcal{M}$ but on $u$ as well, and similarly for the bivector $F$ and
vectors $E$, $B$ and $v$ and also for $D$ and $p$, $m$ and $u$.$\ $

Note that Cross [36] also deals with the \textquotedblleft
hidden\textquotedblright\ quantities relating them to the time-space
components, e.g., $N^{0i}$ ($T^{0i}$ in Cross's notation) of the torque
tensor. He argues: \textquotedblleft .. that the torque predicted in the
moving frame is correct and necessary to balance the \textquotedblleft
hidden\textquotedblright\ angular momentum of the moving dipole rather than
causing a precession of the spin.\textquotedblright\ This means that under
the torque\ he does not understand the whole torque tensor, but only the 3D
torque $\mathbf{T}$ that is related with the rotation in the 3D space.
Indeed, for the time-space components, i.e., for the components of his $%
\mathbf{R}$ it is stated: \textquotedblleft These \textquotedblleft
torque\textquotedblright\ components are connected with the motion of the
center of energy .. .\textquotedblright\

In the 4D spacetime the center of energy is not well-defined quantity. As
seen from [12-14] and as will be seen below such understanding is in a sharp
contrast to ISR in which \emph{only} the 4D geometric quantities are
well-defined in the 4D spacetime, e.g., the torque bivector $N$ and the
torque vectors $N_{s}$ and $N_{t}$, see Eq. (\ref{nls}).

Observe also that Cross [36], in contrast to all others from [34-36], showed
that the offending torque $T^{\prime x}=\gamma vmE$ (in his notation) that
is obtained in a non-covariant analysis, i.e., with the 3-vectors and their
AT for EDM $\mathbf{p}$ and MDM $\mathbf{m}$, Eq. (\ref{plp}), is determined
as the Lorentz transformed time-space component that exists even in the rest
frame. In his notation that nonvanishing component is $R^{y}\equiv
T^{ty}=-mE $. Then, he states: \textquotedblleft Under a Lorentz boost to
the moving frame the space-space and time-space components
mix.\textquotedblright\ Hence, his 3D torque $T^{^{\prime }x}=\Lambda
_{t}^{z}T^{yt}$ is considered to be obtained by the LT. However, the
comparison of his derivation of $T^{^{\prime }x}$ with the derivation of the
AT (\ref{JCB}), i.e., (\ref{ee}), in Sec. 3.1 reveals that both derivations
are exactly the same. Indeed, Cross [36] simply assumes that the components
of the 3-vectors $\mathbf{R}$ and $\mathbf{T}$ are identified with the
components $T^{\alpha \beta }$ of the torque tensor \emph{in both frames}
(this corresponds to Eqs. (\ref{ieb}) and (\ref{eb2})) and accordingly they
transform like the components $N^{\alpha \beta }$ ($T^{\alpha \beta }$ in
Cross's notation) transform. For comparison, the reader can look at the text
in Sec. 3.1 between Eq. (\ref{ee}) and Eq. (\ref{JCB}). Thus, contrary to
the assertion in Cross [36] the transformations for $T^{^{\prime }x}$ are
not the LT, but they are exactly the same as the LPET, i.e., the AT (\ref%
{JCB}). The explicit AT of the components of the 3-vectors $\mathbf{T}$ and $%
\mathbf{R}$ are given, e.g., by Eq. (19) in [13], which is repeated here

\begin{eqnarray}
T_{1} &=&T_{1}^{\prime },\ T_{2}=\gamma (T_{2}^{\prime }-\beta R_{3}^{\prime
}),\ T_{3}=\gamma (T_{3}^{\prime }+\beta R_{2}^{\prime }),  \notag \\
R_{1} &=&R_{1}^{\prime },\ R_{2}=\gamma (R_{2}^{\prime }+\beta T_{3}^{\prime
}),\ R_{3}=\gamma (R_{3}^{\prime }-\beta T_{2}^{\prime }).  \label{tc}
\end{eqnarray}%
They are written for the motion along the $x^{1}$ axis and not along the $%
x^{3}$ axis as in Cross [36] and also the components $T_{t,i}$ from Eq. (19)
in [13] are replaced by $R_{i}$. With these changes, the component in (\ref%
{tc}) that corresponds to $T^{\prime x}$ in Cross [36] is $T_{2}=-\gamma
\beta R_{3}^{\prime }$. It is visible from (\ref{tc}) that the
transformations for $R_{i}$ are the same as the AT for $E_{i}$ in Eq. (\ref%
{ee}) and similarly is for $T_{i}$ and $B_{i}$. The essential point is that,
e.g., \emph{the components} $T_{i}$ \emph{of the torque 3-vector} $\mathbf{T}
$ \emph{in the moving frame} \emph{are expressed by the mixture of the
components of the 3D} \emph{vector} $\mathbf{T}^{\prime }$ \emph{and} \emph{%
of another 3D vector} $\mathbf{R}^{\prime }$ \emph{from the rest frame. }%
This is the reason that the components of the usual 3D torque $\mathbf{T}$
will not vanish in the $S$ frame even if they vanish in the $S^{\prime }$
frame, i.e., that there is the \textquotedblleft charge-magnet
paradox\textquotedblright\ in all usual approaches to special relativity
that deal with the 3-vectors or with components implicitly taken in the
standard basis. The same discussion as in Sec. 6 after Eq. (\ref{ein}) can
be repeated for the 3-vectors $\mathbf{T}$ and $\mathbf{R}$. Again it can be
stated that \emph{as far as relativity is concerned the quantities, e.g.,} $%
\mathbf{T}$ \emph{and} $\mathbf{T}^{\prime }$ \emph{are not related to one
another }and that their identification is simply a \emph{mistaken identity}.

However, as can be seen from Eq. (\ref{ms}) below, the situation is
completely different for the 4D torque $N_{s}$, and the same for $N_{t}$.
The physical 4D torque, the basis-free, abstract, $N_{s}$, simply has
different representations, 4D CBGQs, in different bases.

Very similar procedure as in GH [36] is commonly used for the resolution\ of
the Trouton-Noble paradox, see Refs. [3-7] in [13]. In the Trouton-Noble
paradox, in the similar way as in [34], there is a 3D torque and so a time
rate of change of 3D angular momentum in one inertial frame, but not in
another relatively moving inertial frame. For the usual resolution of that
paradox it is taken that \emph{there is another 3D torque}, which is equal
in magnitude but of opposite direction, giving that the total 3D torque is
zero in order to have agreement with the principle of relativity. Different
explanations have been offered for the existence of that additional 3D
torque, e.g., the nonelectromagnetic forces with their additional torque,
Refs. [3-6] in [13], or the field angular momentum and its rate of change,
i.e., its additional torque, Ref. [7] in [13].

The resolution of Mansuripur's paradox that is presented by McDonald [36] is
almost the same as in GH [36] with the difference that McDonald introduces
the field angular momentum and states: \textquotedblleft ... the
\textquotedblleft paradoxical\textquotedblright\ nonzero torque is needed to
change the \textquotedblleft hidden\textquotedblright\ mechanical angular
momentum of the system, such that this remains equal and opposite to the
field angular momentum, .. .\textquotedblright\ The above objections (i) -
(iii) hold in the same measure for McDonald's approach with an additional
objection that the usual expressions for the electromagnetic momentum that
are given by his Eq. (1) are all with the 3D quantities, which means that
they also are not well-defined in the 4D spacetime; they do not properly
transform under the LT.

Kholmetskii, Missevitch and Yarman [36] exclusively deal with 3D quantities
and introduce the contribution of the hidden momentum as well.

Milton and Meille [36] also deal with the 3-vectors and the AT (\ref{JCB})
for $\mathbf{E}$ and $\mathbf{B}$. Particularly, for a point dipole they
argue that: \textquotedblleft ... the torque is not zero, but is balanced by
the rate of change of the angular momentum of the electromagnetic field, so
there is no mechanical torque on the dipole.\textquotedblright\ Their 3D
quantities also do not properly transform under the LT.

Brachet and Tirapegui [36] combine the usual covariant approach with
components implicitly taken in the standard basis with the formulation in
terms of the 3-vectors writing, e.g., the energy density $u$ and the
Poynting 3-vector $\mathbf{S}$ in terms of the 3-vectors $\mathbf{E}$ and $%
\mathbf{B}$ and the integrations in $\mathcal{P}_{\text{field}}$ and $%
\mathcal{L}_{\text{field}}$ are over the 3D space. In the lab frame they
have the torque, a 3D quantity, but they argue that \textquotedblleft it is
required to account for the motion at uniform speed of the missing momentum $%
\mathcal{P}_{\text{particle}}$.\textquotedblright\ Thus, they also deal with
quantities which do not correctly transform under the LT.

Boyer [36], as almost all others, defines the relativistic conservation
laws, the Lorentz transformations of forces, the Lorentz transformations of
energy and momentum, the angular momentum and torques, etc., all in terms of
3D quantities, e.g., the 3-vectors $\mathbf{E}$, $\mathbf{B}$, $\mathbf{r}$,
$\mathbf{p}$, $\mathbf{F}$, $\mathbf{L}$, etc., considering that their AT
are the LT. It is interesting that only Boyer in his Ref. 6 mentioned my
Comment (arXiv: physics/0505013) to Jacson's paper [51] as that it was
published. How can it be that \emph{in the 4D spacetime} the 3D quantities
are considered to be well-defined, whereas the 4D geometric quantities are
considered to be ill-defined?

For the electromagnetic momentum that correctly transforms under the LT see
Secs. 4, 5 - 5.3 in [52] (only components) and for the more general
expressions with the 4D geometric quantities see Sec. 2.6 in [12]. There, in
[12], an axiomatic geometric formulation of electromagnetism with the
Faraday bivector field $F$ as the primary quantity for the whole
electromagnetism is presented and in Sec. 2.6 the observer independent
expressions with the abstract quantities, Eqs. (37) - (43), for the
stress-energy vector $T(n)$, the energy density $U$, the Poynting vector $S$
and the momentum density $g$, the angular momentum density $M$ and the
Lorentz force $K_{L}$ are directly derived from the field equation for $F$,
Eq. (4) in [12] (the notation is that one from [12]). These quantities are
also written as CBGQs in the standard basis, Eqs. (44) - (47). Furthermore,
the local conservation laws are directly derived from that field equation
for $F$ and presented in Sec. 2.7 in [12], see Eqs. (48) - (51).

In [36], only in Vanzella's paper the CBGQs in the standard basis are used,
but again the formulation with such quantities is combined with the
formulation in terms of the 3-vectors. He \textquotedblleft mimic the magnet
by a neutral ring conducting an electric stationary current $I$%
.\textquotedblright\ Although he does not explicitly use the AT of the 3D
quantities, the assumption that $j^{\mu }u_{\mu }=0$, \textquotedblleft
reflecting the neutrality of the ring according to $\mathcal{O}$%
,\textquotedblright\ where $\mathcal{O}$ is an observer at rest in the frame
of the ring, is equivalent to the Clausius hypothesis. According to the
discussion in Sec. 7.1 that hypothesis is not well justified in the 4D
spacetime. That assumption leads to the result that for a relatively moving
observer the ring is polarized and there is an induced electric dipole
moment 3-vector and a 3D torque $\mathbf{\tau }=\mathbf{d}\times \mathbf{E}$
(his notation) on the ring. That torque is the same as that the AT (\ref{plp}%
) are explicitly used. Furthermore, he declares: \textquotedblleft the
torque exerted by the electric field is not used to rotate the ring but
rather to move its asymmetric distribution induced by the very same the
electric field. No paradox here.\textquotedblright\ In principle, such a
conclusion is very similar to that one as in GH [36] or Cross [36].
Griffiths and Hnizdo [36] and also Cross [36] consider that there is no
paradox because the offending torque (3-vector) is not \textquotedblleft
associated with actual rotation of the object.\textquotedblright\ In the 4D
spacetime their resolution of the paradox is not correct, because it is
based on the use of the 3D quantities and their AT. As shown below the
relativistically correct torques are the bivector $N$ and the vectors $N_{s}$
and $N_{t}$.

Griffiths and Hnizdo [36] assert: \textquotedblleft This \textquotedblleft
paradox\textquotedblright\ was resolved many years ago by Victor Namias
[4].\textquotedblright\ But, it is not true. Namias, as almost all others,
exclusively dealt with the 3D quantities and their AT. Similarly, Vanzella
[36] declares: \textquotedblleft .. there is a very similar and famous
\textquotedblleft paradox\textquotedblright\ (the Trouton-Noble
\textquotedblleft paradox\textquotedblright\ [2]) which was presented and
resolved more than a hundred years ago [3] (see also Refs. [4,
5]).\textquotedblright\ As can be seen from [12-14] that Vanzella's
statement is not correct. The \textquotedblleft
resolutions\textquotedblright\ presented in Vanzella's references [3-5]
dealt with different AT, the Lorentz contraction, the dilation of time and
the LPET (\ref{JCB}) and (\ref{ps}) considering them as that they are the
relativistically correct LT.

In addition, it is worth mentioning that \emph{all treatments from} [34-36]
\emph{are meaningless} if some other basis and not the standard basis is
taken into account, e.g., \emph{if only the Einstein synchronization is
replaced by the} \textquotedblleft \emph{radio}\textquotedblright\ \emph{%
synchronization}. This conclusion simply follows already from Eqs. (\ref{Fr1}%
) and (\ref{FEr}).\emph{\bigskip }

\noindent \textit{9.2 The resolution of Mansuripur's paradox with the 4D
torques\bigskip }

The geometric approach to the \textquotedblleft charge-magnet
paradox\textquotedblright\ that is presented here significantly differs from
all formulations in [34-36]. First of all, in contrast to [34-36], this
approach deals from the outset with 4D geometric quantities, their LT and
equations with them. It is already shown in the similar treatments of the
Trouton-Noble paradox [12, 13] and Jackson's paradox [14] that in the
approach with 4D geometric quantities the principle of relativity is
naturally satisfied and there is no paradox. Every 4D geometric quantity is
the same quantity for all relatively moving inertial observers and for all
bases chosen by them. The equation like (\ref{ein}) holds for any such
quantity. Furthermore, the most important difference is that the torque
bivector is different from zero even in the common rest frame of the
considered charge and magnet from [34]. The reason for that important
difference is that according to the 4D geometric approach, i.e., the ISR, a
stationary permanent magnet possesses not only the magnetization vector $M$
but the polarization vector $P$ as well. Observe that, according to the
preceding discussion, instead of to speak about a permanent magnet one can
equivalently consider a current loop with a steady current.

Let us consider the system from [34], but, without loss of generality, the
electric charge will be substituted by a uniform electric field. The common
rest frame of the source of the electric field (a point charge $q$ in [34])
and of the permanent magnet will be denoted as $S^{\prime }$, whereas the
lab frame, in which the $S^{\prime }$ frame moves with uniform velocity $%
V=V\gamma _{1}$ along the common $x_{1}$, $x_{1}^{\prime }$ axes, will be
denoted as $S$. (Here, such a choice is taken for easier comparison with
[13] and [14].) The treatment of the interaction between a static electric
field and a permanent magnet that is presented here is very similar to the
formulations given in [12-14]. For this example, the Lorentz force density $%
k_{L}$ is given by Eq. (\ref{kj}) in which it is taken that $j^{(C)}=0$,
i.e., $k_{L}=(1/c)F\cdot j^{(\mathcal{M})}$, where $j^{(\mathcal{M})}$ is
given by Eq. (\ref{jm}) and $F=(1/c)E\wedge v$, what is Eq. (\ref{E2}) with $%
B=0$. Remember that \emph{the vector} $B$ \emph{is zero in all relatively
moving inertial frames of reference }and therefore there is no reason for
the appearance of the paradox. This is an essential difference between the
approach with the 4D geometric quantities and their LT and the usual
approach with the 3D quantities and their AT. Compare with Eqs. (11 - 12b)
in [34]. Hence, $k_{L}$ from (\ref{kj}) becomes%
\begin{equation}
k_{L}=(1/c^{2})(E\wedge v)\cdot \lbrack -(\partial \cdot P)u+(u\cdot
\partial )P+(1/c)[u\wedge (\partial \wedge M)]I].  \label{kl}
\end{equation}%
Observe that the expression for $k_{L}$ contains two velocity vectors, $v$ -
the velocity vector of the observers who measure $E$ and $B$ fields and $u$
- the velocity vector of the permanent magnet, i.e., of the electric current
loop. The Lorentz force density $k_{L}$ (\ref{kl}) can be written as a CBGQ
in the standard basis and its form can be easily inferred from the
expression for the torque density $n$ (\ref{n}) that is given below. Here,
it will be presented the expression for $k_{L}$ in the $S^{\prime }$ frame,
i. e., for the case that $u=v=c\gamma _{0}^{\prime }$ and accordingly that $%
E^{\prime 0}=P^{\prime 0}=M^{\prime 0}=0$. Hence, $k_{L}$ as a CBGQ in the $%
\{\gamma _{\mu }^{\prime }\}$ basis is%
\begin{equation}
k_{L}=(-E^{\prime k}\partial _{0}^{\prime }P_{k}^{\prime }+(1/c)\varepsilon
^{0jkl}E_{j}^{\prime }\partial _{k}^{\prime }M_{l}^{\prime })\gamma
_{0}^{\prime }-E^{\prime i}(\partial _{k}^{\prime }P^{\prime k})\gamma
_{i}^{\prime }.  \label{km}
\end{equation}%
In the usual approaches with the 3-vectors and their AT, e.g., in [34] and
in GH [36], the Lorentz 3-force density is zero in the $S^{\prime }$ frame;
there is not the $\gamma _{0}^{\prime }$ term and there is not the
polarization 3-vector $\mathbf{P}$. The components $k_{L}^{\prime \mu }$ in (%
\ref{km}) correspond to the time and spatial components of $f^{\alpha }$
from [36] (Cross), i.e., to $f^{0}=(1/c)\mathbf{E\cdot }(\partial \mathbf{P}%
/\partial t+\mathbf{\nabla }\times \mathbf{M)}$ and $f^{i}=-E^{i}(\mathbf{%
\nabla P})$. But, as already stated, it is not correct to write the
components of the properly defined vector in the 4D spacetime in terms of
the 3-vectors. The components $k_{L}^{\prime \mu }$ are multiplied by the
unit basis vectors $\gamma _{\mu }^{\prime }$, whereas the 3-vector, e.g., $%
\mathbf{E}$ is constructed from the components $E_{x,y,z}$ and \emph{the
unit 3-vectors} $\mathbf{i}$, $\mathbf{j}$, $\mathbf{k}$.

The torque density $n$ is $n=x\wedge k_{L}$, where $k_{L}$ is given by Eq. (%
\ref{kl}). If $n$ is written as a CBGQ in the standard basis it becomes%
\begin{eqnarray}
n &=&(1/c^{2})x^{\lambda }\{(\partial _{\mu }P^{\mu })[(E^{\nu }u_{\nu
})v^{\rho }-(v^{\nu }u_{\nu })E^{\rho }]-(u^{\mu }\partial _{\mu })P^{\nu
}[E_{\nu }v^{\rho }-v_{\nu }E^{\rho }]  \notag \\
&&+(1/c)\varepsilon ^{\mu \nu \alpha \beta }u_{\mu }(\partial _{\alpha
}M_{\beta })[E_{\nu }v^{\rho }-v_{\nu }E^{\rho }]\}(\gamma _{\lambda }\wedge
\gamma _{\rho }).  \label{n}
\end{eqnarray}%
$n$ from (\ref{n}) will be determined in the $S^{\prime }$ frame in the same
way as $k_{L}$ is determined in (\ref{km}), i.e., with $u=v=c\gamma
_{0}^{\prime }$, and $E^{\prime 0}=P^{\prime 0}=M^{\prime 0}=0$,%
\begin{eqnarray}
n &=&[-(\partial _{k}^{\prime }P^{\prime k})x^{\prime 0}E^{\prime
i}+(\partial _{0}^{\prime }P^{\prime k})E_{k}^{\prime }x^{\prime
i}-(1/c)\varepsilon ^{0jkl}E_{j}^{\prime }(\partial _{k}^{\prime
}M_{l}^{\prime })x^{\prime i}](\gamma _{0}^{\prime }\wedge \gamma
_{i}^{\prime })-  \notag \\
&&(\partial _{k}^{\prime }P^{\prime k})x^{\prime j}E^{\prime i}(\gamma
_{j}^{\prime }\wedge \gamma _{i}^{\prime })  \label{nc}
\end{eqnarray}%
It is worth noting that in the approaches with the 3-vectors, as for $k_{L}$
in (\ref{km}), the torque density $n$ is zero in the $S^{\prime }$ frame.

In $S^{\prime }$, the integrated torque $N$ as a CBGQ is given as%
\begin{equation}
N=-(1/c)E^{\prime 1}m^{\prime 2}(\gamma _{0}^{\prime }\wedge \gamma
_{3}^{\prime })-E^{\prime 1}p^{\prime 3}(\gamma _{1}^{\prime }\wedge \gamma
_{3}^{\prime }),  \label{tn}
\end{equation}%
where $m$ is the magnetic dipole moment vector and $p$ is the electric
dipole moment vector. All quantities in (\ref{tn}) are measured in the
common rest frame $S^{\prime }$. They are all properly defined in the 4D
spacetime and they properly transform under the LT. Furthermore, in the
considered case, the electric field vector $E=E^{\prime 1}\gamma
_{1}^{\prime }$, the MDM $m=m^{\prime 2}\gamma _{2}^{\prime }$ and the EDM $%
p=p^{\prime 3}\gamma _{3}^{\prime }$. Using (\ref{tn}), the LT of the
components $N^{\prime \mu \nu }$ and the inverse LT of the basis $\gamma
_{\mu }^{\prime }\wedge \gamma _{\nu }^{\prime }$ from $S^{\prime }$ to $S$
it can be shown that in $S$, the lab frame, the bivector $N$ is given as
\begin{equation}
N=(-E^{1}m^{2}/c+\beta E^{1}p^{3})(\gamma _{0}\wedge \gamma _{3})+(\beta
E^{1}m^{2}/c-E^{1}p^{3})(\gamma _{1}\wedge \gamma _{3}),  \label{m}
\end{equation}%
and it holds that the whole 4D torque $N$ is unchanged
\begin{equation}
N=(1/2)N^{^{\prime }\mu \nu }\gamma _{\mu }^{\prime }\wedge \gamma _{\nu
}^{\prime }=(1/2)N^{\mu \nu }\gamma _{\mu }\wedge \gamma _{\nu },
\label{enc}
\end{equation}%
where the LT of the electric field vector $E^{\prime 1}=\gamma (E^{1}-\beta
E^{0})$, $E^{\prime 0}=\gamma (E^{0}-\beta E^{1})=0$ are used to derive that
$E^{\prime 1}=(1/\gamma )E^{1}$.

All quantities in (\ref{m}) are measured in $S$ in which the permanent
magnet and the source of the electric field move with uniform velocity $%
V=V\gamma _{1}$ along the common $x_{1}$, $x_{1}^{\prime }$ axes. The
relations (\ref{tn}), (\ref{m}) and (\ref{enc}) show that in the approach
with the 4D torque $N$ the principle of relativity is naturally satisfied
and there is no paradox; \emph{the 4D torque} $N$ \emph{is the same 4D
quantity in all relatively moving inertial frames of reference.} Note that $%
N $ will be the same 4D quantity, as in (\ref{enc}), for all bases, e.g.,
the $\{r_{\mu }\}$ basis, and not only for the standard basis.

In the same way as for $F$, Eqs. (\ref{E2}) and (\ref{E1}), or for $\mathcal{%
M}$, Eqs. (\ref{M1}) and (\ref{M2}), the bivector $N$ can be decomposed into
the \textquotedblleft space-space\textquotedblright\ torque $N_{s}$ and the
\textquotedblleft time-space\textquotedblright\ torque $N_{t}$, which \emph{%
together }contain the same physical information as the bivector $N$, and the
velocity vector $v$ of a family of observers who measures $N$. In the
geometric approach, i.e., in the ISR, \emph{both} $N_{s}$ \emph{and} $N_{t}$
\emph{are equally physical 4D torques which} \emph{taken together are
equivalent to the 4D torque, the bivector} $N$,
\begin{eqnarray}
N &=&(v/c)\wedge N_{t}+(v/c)\cdot (N_{s}I),  \notag \\
N_{t} &=&(v/c)\cdot N,\quad N_{s}=I(N\wedge v/c),  \label{nls}
\end{eqnarray}%
with the condition%
\begin{equation}
N_{s}\cdot v=N_{t}\cdot v=0;  \label{cs}
\end{equation}%
only three components of $N_{s}$ and three components of $N_{t}$ are
independent since $N$ is antisymmetric. \emph{Both}, $N_{s}$ \emph{and} $%
N_{t}$ \emph{depend not only} $N$ \emph{but on} $v$ \emph{as well.} The
primary physical quantity with definite physical reality for the torques is
the 4D torque $N$, whereas $N_{s}$ and $N_{t}$ are derived from $N$ \emph{and%
} $v$. The equations. (\ref{nls}) and (\ref{cs}) are Eqs. (13) and (14)
respectively in the first paper in [14], or Eq. (2) in [13]. If $N_{s}$ and $%
N_{t}$ are written as CBGQs in the $\{\gamma _{\mu }\}$ basis they become

\begin{equation}
N_{s}=(1/2c)\varepsilon ^{\alpha \beta \mu \nu }N_{\alpha \beta }v_{\mu
}\gamma _{\nu },\ N_{t}=(1/c)N^{\mu \nu }v_{\mu }\gamma _{\nu },  \label{nq}
\end{equation}%
what is Eq. (3) in [13]. As seen from (\ref{nq}), in the frame of
\textquotedblleft fiducial\textquotedblright\ observers and in the $\{\gamma
_{\mu }\}$ basis, $v^{\mu }=(c,0,0,0)$, $N_{s}^{0}=N_{t}^{0}=0$ and \emph{%
only the spatial components} $N_{s}^{i}$ and $N_{t}^{i}$ remain
\begin{equation}
N_{s}^{0}=0,\ N_{s}^{i}=(1/2)\varepsilon ^{0ijk}N_{jk},\ N_{t}^{0}=0,\
N_{t}^{i}=N^{0i},  \label{n0}
\end{equation}%
which explains the names the \textquotedblleft
space-space\textquotedblright\ torque for $N_{s}$ and the \textquotedblleft
time-space\textquotedblright\ torque for $N_{t}$. The quotation marks stand
because such an identification of the components of the torques $N_{s}$ and $%
N_{t}$ with the components of the bivector $N$ is not possible for some
other bases. In the ISR, \emph{both vectors} $N_{s}$ \emph{and} $N_{t}$
\emph{have to be treated on an equal footing.} It is worth noting that the
whole discussion with the torque can be completely repeated for the angular
momentum replacing $N$, $N_{s}$ and $N_{t}$ by $J$, $J_{s}$ and $J_{t}$, see
Eqs. (17) - (19) in the first paper in [14]. The Trouton-Noble paradox,
Jackson's paradox and Mansuripur's paradox [34] , all of them stem from the
fact that in the usual approaches an independent physical reality is
attributed only to $N_{s}$ and $J_{s}$, or better to say, to the 3D torque $%
\mathbf{T}$ and the 3D angular momentum $\mathbf{L}$, but not to $N_{t}$ and
$J_{t}$. Furthermore, in the usual approaches, the AT of $\mathbf{T}$ and $%
\mathbf{L}$ are considered to be the relativistically correct LT, see, e.g.,
Jackson's paper [51] and the discussion in Sec. 3 in the first paper in [14].

Let us determine $N_{s}$ and $N_{t}$ for our case as 4D CBGQs in the rest
frame $S^{\prime }$. In $S^{\prime }$, one finds from (\ref{tn}) and (\ref%
{nq}) that
\begin{equation}
N_{s}=N_{s}^{\prime \mu }\gamma _{\mu }^{\prime }=(1/c)E^{\prime 1}p^{\prime
3}v^{\prime 0}\gamma _{2}^{\prime },\quad N_{t}=N_{t}^{\prime \mu }\gamma
_{\mu }^{\prime }=-(1/c^{2})E^{\prime 1}m^{\prime 2}v^{\prime 0}\gamma
_{3}^{\prime },  \label{tst}
\end{equation}%
where, in $S^{\prime }$, $v^{\prime \mu }=(c,0,0,0)$. The \textquotedblleft
time-space\textquotedblright\ torque $N_{t}$ in (\ref{tst}), which comes
from the first term in (\ref{tn}), corresponds to the expression $\mathbf{R}=%
\mathbf{m}\times \mathbf{E}=-mE\widehat{\mathbf{y}}$ in Cross [36] that
describes the interaction of the magnetic moment with the electric field in
the rest frame $S^{\prime }$. (Remember that in Cross [36] the rest frame is
with unprimed quantities and the motion is along the $x^{3}$ axis.) The
\textquotedblleft space-space\textquotedblright\ torque $N_{s}$ in (\ref{tst}%
), which comes from the second term in (\ref{tn}), does not appear in any
previous paper since it emerges from the existence of the EDM $p$ for a
stationary permanent magnet, which is first predicted in Sec. 8 here. It
describes the interaction of the EDM $p=p^{\prime 3}\gamma _{3}^{\prime }$
of the stationary permanent magnet with the electric field $E=E^{\prime
1}\gamma _{1}^{\prime }$ in the rest frame $S^{\prime }$. In the usual
formulation with the 3-vectors it would correspond to the usual 3D torque $%
\mathbf{T}=\mathbf{p}\times \mathbf{E}$, but, in contrast to all previous
formulations, this torque is in the rest frame $S^{\prime }$.

Next, we determine $N_{s}$ and $N_{t}$ in $S$, the lab frame. One way is to
start with Eq. (\ref{tst}) and then to transform by the LT\emph{\ all}
quantities which determine $N_{s}$ and $N_{t}$ in (\ref{tst}), i.e., $%
E^{\prime \mu }$, $m^{\prime \mu }$, $p^{\prime \mu }$, $v^{\prime \mu }$
and $\gamma _{\mu }^{\prime }$, from $S^{\prime }$ to $S$. This yields
\begin{equation}
N_{s}=N_{s}^{\mu }\gamma _{\mu }=(1/\gamma )E^{1}p^{3}\gamma _{2},\quad
N_{t}=-(1/c\gamma )E^{1}m^{2}\gamma _{3}.  \label{st}
\end{equation}%
Another way to determine $N_{s}$ and $N_{t}$ in $S$ is to use the expression
for $N$ in $S$ (\ref{m}) and the relations (\ref{nq}). Note that
\textquotedblleft fiducial\textquotedblright\ observers are moving in $S$.
Therefore the components $v^{\mu }$ of their velocity in $S$, which are
obtained by the LT from $v^{\prime \mu }=(c,0,0,0)$, are $v^{\mu }=(\gamma
c,\gamma \beta c,0,0)$. Of course, for the whole CBGQ $v$ it holds that $%
v=v^{\prime \mu }\gamma _{\mu }^{\prime }=v^{\mu }\gamma _{\mu }$.
Similarly, in this geometric approach, e.g., $N_{s}^{\prime \mu }\gamma
_{\mu }^{\prime }$ transforms under the LT as every vector transforms, i.e.,
as in LT (\ref{LTE2}), which means that \emph{components} $N_{s}^{\prime \mu
}$ \emph{of the \textquotedblleft space-space\textquotedblright\ torque} $%
N_{s}$ \emph{transform to the components} $N_{s}^{\mu }$ of the same torque $%
N_{s}$ in the $S$ frame; \emph{there is no mixing with the components of the
\textquotedblleft time-space\textquotedblright\ torque} $N_{t}$

\begin{equation}
N_{s}^{0}=\gamma (N_{s}^{\prime 0}+\beta N_{s}^{\prime 1}),\
N_{s}^{1}=\gamma (N_{s}^{\prime 1}+\beta N_{s}^{\prime 0}),\
N_{s}^{2,3}=N_{s}^{\prime 2,3}  \label{ans}
\end{equation}%
and the same for $N_{t}^{\mu }$. The LT (\ref{ans}) of the components of $%
N_{s}$ and the same for $N_{t}$ are obtained in the same way as the LT (\ref%
{LTE2}) are obtained, i.e., that both $N$ and $v$ from the definitions (\ref%
{nls}) are transformed by the LT $R$, Eq. (\ref{LTR}), as in (\ref{RM}) and (%
\ref{LM3}). This is in a sharp contrast to the AT (\ref{tc}) in which the
transformed components $T_{i}$ are expressed by the mixture of components $%
T_{k}^{\prime }$ of the 3D vector $\mathbf{T}^{\prime }$ and of components $%
R_{k}^{\prime }$ of another 3D vector $\mathbf{R}^{\prime }$ from the rest
frame. The AT (\ref{tc}) can be obtained in the same way as the AT (\ref{J2}%
) and (\ref{B}) in Sec. 6 are obtained, i.e., that only $N$ from the
definitions (\ref{nls}) is transformed by the LT $R$, Eq. (\ref{LTR}), but
not the velocity of the observer $v$. Furthermore, $N_{s}$ and $N_{t}$ are
geometric quantities in the 4D spacetime since the components $N_{s}^{\mu }$
and $N_{t}^{\mu }$ are multiplied by the unit vectors $\gamma _{\mu }$,
whereas the 3D torque $\mathbf{T}$ is a geometric quantity in the 3D space.
It is formed multiplying the components $N^{\mu \nu }$ (i.e., $T_{i}$
determined by the same identification as in (\ref{ieb})) of a 4D geometric
quantity, the bivector $N$, by the unit 3D vectors $\mathbf{i}$, $\mathbf{j}$%
, $\mathbf{k}$.

It can be easily proved from (\ref{tst}) and (\ref{st}) that the CBGQs $%
N_{s}^{\prime \mu }\gamma _{\mu }^{\prime }$ and $N_{s}^{\mu }\gamma _{\mu }$
are \emph{the same quantity} $N_{s}$ in $S^{\prime }$ and $S$ frames, and
the same for $N_{t}$%
\begin{equation}
N_{s}=N_{s}^{\prime \mu }\gamma _{\mu }^{\prime }=N_{s}^{\mu }\gamma _{\mu
},\quad N_{t}=N_{t}^{\prime \mu }\gamma _{\mu }^{\prime }=N_{t}^{\mu }\gamma
_{\mu };  \label{RMP}
\end{equation}%
remember that $E^{\prime 1}=(1/\gamma )E^{1}$. This again shows, as in [13]
and [14], that \emph{in the approach with the 4D torques} $N_{s}$ \emph{and}
$N_{t}$ \emph{the principle of relativity is naturally satisfied and there
is no paradox}. Observe that $N_{s}$ is determined in all relatively moving
inertial frames of reference by the interaction of the EDM $p$ of the
permanent magnet and the electric field $E$, whereas $N_{t}$ is determined
by the interaction of $m$ and $E$.

It is worth mentioning that, in contrast to the 3D torque $\mathbf{T}$, the
4D torque $N_{s}$ is the same quantity for observers in $S^{\prime }$ and $S$
even if they use different bases, e.g., $\left\{ \gamma _{\mu }\right\} $, $%
\{r_{\mu }\}$%
\begin{equation}
N_{s}=N_{s}^{\mu }\gamma _{\mu }=N_{s}^{\prime \mu }\gamma _{\mu }^{\prime
}=N_{s,r}^{\mu }r_{\mu }=N_{s,r}^{\prime \mu }r_{\mu }^{\prime },  \label{ms}
\end{equation}%
where the primed quantities are the Lorentz transforms of the unprimed ones.
The same holds for $N_{t}$. For the $\{r_{\mu }\}$ basis, this can be proved
using the transformation matrix $R_{\;\nu }^{\mu }$ from Sec. 3.1 or, as
discussed in connection with Eq. (\ref{ein}), using the LT in the $\{r_{\mu
}\}$ basis.

Let us suppose for a moment that a permanent magnet possesses only a MDM $m$
and not an EDM $p$. This is as in the usual approaches, but we deal with
correctly defined vectors in the 4D spacetime and with their LT and not with
the 3-vectors and their AT. In that case, as can be seen from (\ref{tst}), (%
\ref{st}) and (\ref{RMP}), (\ref{ms}) in the rest frame $S^{\prime }$ and in
the lab frame $S$ as well, i.e., \emph{in all relatively moving inertial
frames of reference and for all bases in them, the \textquotedblleft
space-space\textquotedblright\ torque} $N_{s}=0$ \emph{and only remains the
\textquotedblleft time-space\textquotedblright\ torque} $N_{t}$,%
\begin{equation}
p=0;\quad N_{s}=0,\ N_{t}\neq 0.  \label{pn}
\end{equation}%
As already stated, in $S^{\prime }$, the torque $N_{t}$ corresponds to the
expression $\mathbf{R}=\mathbf{m}\times \mathbf{E}=-mE\widehat{\mathbf{y}}$
in Cross [36]. In all other papers from [34-36] there is no 3D torque in the
rest frame $S^{\prime }$. However, as a result of the AT of the 3-vectors (%
\ref{plp}), or (\ref{JCB}), i.e., the AT of components (\ref{tc}) as in
Cross [36], there is the usual 3D torque $\mathbf{T}$ in $S$. In the
considered case ($p=0$), in the approach with $N_{s}$ and $N_{t}$, according
to (\ref{st}) and (\ref{RMP}) $N_{s}=0$ and $N_{t}$ is the same as in $%
S^{\prime }$, which means that there is no paradox. In the formulation with $%
N$, the relations (\ref{tn}), (\ref{m}) and (\ref{enc}) again show that for $%
p=0$ the torque $N$ is the same 4D quantity in all relatively moving
inertial frames of reference and for all bases chosen in them and again
there is no paradox. Also, as in the case with $p\neq 0$, there is no need
either for the \textquotedblleft hidden\textquotedblright\ mechanical
angular momentum or for the \textquotedblleft hidden\textquotedblright\
torque.\bigskip

\noindent \textbf{10. On the Aharonov-Bohm effect in terms of fields. Is the
AB}

\textbf{effect purely quantum mechanical in nature?}\textit{\bigskip }

If the existence of the electric fields from a stationary permanent magnet
would be experimentally proved then it would enable a new interpretation of
the particle interference experiments, particularly of the Aharonov - Bohm
(AB) effect [53]. Such electric fields offer a new possibility for the
explanation of the experimentally observed fringe shift for the magnetic AB
effect even in Tonomura's experiments [54] \emph{in terms of forces, which
so far have been overlooked. }In this paper only a qualitative consideration
will be presented.

Regarding the experiments with microscopic solenoids, e.g., [55], and also
the recent experiment with macroscopic solenoid [56], they can be naturally
explained by the fact that \emph{always} there is an electric field outside
stationary resistive conductors carrying constant currents, i.e., by \emph{%
the existence of the electric force} acting on the particle. For the
existence of such external electric fields from resistive conductors see,
e.g., Sec. 4 in [42] and references therein. If the experiments would be
made with superconducting solenoids then again there would be the external
electric field, the second-order electric field (\ref{eovi}). Thus, even in
that case, it cannot be argued that there is no force acting on the particle
and consequently that the observed phase shift is entirely due to nonzero
vector potential.

Let us explain the above assertions in more detail. Aharonov and Bohm [53]
theoretically predicted that there is a relative phase shift between two
electron beams that pass on both sides of an infinitely long, stationary,
coil with a steady current. The magnetic field does not exist outside that
coil. But, using quantum mechanics, Aharonov and Bohm [53] showed that the
interference fringes are displaced proportionally to the magnetic flux $\Phi
_{B}$ flowing inside the coil even though neither electron beam touch the
magnetic flux. They asserted: \textquotedblleft We shall show that, contrary
to the conclusions of classical mechanics, there exist effects of potentials
on charged particles, even in the region where \emph{all the fields} (my
emphasis) (and therefore the forces on the particles)
vanish.\textquotedblright\ Thus, according to them, the electron wave
packets are influenced although \emph{it is supposed} \emph{that they travel
through regions entirely free from electromagnetic fields.} As the vector
potential $\mathbf{A}$ exists in the considered case even in the magnetic
field-free regions Aharonov and Bohm [53] proposed that \textquotedblleft in
quantum mechanics, the fundamental physical entities are the
potentials.\textquotedblright\ Their results became very important in the
forefront of physics. Vector potentials are generalized to gauge fields and
these fields are considered to be the fundamental physical quantities in the
modern theories of gauge fields.

They, [53], have found that the relative phase shift, $\Delta \varphi $, is
produced between the two wave packets due to the vector potential as%
\begin{equation}
\Delta \varphi =(e/\hbar )\oint \mathbf{A}\cdot \mathbf{dl}=(e/\hbar )\int
\mathbf{B}\cdot \mathbf{dS}=(e/\hbar )\Phi _{B},  \label{sAB}
\end{equation}%
where $\mathbf{B}$ is the magnetic field of the solenoid.

Instead of an infinite coil with current the experiments, e.g., [55], dealt
with microscopic solenoids. M\"{o}llenstedt and Bayh, [55], observed a
fringe shift that is in agreement with the relation (\ref{sAB}). The overlap
between the incident electrons and the \emph{magnetic} field strengths in
their experiments was fairly small. Therefore, the experiments [55] are
usually considered as a convincing demonstration of the existence of the AB
effect, i.e., that the relative phase shift is due only to the vector
potential and, according to (\ref{sAB}), due to the quantum action of the
magnetic flux enclosed between the two interfering electron trajectories.
Thus, according to the presently accepted formulation the AB phase shift is
caused by an enclosed magnetic flux and there is no need to examine any
interaction between the passing charges and the sources of the magnetic flux.

However, several authors questioned the existence of the AB effect and tried
to explain the observed fringe shift in a classical way, i.e., in terms of
fields and the Lorentz force. One reason for such a possibility is the
unphysical character of the infinitely long solenoid that is commonly used
in the discussions of the AB effect. Hence, the experimental results with
finite solenoids, e.g., in [55], and also with finite whiskers [57], could
be attributed to the magnetic flux leaking outside such solenoids or
whiskers.

In Tonomura's experiments [54] a tiny toroidal magnet is used instead of
straight solenoids. The magnet was covered with a superconductor layer and
further with a copper layer. The phase shift between two waves passing
through the hole and outside of the toroid was measured by means of electron
holography. The AB phase shift was detected even though the magnetic field
was confined to the toroidal magnet; the Meissner effect prevented any flux
from leaking out. The copper outer layer prevented any electrons from
penetrating the magnet itself, i.e., there was no overlap of the incident
electron wave with the magnetic fields inside the sample. Thus, it is
generally accepted that Tonomura's experiments [54] give conclusive evidence
for the AB\ effect, i.e., that there is no force acting on the particle and
consequently that the observed phase shift is entirely due to nonzero vector
potential. For a general review see, e.g., [58].

It is very interesting that both in the theoretical discussions and in the
experiments with microscopic solenoids, and also in the recent experiment
with macroscopic solenoid [56], it is never noticed that \emph{always} there
is an electric field outside stationary resistive conductors carrying
constant currents. In such ohmic conductors there are quasistatic surface
charges, which generate not only the electric field inside the wire driving
the current, \emph{but also a time independent electric field outside it,}
see, e.g., Sec. 4 in [42] and references therein. There are no analytic
solutions for these surface charges and the electric fields outside the wire
for the case of finite solenoids. It is very difficult to determine the
distribution of surface charges on the conductors of a circuit because it
depends on the geometry of the circuit itself and of its surroundings. In
order to have some qualitative orientation about that external electric
field one can look at Eq. (16) in [42]. It is derived for a cylindrical wire
of finite length. From that equation it is visible that \emph{the electric
field outside the wire is proportional to the current.} The radial component
of that field falls as $1/r$. The proportionality of the external electric
field with the current in the solenoid explains all important features of,
e.g., the experiments from [55]. The magnetic flux $\Phi _{B}$ flowing
inside the solenoid is also proportional to the current and it explains the
generally accepted belief that the fringe shift in the experiments with
microscopic solenoids is determined by the magnetic flux inside the
solenoid. However, as seen from the above discussion, the relative phase
shift \emph{is not} due to the vector potential, i.e., according to (\ref%
{sAB}), due to the quantum action of the magnetic flux, but \emph{it is due
to the existence of the electric field outside stationary solenoids with
steady currents.}

From the viewpoint of the approach with 4D geometric quantities the \emph{%
components} of the electric field 3-vector in Eq. (16) in [42] have to be
understood as the \emph{spatial} \emph{components }in the standard basis of
the electric field vector, a 4D geometric quantity; the rest frame of the
considered wire is taken to be the $\gamma _{0}$-frame and therefore the
temporal component $E^{0}=0$ (also $B^{0}=0$). Similarly, the surface charge
density from Eq. (12) in [42] has to be understood as the \emph{temporal
component }in the standard basis of the current density vector. In the rest
frame of the wire the \emph{spatial} \emph{components are zero},\emph{\ }$%
j^{i}=0$. Also, in our approach, the vector of the Lorentz force $K$ is
given by the expression (\ref{KEB}).

Observe that in [42] two other contributions to the external electric field
are mentioned in Secs. 3 and 5, but they are of no concern for our
consideration. In Sec. 3 in [42] the electric field from the induced charges
is investigated, but it is the same whether or not there is current in the
wire. In Sec. 5 in [42] the electric field proportional to the square of the
current is derived using an action-at-a-distance Weber's electrodynamics,
which is not in agreement with the field theory, i.e., with the special
relativity. On the other hand, the second-order electric field (\ref{eovi})
is proportional to the square of the current and it is derived in a
relativistically correct way. In our approach it replaces that one from Sec.
5 in [42]. The electric field (\ref{eovi}) exists in a stationary resistive
conductor carrying constant current as well, but there it is negligible in
comparison with the external electric field that is caused by the
distribution of surface charges on the conductor, i.e., with the field which
is proportional to the current.

From the consideration in Sec. 8 it follows that, both in the Amp\`{e}rian
approach and in the approach in which the intrinsic EDM vector $d$ is
determined by the new \textquotedblleft time-space\textquotedblright\ spin
vector $Z$,\ there is a time independent electric field outside stationary
permanent magnet. This yields the possibility to explain the fringe shifts
in Tonomura's experiments [54] by the existence of the external electric
fields from a stationary permanent magnet and not, as generally accepted, by
the existence of a nonzero vector potential. The title of the paper [54] is
\textquotedblleft Evidence for Aharonov-Bohm Effect with Magnetic Field
Completely Shielded from Electron Wave\textquotedblright\ and it is written
in the Abstract: \textquotedblleft A toroidal ferromagnet was covered with a
superconductor layer to confine the field, and further with a copper layer
for complete shielding from the electron wave.\textquotedblright\ Thus, in
[54], the experimental setup is designed in such a way that practically the
overlap between the incident electrons and the \emph{magnetic} field
strengths is negligible.

However, both layers that are used in [54] do not prevent the overlap
between the wave functions of the incident electrons and \emph{the electric
field from stationary, toroidal ferromagnet}, which is predicted in Sec. 8.

A strong theoretical argument that supports the interpretation of the
particle interference experiments in terms of forces and not in terms of
potentials comes from the fact that, as already stated, i.e., as shown in
the axiomatic formulation of the electromagnetism [12], the bivector $F$ can
be taken as the primary quantity and the field equation for $F$ (\ref{MEF})
is the basic equation for the whole electromagnetism; the bivector field $F$
yields the complete description of the electromagnetic field and there is no
need to introduce either the field vectors or the potentials.

Furthermore, the qualitative theoretical explanations of the quantum phase
shifts in terms of classical force vectors in the Aharonov-Casher and the R%
\"{o}ntgen effects are already given in [46, 47]. In [48], the dipole
moments are quantized according to (\ref{1}) and it is shown that the
expectation value for the quantum force vector is not zero in the case of
the Aharonov-Casher and the R\"{o}ntgen effects and in the neutron
interferometry. This means that the phase shifts in these experiments are
not due to force-free interaction of the dipole, i.e., they are not
topological phase shifts.\bigskip

\noindent \textbf{11. Particle interference and Lorentz-violating
electrodynamics}\textit{\bigskip }

In the recent paper [37], under the title \textquotedblleft Particle
interference as a test of Lorentz-violating
electrodynamics,\textquotedblright\ it is argued that in a Lorentz-violating
model of electrodynamics [59] a magnetic solenoid generates not only a
static magnetic field but also a static electric field, which acts on
interfering particles producing an extra path-dependent phase. That
nontopological phase is considered to be the Lorentz-violating correction to
the standard topological (path-independent) Aharonov-Bohm phase.

However, as shown in the consideration in Secs. 7.1, 7.2 and 8, there is a
static electric field not only outside a magnetic solenoid but also outside
a stationary permanent magnet and that field is obtained in a consistent
\emph{Lorentz-covariant approach} with 4D geometric quantities. Furthermore,
it is worth mentioning that in [37], and [59] as well as in almost the whole
physical literature, it is believed that Maxwell's equations with the
3-vectors are covariant under the LT, considering that the LPET (\ref{JCB})
and (\ref{ps}) are the relativistically correct LT. But, the LPET (\ref{JCB}%
) and (\ref{ps}) ARE NOT the LT and consequently, as explicitly proved in
[8], Maxwell's equations with 3-vectors ARE NOT\ covariant under the
relativistically correct LT (\ref{LTE1}) and (\ref{LTE2}). Hence, there is
no sense to develop Lorentz-violating model of electrodynamics assuming that
the LPET (\ref{JCB}) and (\ref{ps}) are the LT and that Maxwell's equations
with 3-vectors are covariant under the LT. It would be much more important
that physics community stops to consider that in the 4D spacetime the
Lorentz contraction and the dilation of time are the intrinsic relativistic
effects and that the LPET (\ref{JCB}) and (\ref{ps}) are the
relativistically correct LT.\bigskip

\noindent \textbf{12. Discussion and Conclusions\bigskip }

The whole consideration explicitly shows that the 3D quantities $\mathbf{E(r,%
}t\mathbf{)}$\emph{\ }and $\mathbf{B(r,}t\mathbf{)}$, $\mathbf{P(r,}t\mathbf{%
)}$\emph{\ }and $\mathbf{M(r,}t\mathbf{)}$, $\mathbf{p}$ and $\mathbf{m}$,
their LPET, i.e., the AT, (\ref{JCB}), (\ref{ps}), or (\ref{psa}), (\ref{plp}%
), and the equations with them are not well-defined in the 4D spacetime.
More generally, we can conclude that \emph{the 3D quantities do not have an
independent physical reality in the 4D spacetime. }Contrary to the general
belief, we find that, in the 4D spacetime, \emph{it is not true that
observers in relative motion see different fields; the transformations, Eqs.}
(\ref{JCB}), (\ref{ps}), or (\ref{psa}),\emph{\ or, equivalently, }(\ref{EP1}%
)\emph{\ and }(\ref{J1})\emph{, i.e., in the standard basis Eqs. }(\ref{J2})%
\emph{\ and }(\ref{B}),\emph{\ are not the LT but the LPET, i.e., the AT.
According to the LT; Eqs.} (\ref{RM}) - (\ref{LTE2}), \emph{the electric
field }$E$ \emph{transforms only} \emph{to the electric field }$E^{\prime }$
\emph{and the same holds for the magnetic field }$B$, \emph{for the vectors
of the polarization }$P$ \emph{and the magnetization }$M$ \emph{and for the
EDM} $p$ \emph{and the MDM} $m$.

As already stated, the principle of relativity is automatically included in
the approach with well-defined 4D geometric quantities, i.e., in the ISR,
whereas in Einstein's formulation of the special relativity [1] the
principle of relativity is postulated outside the framework of a correct
mathematical formulation of the theory and it is supposed that it holds for
the equations, the physical laws, which are expressed in terms of the 3D
quantities.

Minkowski's great discovery of the correct LT, Sec. 11.6 in [21], their
generalization and the explicit forms (\ref{LTE1}) and (\ref{LTE2}) that are
found in [6-11] and also the mathematical argument from [19] that space and
time dependent electric and magnetic fields cannot be the usual 3-vectors
strongly suggest the need for further critical examination of the usual
formulation of electromagnetism with 3-vectors $\mathbf{E(r,}t\mathbf{)}$
and $\mathbf{B(r,}t\mathbf{)}$, $\mathbf{P(r,}t\mathbf{)}$\emph{\ }and $%
\mathbf{M(r,}t\mathbf{)}$ and their LPET (\ref{JCB}), (\ref{ps}), or (\ref%
{psa}). It also suggests the possibility for a complete and relativistically
correct formulation of classical and quantum electromagnetism with
multivector fields (as physically real fields), which are defined on the 4D
spacetime and which transform according to the correct LT (\ref{RM}) - (\ref%
{LTE2}).

The advantages of such formulation with 4D geometric quantities, i.e., of
the ISR, are already revealed in the cases of the interaction between the
dipole moment tensor $D^{ab}$ and the electromagnetic field $F^{ab}$ in [48]
and in much more detail in [33], in the discussion of quantum phase shifts
in [46, 47], in the discussion of shortcomings in the current EDM searches
in [49] and in the formulation of Majorana form of the Dirac-like equation
for the free-photon [60].

Particularly important results of the 4D geometric approach, i.e., of the
ISR, that are reported in this paper, Secs. 7.1, 7.2, 8, refer to the
existence of the second-order electric field outside a superconducting loop
with steady current and to the new prediction of the electric field outside
a stationary permanent magnet, i.e. to the prediction that a stationary
permanent magnet possesses an intrinsic polarization, which induces the
external electric field. Also, it is suggested that the measurements of that
external electric field from a stationary permanent magnet could be
performed by the same method with cold ions as in [31] and possibly as in
[43].

The investigation of the \textquotedblleft charge-magnet
paradox\textquotedblright\ from Secs. 9.1 and 9.2 again shows that the
relativistically correct description of physical phenomena without any
paradoxes can be achieved in the consistent way with 4D geometric quantities
as physical quantities in the 4D spacetime. On the other hand, the use of 3D
quantities and their AT necessarily leads to different ambiguities and
inconsistencies.

A qualitative explanation of the AB\ effect from Sec. 10, together with
possibly positive experimental results for the existence of the electric
field vector from stationary permanent magnet, will surely be important for
better understanding of the classical limit of the quantum physics. A more
quantitative calculation of the AB effect in terms of fields will be done,
e.g., using the method from [61] and the decomposition of $F$ (\ref{E2}%
).\bigskip

\noindent \textbf{Acknowledgments\bigskip }

I am cordially thankful to Zbigniew Oziewicz for numerous and very useful
discussions during years, which, among others, helped me to better
understand that the relativistically correct mathematical formulation of the
electromagnetism \emph{requires} the representation of the electric and
magnetic fields by the 4D geometric quantities and for the continuos support
of my work. It is a pleasure to acknowledge to Larry Horwitz for inviting me
to the IARD conferences, for the valuable discussions and for the continuos
support of my work. I am also grateful to Alex Gersten for useful
discussions and for the continuos support of my work.\bigskip

\noindent \textbf{Appendix\bigskip }

In this Appendix we shall briefly describe the essential differences between
the 4D geometric approach, the ISR, and Einstein's definition of the Lorentz
contraction, e.g., for a moving rod. This is explained in detail in Secs. 2
- 2.3 in [27] and Secs. 3.1, 4.1 and Figs. 1 and 3 in [16]. Here, the
notation is slightly different than in [27] and [16]. In the geometric
approach one deals with the abstract 4D geometric quantities, i.e., with the
position vectors $x_{A}$, $x_{B},$ of the events $A$ and $B$, respectively,
with the distance vector $l_{AB}=x_{B}-x_{A}$ and with the spacetime length,
$l=L_{0}$, see (\ref{sl}). The essential feature of the geometric approach
is that \emph{any abstract 4D geometric quantity}, e.g., the distance vector
$l_{AB}=x_{B}-x_{A}$, \emph{is} \emph{only one quantity, the same quantity
in the 4D spacetime} for all relatively moving frames of reference and for
all systems of coordinates that are chosen in them. The abstract vector $%
l_{AB}$ can be decomposed in different bases and then these representations,
the CBGQs, of the same abstract 4D geometric quantity $l_{AB}$ contain both
the basis components and the basis vectors. Let us explain it taking a
particular choice for $l_{AB}$, which in the usual \textquotedblleft
3+1\textquotedblright\ picture corresponds to a rod that is at rest in an
inertial frame of reference (IFR) $S$ (with the standard basis in it) and
situated along the common $x^{1}$, $x^{\prime 1}$ $-$ axes. Its rest length
is denoted as $L_{0}$. The situation is depicted in Fig. 1 in [16]. $l_{AB}$
is decomposed, i.e., it is written as a CBGQ, in the standard basis and in $%
S $ and $S^{\prime }$, where the rod is moving, as
\begin{equation}
l_{AB}=l_{AB}^{\mu }\gamma _{\mu }=0\gamma _{0}+L_{0}\gamma
_{1}=l_{AB}^{\prime \mu }\gamma _{\mu }^{\prime }=-\beta \gamma L_{0}\gamma
_{0}^{\prime }+\gamma L_{0}\gamma _{1}^{\prime },  \label{dv}
\end{equation}%
As already stated several times, the components $l_{AB}^{\mu }$ are
transformed by the LT and the basis vectors $\gamma _{\mu }$ by the inverse
LT leaving the whole CBGQ unchanged. In $S$, the position vectors $x_{A,B}$
are determined simultaneously, $x_{B}^{0}-x_{A}^{0}=l_{AB}^{0}=0$, i.e., the
temporal part of $l_{AB}^{\mu }$ is zero. In the standard basis, which is
commonly used in the usual approaches, there is a dilation of the spatial
part $l_{AB}^{\prime 1}=\gamma L_{0}$ with respect to $l_{AB}^{1}=L_{0}$ and
not the Lorentz contraction as predicted in Einstein's formulation of
special relativity. Similarly, as explicitly shown in [27] and [16], in the $%
\{r_{\mu }\}$ basis, i.e., with the \textquotedblleft r\textquotedblright\
synchronization, if only spatial parts of $l_{AB,r}^{\mu }$ and $%
l_{AB,r}^{\prime \mu }$ are compared then one finds the dilation $\infty
\succ l_{AB,r}^{\prime 1}\geq L_{0}$ for all $\beta _{r}$. However, the
comparison of only spatial parts of the components of the distance vector $%
l_{AB}$ in $S$ and $S^{\prime }$ is physically meaningless in the geometric
approach,\ since \emph{some components of the tensor quantity, when they are
taken alone, do not correspond to some definite 4D physical quantity.} Note
that if $l_{AB}^{0}=0$ then the LT yield that $l_{AB}^{\prime \mu }$ in any
other IFR $S^{\prime }$ contains the time component as well, $l_{AB}^{\prime
0}=x_{B}^{\prime 0}-x_{A}^{\prime 0}=-\beta \gamma L_{0}\neq 0$. Hence,
\emph{the LT yield that the spatial ends of the rod are not determined
simultaneously in} $S^{\prime }$, \emph{i.e., the temporal part of} $%
l_{AB}^{\prime \mu }$ \emph{is not zero.} For the spacetime length $l$ it
holds that%
\begin{equation}
l^{2}=\mid l_{AB}^{\mu }l_{AB,\mu }\mid =\mid l_{AB}^{\prime \mu }l_{AB,\mu
}^{\prime }\mid =\mid l_{AB,r}^{\mu }l_{AB,r,\mu }\mid =L_{0}^{2}.
\label{sl}
\end{equation}%
In $S$, the rest frame of the rod, where the temporal part of $l_{AB}^{\mu }$
is $l_{AB}^{0}=0,$ the spacetime length $l$ is a measure of the spatial
distance, i.e., of the rest spatial length of the rod, as in the
prerelativistic physics. \emph{The observers in all other IFRs will
\textquotedblleft look\textquotedblright\ at the same events} $A$ \emph{and}
$B$, \emph{the same distance vector} $l_{AB}$ \emph{and the same spacetime
length} $l$, \emph{but associating with them different coordinates; it is
the essence of the geometric approach. They all obtain the same value} $l$
\emph{for the spacetime length,} $l=L_{0}$.

It is worth mentioning, once again, that the 4D geometric treatment with $%
l_{AB}$ and $l$ is a generalization and a mathematically better founded
formulation of the ideas expressed by Rohrlich [15] and Gamba [40]. Indeed,
Rohrlich [15] states: \textquotedblright A quantity is therefore physically
meaningful (in the sense that it is of the same nature to all observers) if
it has tensorial properties under Lorentz
transformations.\textquotedblright\ Similarly Gamba [40], when discussing
the sameness of a physical quantity (for example, \emph{a nonlocal quantity}
$A_{\mu }(x_{\lambda },X_{\lambda })$, which is a function of two points in
the 4D spacetime $x_{\lambda }$ and $X_{\lambda }$) for different inertial
frames of reference $S$ and $S^{\prime }$, declares: \textquotedblright The
quantity $A_{\mu }(x_{\lambda },X_{\lambda })$ for $S$ is the same as the
quantity $A_{\mu }^{\prime }(x_{\lambda }^{\prime },X_{\lambda }^{\prime })$
for $S^{\prime }$ when all the primed quantities are obtained from the
corresponding unprimed quantities through Lorentz transformations (tensor
calculus).\textquotedblright\ Rohrlich and Gamba worked with the usual
covariant approach, i.e., with the components implicitly taken in the
standard basis, which means that only Einstein's synchronization is
considered to be physically admissible. The quantities $A_{\mu }(x_{\lambda
},X_{\lambda })$ and $A_{\mu }^{\prime }(x_{\lambda }^{\prime },X_{\lambda
}^{\prime })$ refer to the same physical quantity, but they are not
mathematically equal quantities since bases are not included. In the
approach with the 4D geometric quantities, i.e., in the ISR, one deals with
mathematically equal quantities, e.g., for a nonlocal quantity $%
l_{AB}=x_{B}-x_{A}$ it holds that

\begin{equation}
l_{AB}=l_{AB}^{\mu }\gamma _{\mu }=l_{AB}^{\prime \mu }\gamma _{\mu
}^{\prime }=l_{AB,r}^{\mu }r_{\mu }=l_{AB,r}^{\prime \mu }r_{\mu }^{\prime
}=..,  \label{dv1}
\end{equation}%
where the primed quantities are the Lorentz transforms of the unprimed ones.
In order to treat different systems of coordinates on an equal footing we
have derived a form of the LT that is independent of the chosen system of
coordinates, including different synchronizations, see Eq. (2) in [27], or
Eq. (1) in [16]. Also, Eq. (4) in [16], it is presented the transformation
matrix that connects Einstein's system of coordinates with another system of
coordinates in the same reference frame.

On the other hand, as shown in Sec. 2.2 in [27] and Sec. 4.1 and Fig. 3 in
[16], in Einstein's formulation of SR, instead of to work with geometric
quantities $x_{A,B}$, $l_{AB}$ and $l$ one deals only with the spatial, or
temporal, \emph{components} of their coordinate representations $x_{A}^{\mu
} $, $x_{B}^{\mu }$ and $l_{AB}^{\mu }$ \emph{in the standard basis}. The
geometric character of physical quantities, i.e., the basis vectors, and
some asymmetric synchronization, e.g., the \textquotedblleft
r\textquotedblright\ synchronization, which is equally physical as the
Einstein synchronization, are never taken into account. According to
Einstein's definition [1] of the spatial length the spatial ends of the rod
must be taken simultaneously for the observer, i.e., \emph{he} \emph{defines
length as the spatial distance between two spatial points on the (moving)
object measured by simultaneity in the rest frame of the observer.} In the
4D (here, for simplicity, as in [27] and [16], we deal only with 2D)
spacetime and in the $\{\gamma _{\mu }\}$ basis the simultaneous events $A$
and $B$ (whose spatial parts correspond to the spatial ends of the rod) are
the intersections of $x^{1}$ axis (that is along the spatial basis vector $%
\gamma _{1}$) and the world lines of the spatial ends of the rod that is at
rest in $S$ and situated along the $x^{1}$ axis. The components of the
distance vector are $l_{AB}^{\mu }=x_{B}^{\mu }-x_{A}^{\mu }=(0,L_{0})$; for
simplicity, it is taken that $t_{B}=t_{A}=a=0$. Then in $S$, the rest frame
of the object, the spatial part $l_{AB}^{1}=L_{0}$ of $l_{AB}^{\mu }$ is
considered to define the rest spatial length. Furthermore, one uses the
inverse LT to express $x_{A}^{\mu }$, $x_{B}^{\mu }$ and $l_{AB}^{\mu }$ in $%
S$ in terms of the corresponding quantities in $S^{\prime }$, in which the
rod is moving. This procedure yields
\begin{eqnarray}
l_{AB}^{0} &=&ct_{B}-ct_{A}=\gamma (l_{AB}^{\prime 0}+\beta l_{AB}^{\prime
1}),  \notag \\
l_{AB}^{1} &=&x_{B}^{1}-x_{A}^{1}=\gamma (l_{AB}^{\prime 1}+\beta
l_{AB}^{\prime 0}).  \label{lc}
\end{eqnarray}%
Now, instead of to work with 4D tensor quantities and their LT, as in the 4D
geometric approach, in the usual formulation one forgets about the
transformation of the temporal part $l_{AB}^{0}$, the first equation in (\ref%
{lc}), and considers only the transformation of the spatial part $l_{AB}^{1}$%
, the second equation in (\ref{lc}). Furthermore, \emph{in that relation for}
$l_{AB}^{1}$ \emph{one assumes that} $t_{B}^{\prime }=t_{A}^{\prime
}=t^{\prime }=b$, i.e., \emph{that} $x_{B}^{\prime 1}$ \emph{and} $%
x_{A}^{\prime 1}$ \emph{are simultaneously determined at some arbitrary} $%
t^{\prime }=b$ in $S^{\prime }$. However, \emph{in 4D} (at us 2D) spacetime
such an assumption means that \emph{in} $S^{\prime }$ \emph{one does not
consider the same events} $A$ \emph{and} $B$ \emph{as in} $S$ but some other
two events $C$ and $D$, which means that $t_{B}^{\prime }=t_{A}^{\prime }$
has to be replaced with $t_{D}^{\prime }=t_{C}^{\prime }=b$. The events $C$
and $D$ are the intersections of the line (the hypersurface $t^{\prime }=b$
with arbitrary $b$) parallel to the spatial axis $x^{\prime 1}$ (which is
along the spatial base vector $\gamma _{1}^{\prime }$) and of the above
mentioned world lines of the spatial end points of the rod. Then, in the
above transformation for $l_{AB}^{1}$ (\ref{lc}) one has to write $%
x_{D}^{\prime 1}-x_{C}^{\prime 1}=l_{CD}^{\prime 1}$ instead of $%
x_{B}^{\prime 1}-x_{A}^{\prime 1}=l_{AB}^{\prime 1}$. The spatial parts $%
l_{AB}^{1}$ and $l_{CD}^{\prime 1}$ are the \emph{spatial distances }between
the events $A$, $B$ and $C$, $D$, respectively. \emph{In Einstein's
formulation, the spatial distance}\textit{\ }$%
l_{AB}^{1}=x_{B}^{1}-x_{A}^{1}=L_{0}$ \emph{defines the spatial length of
the rod at rest in }$S$,\emph{\ while} $l_{CD}^{\prime 1}=x_{D}^{\prime
1}-x_{C}^{\prime 1}$ \emph{is considered to define the spatial length of the
moving rod in}\textit{\ }$S^{\prime }$. Hence, from the equation for $%
l_{AB}^{1}$ (\ref{lc}) one finds the relation between $l^{\prime
1}=l_{CD}^{\prime 1}$ and $l^{1}=l_{AB}^{1}=L_{0}$ as the famous formula for
the Lorentz contraction of the moving rod
\begin{equation}
l^{\prime 1}=x_{D}^{\prime 1}-x_{C}^{\prime 1}=L_{0}/\gamma
=(x_{B}^{1}-x_{A}^{1})/\gamma ,\,\,\,\mathrm{with}\,\,t_{C}^{\prime
}=t_{D}^{\prime },\,\,\,\mathrm{and}\,\,\,t_{B}=t_{A},  \label{contr}
\end{equation}%
where $\gamma =(1-\beta ^{2})^{-1/2}$, $\beta =U/c$ and $U=\left\vert
\mathbf{U}\right\vert $; $\mathbf{U}$ is the 3-velocity of $S^{\prime }$
relative to $S$. As can be nicely seen from Fig. 3 in [16], the spatial
lengths $L_{0}$ and $l_{CD}^{\prime 1}$ refer not to the same 4D tensor
quantity,\ as in the 4D geometric approach, see Fig. 1 in [16], but to two
different quantities in the 4D spacetime. These quantities are obtained by
the same measurements in $S$ and $S^{\prime }$; the spatial ends of the rod
are measured simultaneously at some $t=a$ in $S$ and also at some $t^{\prime
}=b$ in $S^{\prime }$, and $a$ in $S$ and $b$ in $S^{\prime }$ are not
related by the LT or any other coordinate transformation. This means that
the Lorentz contraction, as already shown by Rohrlich [15] and Gamba [40],
is a typical example of an AT. \emph{It has nothing in common with the LT of
the 4D geometric quantities.} We see that in Einstein's approach [1] the
spatial and temporal parts of events are treated separately, and moreover
the time component is not transformed in the transformation that is called -
the Lorentz contraction. Thus, contrary to the generally accepted opinion,
\emph{the Lorentz contraction is not a well-defined relativistic effect in
the 4D spacetime.}\bigskip

\noindent \textbf{References\bigskip }

\noindent \lbrack 1] A. Einstein, Annalen der Physik \textbf{17,} 891
(1905); Translated by W. Perrett

and G. B. Jeffery in: \textit{The Principle of Relativity} (Dover, New York,
1952).

\noindent \lbrack 2] J. D. Jackson, \textit{Classical Electrodynamics} 3rd
ed. (Wiley, New York, 1998 ).

\noindent \lbrack 3] H. A. Lorentz, Proceedings of the Royal Netherlands

Academy of Arts and Sciences \textbf{6,} 809 (1904).

\noindent \lbrack 4] H. Poincar\'{e}, Rend. del Circ. Mat. di Palermo
\textbf{21,} 129 (1906).

\noindent \lbrack 5] A. A. Logunov, Hadronic J. \textbf{19,} 109 (1996).

\noindent \lbrack 6] T. Ivezi\'{c}, Found. Phys\textit{.} \textbf{33,} 1339
(2003).

\noindent \lbrack 7] T. Ivezi\'{c}, Found. Phys. Lett. \textbf{18, }301
(2005).

\noindent \lbrack 8] T. Ivezi\'{c}, Found. Phys. \textbf{35,} 1585 (2005).

\noindent \lbrack 9] T. Ivezi\'{c}, Fizika A\textit{\ }\textbf{17,} 1 (2008).

\noindent \lbrack 10] T. Ivezi\'{c}, arXiv: 0809.5277.

\noindent \lbrack 11] T. Ivezi\'{c}, Phys. Scr. \textbf{82,} 055007 (2010).

\noindent \lbrack 12] T. Ivezi\'{c}, Found. Phys. Lett. \textbf{18,} 401
(2005).

\noindent \lbrack 13] T. Ivezi\'{c}, Found. Phys. \textbf{37}, 747 (2007).

\noindent \lbrack 14] T. Ivezi\'{c}, Found. Phys. \textbf{36}, 1511 (2006);

T. Ivezi\'{c}, Fizika A \textbf{16,} 207 (2007).

\noindent \lbrack 15] F. Rohrlich, Nuovo Cimento B \textbf{45}, 76 (1966).

\noindent \lbrack 16] T. Ivezi\'{c}, Found. Phys. \textbf{31,} 1139 (2001).

\noindent \lbrack 17] T. Ivezi\'{c}, Found. Phys. Lett. \textbf{15,} 27
(2002); arXiv: physics/0103026;

arXiv: physics/0101091.

\noindent \lbrack 18] D. Hestenes, \textit{Space-Time Algebra (}Gordon \&
Breach, New York, 1966);

D. Hestenes and G. Sobczyk, \textit{Clifford Algebra to }

\textit{Geometric Calculus }(Reidel, Dordrecht, 1984);

C. Doran and A. Lasenby, \textit{Geometric algebra for physicists}

(Cambridge University Press, Cambridge, 2003 ).

\noindent \lbrack 19] Z. Oziewicz, J. Phys.: Conf. Ser. \textbf{330}, 012012
(2011);

Z. Oziewicz, Rev. Bull. Calcutta Math. Soc.\textit{\ }\textbf{16,} 49 (2008);

Z. Oziewicz and C. K. Whitney, Proc. Nat. Phil. Alliance\textit{\ }

\textit{(NPA)} \textbf{5,} 183 (2008) (also at http://www.worldnpa.org/php/).

\noindent \lbrack 20] T. Ivezi\'{c}, arXiv: 1101.3292.

\noindent \lbrack 21] H. Minkowski, Nachr. Ges. Wiss. G\"{o}ttingen, 53
(1908);

Reprinted in: Math. Ann. \textbf{68,} 472 (1910);

English translation in: M. N. Saha and S. N. Bose \textit{The Principle }

\textit{of Relativity: Original Papers by A. Einstein and H. Minkowski}

(Calcutta University Press, Calcutta, 1920).

\noindent \lbrack 22] W. G. W. Rosser, \textit{Classical Electromagnetism
via Relativity}

(Plenum, New York, 1968).

\noindent \lbrack 23] W. K. H. Panofsky and M. Phillips, \textit{Classical
electricity and magnetism}

2nd ed. (Addison-Wesley, Reading, 1962).

\noindent \lbrack 24] E.M. Purcell, \textit{Electricity and Magnetism }2nd
ed. (McGraw-Hill, New York,

1985\textit{)}.

\noindent \lbrack 25] R. P. Feynman, R. B. Leighton and M. Sands, \textit{%
The Feynman Lectures on}

\textit{Physics} Volume II (Addison-Wesley, Reading, 1964).

\noindent \lbrack 26] D. J. Griffiths, \textit{Introduction to
Electrodynamics} 3rd ed. (Prentice-Hall,

Upper Saddle River, 1999).

\noindent \lbrack 27] T. Ivezi\'{c}, Found. Phys. Lett. \textbf{12,} 507
(1999); arXiv: physics/0102014.

\noindent \lbrack 28] W. F. Edwards, C. S. Kenyon and D. K. Lemon,

Phys. Rev. D \textbf{14,} 922 (1976); D. K. Lemon, W. F. Edwards

and C. S. Kenyon, Phys. Lett. A\textit{\ \textbf{62,} 105 (1992).}

\noindent \lbrack 29] G. G. Shishkin et al., J. Phys. D: Appl. Phys. \textbf{%
35} 497 (2002).

\noindent \lbrack 30] U. Bartocci, F. Cardone and R. Mignani, Found. Phys.
Lett. \textbf{14,} 51 (2001);

F. Cardone, R. Mignani and R. Scrimaglio, Found. Phys.\textit{\ }\textbf{36,}
263 (2006).

\noindent \lbrack 31] R. Folman, arXiv: 1109.2586.

\noindent \lbrack 32] T. Ivezi\'{c}, Phys. Lett. A \textbf{144}, 427 (1990).

\noindent \lbrack 33] T. Ivezi\'{c}, Phys. Scr.\textit{\ }\textbf{81,}
025001 (2010).

\noindent \lbrack 34] M. Mansuripur, Phys. Rev. Lett. \textbf{108}, 193901
(2012).

\noindent \lbrack 35] A. Cho, Science \textbf{336}, 404 (2012).

\noindent \lbrack 36] D. A. T. Vanzella, arXiv: 1205.1502; C. S.
Unnikrishnan, arXiv: 1205.1080;

D. J. Griffiths and V. Hnizdo, arXiv: 1205.4646; K. T. McDonald,

www.physics.princeton.edu/mcdonald/examples/mansuripur.pdf;

D. J. Cross, arXiv: 1205.5451; P. L. Saldanha, arXiv: 1205.6858;

T. H. Boyer, arXiv: 1206.5322; M. Brachet and E. Tirapegul, arXiv: 1207.4613;

Kimball A. Milton, Giulio Meille, arXiv: 1208.4826; A. L. Kholmetskii,

O. V. Missevitch, T. Yarman, 1208.5296.

\noindent \lbrack 37] A. Kobakhidze and B. H. J. McKellar, Phys. Rev. D
\textbf{76}, 093004 (2007).

\noindent \lbrack 38] C. Leubner, K. Aufinger and P. Krumm, Eur. J. Phys.
\textbf{13,} 170 (1992).

\noindent \lbrack 39] J. Van Bladel, \textit{Relativity and Engineering}
(Springer-Verlag, Berlin, 1984).

\noindent \lbrack 40] A. Gamba, Am. J. Phys. \textbf{35}, 83 (1967).

\noindent \lbrack 41] T. Ivezi\'{c}, Int. J. Mod. Phys. B 26, 1250040 (2012).

\noindent \lbrack 42] A. K. T. Assis, W. A. Rodrigues Jr. and A. J. Mania,

Found. Phys. \textbf{29}, 729 (1999).

\noindent \lbrack 43] W. J. Kim, A. O. Sushkov, D. A. R. Dalvit and S. K.
Lamoreaux, Phys.

Rev. Lett. \textbf{103}, 060401 (2009).

\noindent \lbrack 44] A. Romannikov, Found. Phys.\textit{\ }\textbf{41,}
1331 (2011).

\noindent \lbrack 45] T. Ivezi\'{c}, Phys. Lett. A \textbf{156}, 27 (1991).

\noindent \lbrack 46] T. Ivezi\'{c}, Phys. Rev. Lett. \textbf{98,} 108901
(2007).

\noindent \lbrack 47] T. Ivezi\'{c}, Phys. Rev. Lett. \textbf{98, }158901
(2007).

\noindent \lbrack 48] T. Ivezi\'{c}, arXiv: hep-th/0705.0744.

\noindent \lbrack 49] T. Ivezi\'{c}, arXiv: 1005.3037; arXiv: 1006.4154.

\noindent \lbrack 50] F. T. Trouton and H. R. Noble, Proc. Royal Soc.
\textbf{74,} 132 (1903).

\noindent \lbrack 51] J. D. Jackson, Am. J. Phys. \textbf{72}, 1484 (2004).

\noindent \lbrack 52] T. Ivezi\'{c}, Found. Phys. Lett. \textbf{12,} 105
(1999).

\noindent \lbrack 53] Y. Aharonov and D. Bohm, Phys. Rev. \textbf{115}, 485
(1959).

\noindent \lbrack 54] A. N. Tonomura, T. Osakabe, T. Matsuda, T. Kawasaki,
J. Endo,

S. Yano, and H. Yamada, Phys. Rev. Lett. \textbf{56}, 792 (1986).

\noindent \lbrack 55] G. M\"{o}llenstedt and W. Bayh, Naturwissenschaften
\textbf{49}, 81 (1962).

\noindent \lbrack 56] A. Caprez, B. Barwick and H. Batelaan, Phys. Rev.
Lett. \textbf{99}, 210401 (2007).

\noindent \lbrack 57] R. G. Chambers, Phys. Rev. Lett. \textbf{5}, 3 (1960).

\noindent \lbrack 58] M. Peshkin and A. Tonomura, \textit{The Aharonov-Bohm
Effect} (Springer, New York, 1989).

\noindent \lbrack 59] V. A. Kostelecky and M. Mewes, Phys. Rev. D \textbf{66}%
, 056005 (2002).

\noindent \lbrack 60] T. Ivezi\'{c}, EJTP \textbf{10,} 131 (2006).

\noindent \lbrack 61] J. Anandan, Int. J. Theor. Phys. \textbf{19}, 537
(1980).

\end{document}